\documentclass[acmsmall]{acmart}

\usepackage{color}
\usepackage{ulem}
\usepackage{multirow}
\usepackage{graphicx}
\usepackage{enumitem}
\usepackage{sistyle}
\usepackage{algorithm}
\usepackage{algorithmic}
\usepackage{booktabs}
\usepackage{caption}
\usepackage{subfigure}
\usepackage{mathrsfs}
\usepackage{longtable}
\usepackage{filecontents}
\usepackage{bbm}
\usepackage{pgfplots}
\usetikzlibrary{patterns}
\usepackage{adjustbox}

\definecolor{c1}{RGB}{255,255,157}
\definecolor{c2}{RGB}{190, 235, 159}
\definecolor{c3}{RGB}{94,180,210}

\newcommand\blfootnote[1]{%
  \begingroup
  \renewcommand\thefootnote{}\footnote{#1}%
  \addtocounter{footnote}{-1}%
  \endgroup
}
\AtBeginDocument{%
  \providecommand\BibTeX{{%
    \normalfont B\kern-0.5em{\scshape i\kern-0.25em b}\kern-0.8em\TeX}}}

\setcopyright{acmcopyright}
\copyrightyear{*}
\acmYear{*}
\acmDOI{*}

\acmJournal{TOIS}
\acmVolume{*}
\acmNumber{*}
\acmArticle{*}
\acmYear{*}
\acmMonth{2}
\setcopyright{rightsretained}

\begin{document}

\title{Are Neural Ranking Models Robust?}

\author{Chen Wu, Ruqing Zhang, Jiafeng Guo$^{*}$, Yixing Fan, \lowercase{and} Xueqi Cheng$^{*}$}
\blfootnote{$^{*}$ Jiafeng Guo and Xueqi Cheng are corresponding authors.}
\email{{wuchen17z,zhangruqing,guojiafeng,fanyixing,cxq}@ict.ac.cn}
\affiliation{%
  \institution{
  CAS Key Lab of Network Data Science and Technology, Institute of Computing
Technology, Chinese Academy of Sciences, Beijing, China; University of Chinese Academy of Sciences, Beijing}
  \streetaddress{NO. 6 Kexueyuan South Road, Haidian District}
  \city{Beijing}
  \country{China}
  \postcode{100190}
}

\orcid{1234-5678-9012}


\renewcommand{\shortauthors}{C.Wu, et al.}

\begin{abstract}

Recently, we have witnessed the bloom of neural ranking models in the information retrieval (IR) field. So far, much effort has been devoted to developing effective neural ranking models that can generalize well on new data. 
There has been less attention paid to the robustness perspective. Unlike the effectiveness which is about the average performance of a system under normal purpose, robustness cares more about the system performance in the worst case or under malicious operations instead. 
When a new technique enters into the real-world application, it is critical to know not only how it works in average, but also how would it behave in abnormal situations. 
So we raise the question in this work: Are neural ranking models robust? To answer this question, firstly, we need to clarify what we refer to when we talk about the robustness of ranking models in IR. We show that robustness is actually a multi-dimensional concept and there are  three ways to define it in IR: 1) The \textit{performance variance} under the independent and identically distributed (I.I.D.) setting; 2) The \textit{out-of-distribution (OOD) generalizability}; and 3) The \textit{defensive ability} against adversarial operations. 
The latter two definitions can be further specified into two different perspectives respectively, leading to 5 robustness tasks in total. 
Based on this taxonomy, we build corresponding benchmark datasets, design empirical experiments, and systematically analyze the robustness of several representative neural ranking models against traditional probabilistic ranking models and learning-to-rank (LTR) models.  
The empirical results show that there is no simple answer to our question. While neural ranking models are less robust against other IR models in most cases, some of them can still win 2 out of 5 tasks. 
This is the first comprehensive study on the robustness of neural ranking models.
We believe the way we study the robustness as well as our findings would be beneficial to the IR community.
We will also release all the data and codes to facilitate the future research in this direction.

\end{abstract}



\maketitle

\section{Introduction}
\label{sec:intro}
Relevance ranking is a core problem of information retrieval (IR). 
Given a query and a set of candidate documents, a scoring function is usually learned to determine the relevance degree of a document with respect to the query. 
With the advance of deep learning technology, we have witnessed substantial growth of neural ranking models \cite{DRMM, KNRM, Conv-KNRM, Duet, BERT_FIRSTP}, achieving promising results in learning query-document relevance patterns.  
Yet, due to the hype surrounding neural ranking models, some researchers have doubted whether they are actually effective for the classic ad-hoc retrieval problem without vast quantities of training data \cite{neural_hype}. 
Recently, pre-trained language representation models such as BERT \cite{BERT} have brought breakthrough to various downstream natural language processing (NLP) tasks \cite{bert_cls_app, bert_st_app, sabert_app, bert_qa_app}. 
The success of pre-trained models in NLP has also attracted a lot of attention in the IR community.
Researchers have applied the popular models, e.g., BERT and ELMo \cite{elmo}, for ad-hoc document ranking, and shown that they can achieve  remarkable success on a broad range of ranking problems \cite{B-PROP,spanbert,sabert_app}.

However, up to now, there has been little attention paid to the robustness, an important issue beyond the effectiveness, of neural ranking models.
Unlike the effectiveness which is about the average performance of a system, robustness cares more about the worst-case performance instead. 
When a new technique enters into the real-world retrieval scenarios, it is critical to know not only how it works in average, but also how would it behave in abnormal situations.   
Therefore, in this work, we raise the following question: Are neural ranking models robust?

To answer this question, we need to first clarify what we refer to when we talk about the robustness of ranking models in IR.   
Although the robustness of neural ranking models has not been analyzed so far, there have been some related studies on the robustness of traditional probabilistic ranking models.  
For example, \citet{goren2018SIGIR} analyzed the robustness of LTR models under adversarial document manipulation, and found that increased regularization of linear ranking functions results in increased ranking robustness.   
\citet{trec2004overview} proposed to measure the robustness of traditional probabilistic ranking models by focusing on poorly performing queries. 
As we can see, the existing studies on the robustness of ranking models took quite different perspectives.
In fact, the robustness is not a simple concept, but with multi-dimensional definition as shown in the research from the machine learning (ML) community \cite{shafique2020robust, zhang2020robustnessml}. 
Therefore, if one only picks up one perspective, he/she is very likely to obtain misleading conclusions with respect to robustness of neural ranking models.
Unfortunately, there has been no comprehensive study on this question yet.

Therefore, in this paper, based on the previous work in IR and also inspired by the study of robustness in ML, we propose to define the robustness of ranking models via a comprehensive taxonomy which contains three major perspectives as shown in Figure \ref{figure: robustness taxonomy}.

Before introducing the taxonomy of the robustness of ranking models, we first describe the key notations and the formulation of the effectiveness in this work.
Suppose that $Y = \{r_1, r_2, ..., r_l\}$ is the set of ranks, where $l$ denotes the number of ranks. There exists a total order between the ranks $r_l \succ r_{l-1} \succ ... \succ r_1$, where $\succ$ denotes the order. 
Specifically, $r_1$ usually takes 0 as value, representing being irrelevant. 
Suppose that $Q = \{q_1, q_2, ..., q_m\}$ is the set of queries in training. 
Each query $q_i$ is associated with a list of documents $\mathbf{d}_i=\{d_{i1}, d_{i2}, ..., d_{i,n(q_i)} \}$ and a list of corresponding labels $\mathbf{y}_i = \{y_{i1}, y_{i2}, ... ,y_{i, n(q_i)}\}$, where $n(q_i)$ denotes the size of list $\mathbf{d}_i$, $d_{ij} \in \mathbf{D}$ denotes the $j^{th}$ document in $\mathbf{d}_i$, and $y_{ij} \in Y $ denotes the label of document $d_{ij}$.

We consider the ranking model $f$ learned on the examples $\{q_i,\textbf{d}_i,\textbf{y}_i\}_{i=1}^{m}$, which are drawn from the training distribution $\mathcal{G}$. 
In this way, for each query $q_i$, the elements in its corresponding document list $\textbf{d}_i$ can be assigned relevance scores using the model and then be ranked according to the scores. 
Let $\mathbf{d}_i$ be identified by the list of integers $\{1,2,...,n(q_i)\}$, then permutation $\pi(q_i, \mathbf{d}_i, f)$ is defined as a bijection from $\{1,2,...,n(q_i)\}$ to itself.
Ranking is nothing but to produce a permutation $\pi(q_i, \mathbf{d}_i, f)$ for the given query $q_i$ and the associated list of documents $\mathbf{d}_i$ using the ranking model. 
Given an effectiveness evaluation metric $M$, the ranking models are usually evaluated by the average performance over the test queries under the I.I.D. setting, i.e.,
\begin{equation}
	\mathbb{E}_{(q_t,\mathbf{d}_t,\mathbf{y}_t)\sim\mathcal{G}} M(\pi(q_t, \mathbf{d}_t, f), \mathbf{y}_t),
\label{equ: mean_performance}
\end{equation} 
where $q_t,\mathbf{d}_t$ and $\mathbf{y}_t$ denotes the query, the document list and the label in the test set, respectively. Specifically, the test samples are supposed to be drawn from the same distribution as the training distribution $\mathcal{G}$. 

We define the robustness of ranking models from three major perspectives as follows:

\begin{figure}[t]
\centering
\includegraphics[scale=0.52]{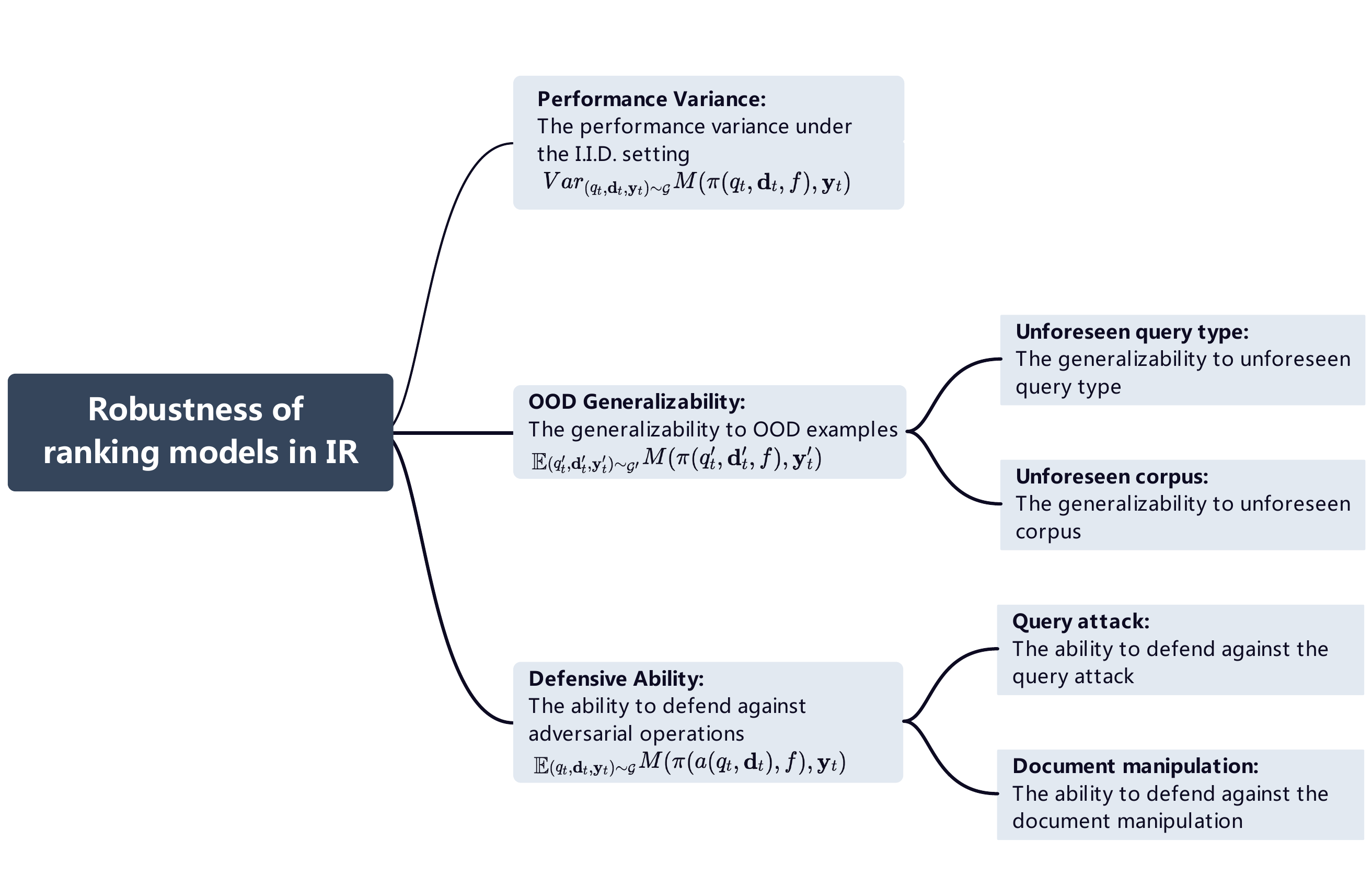}
\caption{The taxonomy of the robustness of ranking models in IR.}
\label{figure: robustness taxonomy}
\end{figure}

\begin{itemize}
\item \textbf{Performance Variance}:
The performance variance aims to analyze the robustness of ranking models by emphasizing the worst-case performance across different individual queries under the I.I.D. setting. Formally, the performance variance is defined as 
\begin{equation}
	Var_{(q_t,\mathbf{d}_t,\mathbf{y}_t)\sim\mathcal{G}} M(\pi(q_t, \mathbf{d}_t, f), \mathbf{y}_t),
\label{equ: performance variance}
\end{equation}
where $Var(\cdot)$ denotes the variance of effectiveness over all the test queries.
Besides, a special case of performance variance, i.e., the poorly-performing queries, is emphasized to analyze the ranking  robustness in the worse case.

\item \textbf{OOD Generalizability}:
The OOD generalizability aims to analyze the robustness of ranking models according to the transfer effectiveness on OOD examples.
Formally, suppose that OOD examples $q_t',\mathbf{d}_t'$ and $\mathbf{y}_t'$ are drawn from a new distribution $\mathcal{G}'$. 
The OOD generalizability is defined as 
\begin{equation}
	\mathbb{E}_{(q_t',\mathbf{d}_t',\mathbf{y}_t')\sim\mathcal{G}'} M(\pi(q_t', \mathbf{d}_t', f), \mathbf{y}_t'),
\label{equ: OOD generalizability}
\end{equation}

Specifically, the OOD generalizability can be further defined in two ways, i.e., OOD generalizability on unforeseen query types and OOD generalizability on unforeseen corpus.

\item \textbf{Defensive Ability}: 
The defensive ability aims to analyze the robustness of ranking models according to their ability to defend against adversarial operations.
Given an adversarial attack function $a$ for the query and document, the defensive ability is formalized as
\begin{equation}
	\mathbb{E}_{(q_t,\mathbf{d}_t,\mathbf{y}_t)\sim\mathcal{G}} M(\pi(a(q_t, \mathbf{d}_t), f), \mathbf{y}_t), 
\label{equ: defensive ability}
\end{equation} 
Specifically, the defensive ability can be measured with respect to two types of adversarial operations, i.e., query attack and document manipulation. 
\end{itemize}

Based on this taxonomy, we design the corresponding experiments, and conduct empirical studies to analyze the robustness of several representative neural ranking models against traditional probabilistic ranking models and LTR models. All the experimental datasets and codes\footnote{The experimental datasets and codes will be available at URL.} used in our study would be publicly available for the research community.
Specifically, the design of our empirical experiments for each task are as follows. 
\begin{itemize}
\item  To facilitate the study of the performance variance, we conduct experiments on three ad-hoc retrieval datasets, i.e., MQ2007 \cite{letor4.0}, Robust04, and MS MARCO, and follow the previous works \cite{trec2004overview,zhang2013bias} to measure the robustness of different ranking models.

\item To facilitate the study of the OOD generalizability, we first build two  benchmark datasets for unforeseen query type and unforeseen corpus, respectively. 
Specifically, we construct the benchmark dataset for unforeseen query type by splitting the MS MARCO dataset into 5 sub-datasets, with respect to the official query type.   
Besides, we construct the benchmark dataset for unforeseen corpus based on three  ad-hoc retrieval corpora, i.e., MQ2007, Robust04, and MS MARCO. 
Furthermore, we propose the drop rate metric to measure the OOD generalizability of different ranking models.

\item To facilitate the study of the defensive ability, we first build a  benchmark dataset for query attack. Specifically, we introduce four types of character-level edits \cite{jones2020NLProbustEncoding, pruthi2019combating} and three types of word-level edits to attack the query in the MS MARCO dataset.  
Then, we propose the drop rate metric to measure the defensive ability against  query attack.
For document manipulation, we conduct experiments on the ASRC dataset \cite{raifer2017SIGIR} and follow the previous work \cite{goren2018SIGIR} to measure the robustness of different ranking models.

\end{itemize}

The empirical results demonstrate that:
1) Neural ranking models are in general not robust as compared with other IR models. The finding is consistent with the previous study \cite{ml_complex_robust} in ML which demonstrates that the complexity in the model could lead to robustness issues.
2) Pre-trained ranking models exhibit the best robustness  
  against all the other models from the perspective of the performance variance. 
3) Some neural ranking models show the superiority of the defensive ability. For example, DSSM, Duet and Conv-KNRM are robust to the query attack. 
4) Based on the empirical evaluation results, there remains clearly room for future robustness improvements.

We organize this work as follows. 
We first introduce the ranking models which are evaluated through all the robustness metrics, including traditional probabilistic ranking models, LTR models and neural ranking models in Section \ref{sec:model}.
We then analyze the robustness of ranking models in terms of the performance variance, the OOD generalizability, and the defensive ability, in Section \ref{sec:performance variance}, Section \ref{sec:OOD} and Section \ref{sec:adv} respectively.   
Finally, we briefly review the related work in Section \ref{sec:related work} and conclude this work in Section \ref{sec:con}.

\section{Ranking Models}
\label{sec:model}

In this paper, we aim to investigate whether the neural ranking models are robust. 
Specifically, we adopt three types of representative neural ranking models for analysis, i.e., representation-focused deep matching models,
 interaction-focused deep matching models, hybrid deep matching models and advanced pre-trained models. 

\begin{itemize} 

\item \textbf{Representative-focused Deep Matching Models} include,

\begin{itemize}
	\item \textbf{DSSM}: DSSM \cite{DSSM} is a representative-focused deep matching model designed for Web search, which contains a letter n-gram based word hashing layer, two non-linear hidden layers and an output layer.
\end{itemize}

\item \textbf{Interaction-focused Deep Matching Models} include,

\begin{itemize}
 \item \textbf{DRMM}: DRMM \cite{DRMM} is an interaction-focused deep matching model designed for ad-hoc retrieval. It consists of a matching histogram mapping, a feed forward matching network and a term gating network. In this paper, we use LogCount-based Histogram (LCH) as the matching histogram mapping and Term Vector (TV) as the term gating function.

 \item \textbf{Conv-KNRM}: Conv-KNRM \cite{Conv-KNRM} is another popular interaction-focused deep matching model, which models n-gram soft matches for ad-hoc retrieval based on convolutional neural networks (CNN) and kernel-pooling. 

\end{itemize}
 
\item \textbf{Hybrid Deep Matching Models} include, 
\begin{itemize}
 \item \textbf{Duet}: Duet \cite{Duet} is a hybrid deep matching model which combines both the representation-focused architecture and the interaction-focused architecture. Specifically, it contains two separate deep neural networks, one that matches using a local representation of text, and another that learns a distributed representation before matching. 
 
\end{itemize}

\item \textbf{Pre-trained Models} include, 

\begin{itemize}
 \item \textbf{BERT}: The key technical innovation of BERT \cite{BERT} is applying the multi-layer bidirectional Transformer encoder architecture for language modeling.
BERT uses two different types of pre-training objectives including Masked Language Model (MLM) and Next Sentence Prediction (NSP). 
In this paper, the query and the document are concatenated as the input to BERT with special tokens delimiting them, i.e., [CLS] and [SEP]. 
To obtain the relevance score of the document to a given query, we apply a sigmoid function over the representation of [CLS] following previous studies \cite{monoBERT, BERT_FIRSTP}.

 \item \textbf{ColBERT}: ColBERT \cite{ColBERT} employs contextualized late interaction over deep language models (in particular, BERT) for efficient retrieval. It independently encodes queries and documents into fine-grained representations that interact via cheap and pruning-friendly computations. 
Here, we analyze ColBERT's ability to re-rank documents following  the original paper. 

\end{itemize}

\end{itemize}

Specifically, we use the implementations of DRMM, Conv-KNRM, and Duet from MatchZoo \cite{matchzoo}.   
We adopt the parameters of BERT$_{base}$ released by Google\footnote{https://github.com/google-research/bert} to initialize BERT. 
For ColBERT, we use the original code\footnote{https://github.com/stanford-futuredata/ColBERT/} released by the authors for implementation.

In order to better analyze the robustness of neural ranking models, we adopt 
three types of LTR models for comparison, including pointwise LTR models, 
pairwise LTR models and listwise LTR models.

\begin{itemize}
	\item  \textbf{Pointwise LTR Models} include, 
	\begin{itemize}
	\item  \textbf{Prank}: Prank \cite{Prank} is a famous pointwise LTR model on ordinal regression. 
	\end{itemize}
\end{itemize}

\begin{itemize} 
\item \textbf{Pairwise LTR Models} include, 
\begin{itemize} 
 \item \textbf{RankSVM}: RankSVM \cite{RankSVM} is a representative pairwise LTR model based on Structural Support Vector Machine (SVM). 
\end{itemize} 
\end{itemize}

\begin{itemize} 
\item \textbf{Listwise LTR Models} include, 
\begin{itemize}
 \item \textbf{LambdaMART}: LambdaMART \cite{LambdaMART} is a state-of-the-art listwise LTR algorithm which uses  gradient boosting to produce an ensemble of retrieval models. 
\end{itemize} 
\end{itemize}

For Prank, we implement it according to the original paper since there is no publicly available code.
For RankSVM, we directly use the implementation in SVM$^{rank}$  \cite{svm_rank}.    
LambdaMART is implemented using RankLib\footnote{https://sourceforge.net/p/lemur/wiki/RankLib/}, which is a widely used LTR tool.

Besides, we compare the above neural models with two representative traditional probabilistic ranking models, including, 

\begin{itemize} 
 \item \textbf{QL}: Query likelihood model based on Dirichlet smoothing \cite{ql_dir} is one of the best performing language models. 
 \item \textbf{BM25}: The BM25 formula \cite{bm25} is another highly effective retrieval model that represents the classical probabilistic retrieval model. 
\end{itemize}

Specifically, we use the implementations of QL and BM25 from Anserini  \cite{yang2018anserini} toolkit\footnote{http://anserini.io/}.



\section{Performance Variance under the I.I.D. Setting}
\label{sec:performance variance}

Most ranking models are designed under the simple assumption that the observations are from I.I.D. random variables, and focus on improving the average effectiveness of retrieval results. 
Recently, it has been recognized that, when we attempt to improve the mean retrieval effectiveness over all queries, the stability of performance across different individual queries could be hurt \cite{zhang2013bias}.   
Therefore, in this section, we analyze the robustness of ranking models by emphasizing the worst-case performance across different individual queries, i.e., the performance variance under the I.I.D. settings. 
In the following, we first introduce the definition and metric of the performance variance, and then conduct experiments to analyze the robustness of ranking models.

\subsection{Definition of Performance Variance}
The performance variance of ranking models refers to the variance of effectiveness across different individual queries, which has been formulated in Eq. (\ref{equ: performance variance}). 
When a ranking model achieves improvements in mean retrieval effectiveness (e.g., mean average precision (MAP) \cite{MAP}), the performance of some individual queries could be hurt \cite{zhang2013bias}. 
Although failures on a small number of queries may not have a significant effect on the average performance, users who are interested in such queries are unlikely to be tolerant of this kind of deficiency. 
Accordingly, an ideal ranking model is expected to balance the trade-off between effectiveness and robustness by achieving high average effectiveness and low variance of effectiveness \cite{zhang2014bias_qikan}.

\subsection{Metric of Performance Variance}

Here, we propose the variance of normalized average precision (VNAP) to measure the performance variance, which is defined as,
\begin{equation}
  VNAP =\mathbb{E}(NAP(q_t)-\mathbb{E}(NAP(q_t)))^2, 
\end{equation}
where the expectation $\mathbb{E}(\cdot)$ is over a set of queries that are assumed to be uniformly distributed. $NAP(q_t)$ denotes the normalized average precision with respect to the query $q_t$, which is defined as,
\begin{equation}
	NAP(q_t) = \frac{AP(q_t)}{\mathbb{E}(AP(q_t))},
\end{equation}
where $AP(q_t)$ denotes the average precision with respect to the query $q_t$, which is defined as,
\begin{equation}
 AP(q_t) =\frac{1}{R_{q_t}}\sum_{k=1}^{R_{q_t}}\frac{1}{o_k}\sum_{n=1}^{o_k}\delta(y_{tn}>0),
 \label{equ:AP}
\end{equation}   
where $R_{q_t}$ denotes the number of relevant documents to the query $q_t$. 
$o_k$ is the rank of the relevant document $k$ predicted by the ranking model, which ranges from 1 to the size of document list. 
$y_{tn}$ denotes the ground-truth label of document $d_{tn}$. 
$\delta$ is the indicator function, which aims to count the number of relevant documents. 

Note that VNAP is similar to VAP \cite{zhang2013bias}. The difference between them is that we have normalized the average precision to eliminate the influence of the mean performance. Since different models are likely to have different mean performance, it is necessary to eliminate the influence of the mean performance to better measure the variance of ranking models.
The ranking model would be more robust with a lower VNAP value.

\subsection{Experimental Settings}

In this section, we introduce our experimental settings, including datasets and implementations details.

\subsubsection{Datasets}
\label{MQ2007_data}

To evaluate the performance variance of different ranking models, we conduct experiments on several representative ad-hoc retrieval datasets, i.e., 
\begin{itemize}
\item \textbf{MQ2007}.  \textbf{M}illion \textbf{Q}uery Track \textbf{2007} (MQ2007) is a LETOR~\cite{letor4.0} benchmark dataset based on the GOV2 collection which includes 25 million documents in 426 gigabytes. Queries for MQ2007 are divided into 5 folds as described in LETOR4.0. 
\item \textbf{Robust04}. Robust04 contains 250 queries and 528k news articles, whose queries are collected from TREC 2004 Robust Track\footnote{https://trec.nist.gov/data/robust.html}. There are about 70 relevant documents (news articles) for each query.
\item \textbf{MS MARCO}. \textbf{M}icro\textbf{S}oft \textbf{MA}chine \textbf{R}eading \textbf{CO}mprehension Document Ranking Dataset (MS MARCO)\footnote{https://github.com/microsoft/MSMARCO-Document-Ranking} is a large-scale benchmark dataset for web document retrieval. MS MARCO contains 4 million documents and 0.37 million training queries, where each query has one relevant document. Evaluation was on the dev set, which contains 5193 queries\footnote{The relevance assessments for the official test set were not public at the time the paper was written.}.
 \end{itemize}

\begin{table}[t]
\centering 
 \renewcommand{\arraystretch}{1.6}
 \setlength\tabcolsep{5pt}
 \caption{Statistics of datasets used for evaluating the performance variance.}
\begin{tabular}{lccc} 
\toprule \hline
 & MQ2007  & Robust04& MS MARCO  \\
\hline
\#Queries & 1692 & 250   & 0.37M \\
\#Documents & 65,323 &  528k & 3.21M \\
Avg document length & 3587 &  471 & 1129 \\
Avg \#Relevant Documents & 10 & 70 & 1  \\

\hline \bottomrule
\end{tabular}
\label{table:PV dataset}
\end{table}
 
The detailed statistics of datasets are shown in Table \ref{table:PV dataset}. 
We choose these three datasets according to three criteria:
1) The datasets are public and the original document contents are available, 
2) The query numbers are diverse (e.g., 250 queries in Robust04 and 0.37 million queries in MS MARCO), and  
3) The literary style of the documents ranges from newswire articles to Web document.

\subsubsection{Implementation Details}

\label{Implementation}

In preprocessing, all the words in the documents and queries are white-space tokenized, lower-cased, and stemmed using the Krovetz stemmer \cite{krovetz_stemmer}. 
For the MQ2007 dataset, we use the data partition and top 40 ranked documents released by the official LETOR4.0 \cite{letor4.0}. 
For the Robust04 dataset, an initial retrieval is performed using the Anserini toolkit with the QL model to obtain the top 100 ranked documents.  
For the MS MARCO dataset, we use the official top 100 ranked documents retrieved by the QL model.

For traditional probabilistic ranking models,   
we adopt the static parameters suggested by \cite{huston2014parameters} on the Robust04 dataset and tune the parameters on the corresponding validation set on the MQ2007 and MS MARCO dataset\footnote{$\mu$ in 1-2000 (by 10) for QL and $k_1$ in 0.1-5 (by 0.1), $b$ in 0.1-1 (by 0.1) for BM25. }.
For LTR models, we use standard features released by LETOR 4.0 for MQ2007. 
In LETOR, there are 46 hand-crafted features in total, among which 42 features are constructed based on the textual elements (e.g., term frequencies, BM25 and language model scores based on title, body, anchor texts, and URL), and 4 features are based on link analysis (e.g., PageRank, inlink number, outlink number, and number of child page).
Due to the lack of officially released LTR features, we construct the LTR features for Robust04 and MS-MARCO respectively, by computing the TF, IDF, TF-IDF, Document Length, BM25 score, QL-DIR score and QL-JM scores on the title, url, body and document respectively as the final 28-dimensional LTR features following \cite{letor4.0}. 
 For RankSVM, we use the linear kernel with the hyper-parameter C selected from the range [0.0001, 10] on the validation set for three datasets following \cite{Conv-KNRM}.
For LambdaMART, we select the number of trees (in 100-500, by 10), leafs (in 2-20, by 2) and shrinkage coefficient (in \{0.001, 0.005, 0.01, 0.05, 0.1\}) on the validation set for three datasets.

For neural ranking models, 
  the model-specific hyper-parameters are as follows.
For DSSM, we use a three-layer DNN and set the node number of each layer as 100, 100 and 50 to avoid overfitting on Robust04 and MQ2007 datasets following \cite{deeprank}. We set the node number of each layer as 300, 300, 128 on the MS MARCO dataset.
For DRMM, we use a four-layer architecture and set the node number of each layer as 30 (histogram bins), 5, 1 and 1 on Robust04 and MQ2007 datasets following \cite{DRMM, deeprank}. We set the node number of each layer as 30 (histogram bins), 10, 1 and 1 on MS MARCO dataset.
For Conv-KNRM, we set the n-gram size as 3 and the number of CNN filters as 128. For kernel-pooling, we set the number of kernels to 11 with the mean values $\mu$ of the Gaussian kernels varying from $-1$ to $+1$, and standard deviation $\sigma$ of 0.1 for all kernels (the first kernel's $\sigma$ is set as 0.001 for exact matching ) on three datasets following \cite{Conv-KNRM, hofstatter2020interpretable}.
For Duet, we used 10 filters with both the local and distributed model, with hidden dimensions set to 30 and 699, respectively on Robust04 dataset following \cite{yates2020capreolus}.
We set the filter size as 10 in both local model and distributed model, and the hidden size as 20 in the fully-connected layer on MQ2007 dataset following \cite{HiNT}. 
We set the filter size as 32 in both local and distributed model and the hidden size as 32 in the fully-connected layer on MS MARCO dataset.
The learning rates of above models are tuned on the corresponding validation set in the range of [1e-5, 5e-3].
For BERT, we choose the fine-tuning learning rate from \{5e-5, 3e-5, 2e-5\} as recommended in \cite{BERT} on three datasets.
For ColBERT, we set the fine-tuning learning rate as 3e-6 on three datasets following \cite{ColBERT}.
For all neural ranking models, we use the Adam \cite{kingma2014adam} optimizer.
We use LTR and neural ranking models to re-rank the top candidate documents.

\subsection{Empirical Results on Performance Variance}

In this section, we analyze the empirical results on the performance variance under the I.I.D. setting. 
Specifically, we first analyze the average ranking effectiveness and the variance of effectiveness of different ranking models, respectively. 
Furthermore, we analyze the relationship between the average ranking effectiveness and the variance of effectiveness. 

\begin{table*}[t]
\small
\centering 
 \renewcommand{\arraystretch}{2}
 \setlength\tabcolsep{1pt}
 \caption{  The average ranking effectiveness over all the queries on the Robust04, MQ2007 and MS MARCO dataset in terms of MAP, NDCG@10, NDCG@20, P@10, P@20, R@10, R@20, MRR@10 and MRR@100.}
\begin{tabular}{c  c c c c  c c c c  c c c }  \toprule \hline
\multirow{4}{*}{Model}  & \multicolumn{4}{c}{Robust04} & \multicolumn{4}{c}{MQ2007} & \multicolumn{3}{c}{MS MARCO} \\
\cmidrule(r){2-5} \cmidrule(r){6-9} \cmidrule(r){10-12} 
 & MAP & NDCG@20 & P@20 & R@20 & MAP & NDCG@10 & P@10 & R@10 & MRR@10 & MRR@100 & R@10 \\
 \hline
 QL & 0.2109 & 0.4113 & 0.3516  & 0.2094 & 0.3928 & 0.3891 & 0.3348 & 0.3192 & 0.2144 & 0.2271 & 0.4489  \\  
BM25 & 0.2162 & 0.4200 & 0.3594 & 0.2105 & 0.3991 & 0.4000 & 0.3435 & 0.3293 & 0.2184 & 0.2301 & 0.4439    \\ \hline
Prank & 0.1190 & 0.2181 & 0.2157 & 0.1168 & 0.3650 & 0.3162 & 0.3008 & 0.2665 & 0.2184 & 0.2311 & 0.4645\\
RankSVM & \textbf{0.2179} & 0.4234 & 0.3610  & 0.2120 & 0.4637 & \textbf{0.4428} & 0.3811 & 0.3657 & 0.2398 & 0.2522 & 0.4818 \\                    
LambdaMart & 0.2171 & 0.4218  & 0.3584  & 0.2116 & \textbf{0.4646}  & 0.4401  &  0.3741 & 0.3633 & 0.2602 & 0.2710 & 0.4986 \\   \hline
DSSM & 0.1216 & 0.2306 & 0.2283 & 0.1198 & 0.3998 & 0.3606 & 0.3421 & 0.3070 & 0.1051 & 0.1220 & 0.2070 \\
DRMM & 0.1649 & 0.3096 & 0.2748 & 0.1585  & 0.4317  & 0.4009 & 0.3662  & 0.3338 & 0.1168 & 0.1336 & 0.3046 \\                       
Conv-KNRM & 0.1332  &  0.2526  &  0.2426 & 0.1220 & 0.3957  & 0.3553  & 0.3436  & 0.3047 & 0.2075 & 0.2209 & 0.4435 \\                                       
Duet  &  0.1308  &  0.2503  & 0.2421  & 0.1299 & 0.4089  & 0.3727  & 0.3561  & 0.3223 & 0.1233 & 0.1402 & 0.3054  \\                       
BERT  & 0.2108   & \textbf{0.4330}   & \textbf{0.3808}  & \textbf{0.2127} & 0.4620  & 0.4420  & \textbf{0.3878}  & \textbf{0.3708} & 0.3093  & 0.3181 & 0.5748 \\                      
ColBERT &  0.1989  & 0.4010   & 0.3594  & 0.2043 & 0.4420  & 0.4153  & 0.3752  & 0.3529 & \textbf{0.3469} & \textbf{0.3541} & \textbf{0.6212}  \\ 
 
 \hline \bottomrule
    \end{tabular}

\label{table: sec3_effectiveness}
\end{table*}

\subsubsection{\textbf{How does different ranking models perform in terms of the average effectiveness?}} 
\label{average effectiveness}

 To better understand the performance variance, we first analyze the average ranking effectiveness over the queries.  
Following \cite{DRMM, HiNT}, we take the topic ``title'' as queries and conduct 5-fold cross-validation on Robust04 and MQ2007 datasets to minimize over-fitting without reducing the number of learning instances.
For Robust04 dataset, we use the mean average precision (\textbf{MAP}) \cite{MAP}, normalized discounted cumulative gain at rank 20 (\textbf{NDCG@20}) \cite{nDCG}, and precision at position 20 (\textbf{P@20}) \cite{Precision} as the evaluation metrics following \cite{DRMM}.
For MQ2007 dataset, we report the \textbf{P@10} and \textbf{NDCG@10} following \cite{HiNT}.
Moreover, we also use the recall at position 20 (\textbf{R@20}) \cite{Recall} and recall at position 10 (\textbf{R@10}) as the recall oriented evaluation metrics for Robust04 and MQ2007, respectively.
For MS MARCO dataset, we report the Mean Reciprocal Rank at 100 (\textbf{MRR@100} \cite{MRR}), which is suggested in the official instructions. We also report the \textbf{MRR@10} and \textbf{R@10} for MS MARCO dataset.

The main results are shown in Table \ref{table: sec3_effectiveness}.  From the results, we can find that: 
(1) For the two traditional probabilistic ranking models, we can see that \textit{BM25} is a strong baseline which performs better than \textit{QL} in most cases.  
(2)   For LTR models, RankSVM and LambdaMart perform better than the traditional probabilistic ranking models.  
It is not surprising since LTR models combine various features including the two  traditional probabilistic ranking models (i.e., BM25 and QL scores), which are capable of characterizing the relevance between a document and a query. 
However,   Prank performs worse than the traditional probabilistic ranking models on the Robust04 and MQ2007 dataset.
The reason might be that pointwise LTR ignores the fact that some documents are associated with the same query and some others are not \cite{liu2011learning}. 
(3) Most neural ranking models perform worse than traditional probabilistic ranking models and LTR models.  
The reason might be that it is difficult for a deep neural model to train from scratch with such a few supervised pairs \cite{qin2021neural}. 
 (4) Pre-trained models (i.e., \textit{BERT} and \textit{ColBERT}) perform the best on   all the three datasets. 
These results indicate that the rich language information from text captured by the pre-trained model is useful for relevance modeling.


\begin{figure}[t]
\centering
\includegraphics[scale=0.6]{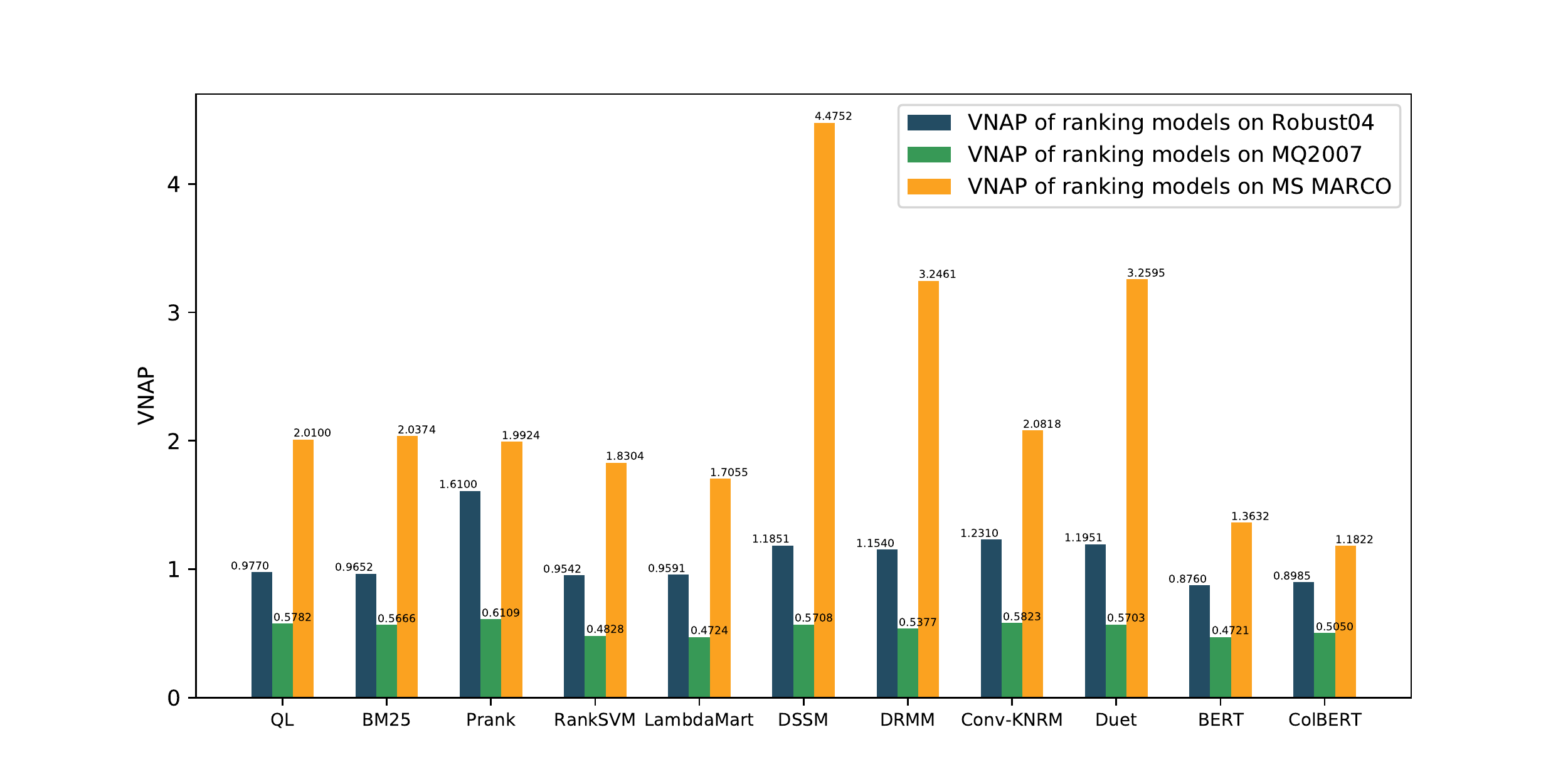}
\caption{ The performance variance of different ranking models on the Robust04, MQ2007 and MS MARCO dataset in terms of VNAP.}
\label{figure: sec3_VNAP}
\end{figure}

\subsubsection{ \textbf{How does different ranking models perform in terms of the variance of normalized average precision?}}
 
The performance variance comparisons among different ranking models in terms of VNAP are shown in Figure \ref{figure: sec3_VNAP}. Recall that high values of VNAP attest to decreased ranking robustness.

From the results, we can observe that:
(1) For the two traditional probabilistic ranking models, BM25 achieves lower VNAP values than QL on the Robust04 and MQ2007 datasets, indicating that BM25 is more robust than QL on these datasets. 
(2) For the LTR models, RankSVM and LambdaMart achieve lower VNAP than traditional probabilistic ranking models on the three datasets, indicating that LTR models are more robust than traditional probabilistic ranking models.
(3) For neural ranking models, DSSM achieves the highest VNAP on the MS MARCO dataset. The reason might be that DSSM achieves poor mean performance on the MS MARCO dataset and the corresponding VNAP is relatively higher.
(4) When we look at the pre-trained models (i.e., BERT and ColBERT), we find they achieve the lowest VNAP among all the ranking models on the three datasets.
It is interesting to find that pre-trained models exhibit the strongest robustness in terms of VNAP. 
As shown in Table \ref{table: sec3_effectiveness}, the pre-trained models have shown great success on the average effectiveness. 
The result indicates that pre-trained models present a great potential since it performs well in terms of the mean performance and the robustness (e.g, in terms of the variance).

\subsubsection{\textbf{What is the relationship between the average effectiveness and the variance of effectiveness?}}
Here, we analyze the relationship between the average effectiveness and the robustness.  
Specifically, we consider the variance of effectiveness in Figure \ref{figure: sec3_VNAP} and the average effectiveness in Table \ref{table: sec3_effectiveness} simultaneously. 
Overall, the robustness of different models generally increases (i.e., the VNAP decreases) with the increase of the average ranking effectiveness (i.e., the MAP increases). 
In addition, we have computed the correlation between the effectiveness and the variance.
For example, the Pearson correlation coefficient between the MAP and VNAP is $-0.9869$.
The result indicates that there is a strong correlation between the average effectiveness and the variance of effectiveness.

\subsection{Analysis on Poorly-performing Queries}

The previous experiments have analyzed the performance variance over all the queries. 
 A natural question is how the ranking models perform in the worse case, which can be well evaluated on the poorly-performing queries.  
Therefore, in this section, we emphasize on the poorly-performing queries to analyze the robustness of the ranking models, which refer to the queries that have the poor ranking performance among all the test queries. 
 We first introduce the evaluation metric of the poorly-performing queries, and then analyze the experimental results.

\subsubsection{Metric of Poorly-performing Queries}

To evaluate the robustness of ranking models in terms of the poorly-performing queries, we use two metrics following the previous works \cite{trec2004overview}.
 
\begin{itemize}
 \item \textbf{\%no} denotes the percentage of queries with no relevant documents in the top 10 retrieved, i.e.,   
 \begin{equation}
  \%no=\frac{1}{|Q|}\sum_{t=1}^{|Q|}\prod_{n=1}^{10}\delta(y_{tn}=0),
 \end{equation}
where $y_{tn} \in Y$ is the relevance label (i.e., the larger the relevance label, the more relevant the query-document pair) of the top $n \in [1,10]$  document with respect to the query $q_t$.
 $\delta$ is the indicator function and $|Q|$ is the total number of evaluated queries. The ranking model would be more robust with a lower  \%no value.

 \item \textbf{gMAP} denotes the geometric mean average precision, which is defined as,  
 \begin{equation}
  gMAP =\text{exp}(\frac{1}{|Q|}\sum_{t=1}^{Q}\text{log}(AP(q_t)+\epsilon)) - \epsilon, \\
  \label{equ:gmap}
 \end{equation}
 where $AP(q_t)$ is formulated as in Eq. (\ref{equ:AP}).
$\epsilon$ is a minimal positive since the average precision score for a single query may approximate to zero \cite{trec2005overview}. The ranking model would be more robust with a higher gMAP value. 

 \end{itemize}

 \begin{table*}[t]
\centering 
 \renewcommand{\arraystretch}{1.6}
 \setlength\tabcolsep{13pt}
 \caption{ The performance of poorly-performing queries on the Robust04, MQ2007, and MS MARCO dataset in terms of \%no and gMAP.}
 
\begin{tabular}{c  c c  c c  c c }  \toprule \hline
\multirow{4}{*}{Model}  & \multicolumn{2}{c}{Robust04} & \multicolumn{2}{c}{MQ2007} & \multicolumn{2}{c}{MS MARCO} \\
\cmidrule(r){2-3} \cmidrule(r){4-5} \cmidrule(r){6-7} 
 & \%no & gMAP & \%no & gMAP & \%no & gMAP \\
 \hline
 QL & 11.24  & 0.0933  & 26.25 & 0.0802 & 55.11 & 0.0144 \\  
BM25 & \textbf{8.02} & 0.0952  & 26.84 & 0.0805 & 55.61 & 0.0108 \\ \hline
Prank & 9.22 & 0.0947 & 28.43 & 0.0760 & 53.55 & 0.0141 \\
RankSVM & 9.63  & 0.0965  & 22.39 & 0.0972 & 51.82 & 0.0154\\                    
LambdaMart &  8.42 & 0.0961  & 22.28 & 0.0978  & 50.14  & 0.0160\\   \hline
DSSM & 22.11 & 0.0631 & 25.71 & 0.0821 & 79.30 & 0.0065\\
DRMM &  20.51 & 0.0711  & 24.58  & 0.0889  & 69.54 & 0.0086\\                       
Conv-KNRM &  23.29  &  0.0609  & 25.41  & 0.0814 & 55.65  & 0.0135 \\                                       
Duet  &  26.92  &  0.0585  & 24.35  & 0.0840 & 69.46  & 0.0087 \\                       
BERT  &  9.23  &  \textbf{0.1067}  & \textbf{20.98}  & \textbf{0.0979}  & 42.52  & 0.0205\\                      
ColBERT & 14.86  & 0.0949  & 22.70  & 0.0919  & \textbf{37.88}  & \textbf{0.0237}\\ 
 
 \hline \bottomrule
    \end{tabular}

\label{table: sec3_Poorly-performing queries evaluation}
\end{table*}

\subsubsection{Empirical Results on Poorly-performing Queries}

Here, we first measure the performance of the poorly-performing queries in terms of \%no and gMAP. 
Then, we analyze the relationship between the performance of all the queries and the performance of the poorly-performing queries.

\begin{itemize}

\item {\textbf{How does different ranking models perform over poorly-performing queries?}} We first focus on the poorly-performing queries to analyze the robustness of different ranking models in terms of \%no and gMAP. The main results are shown in Table \ref{table: sec3_Poorly-performing queries evaluation}. Recall that low values of \%no and high values of gMAP attest to increased ranking robustness. 
From the results, we can observe that: 
(1) For traditional probabilistic ranking models, \textit{QL} achieves a lower \%no value than BM25 on the   MQ2007 and MS MARCO, demonstrating that \textit{QL} is more robust than BM25 over the poorly-performing queries.
(2)   For LTR models, RankSVM and LambdaMart achieve lower \%no values and higher gMAP values as compared with traditional probabilistic ranking models on the MQ2007 and MS MARCO.     
The results indicate that combining different kinds of human knowledge (i.e., relevance features) could achieve good improvements on the ranking robustness over the poorly-performing queries. 
 (3) When we look at the pre-trained models (i.e., \textit{BERT} and \textit{ColBERT}), we find that \textit{BERT} achieves the lowest \%no value and the highest gMAP value among all the models on the MQ2007 dataset, while \textit{ColBERT} achieves the lowest \%no value  and the highest gMAP value on the MS MARCO dataset.    
It is interesting that pre-trained models exhibit the strongest robustness over the poorly-performing queries under the I.I.D. setting.

\begin{figure*}[h]
\centering

\subfigure[gMAP against map]{  
\begin{minipage}{6.8cm}      
\centering 
\includegraphics[scale=0.44]{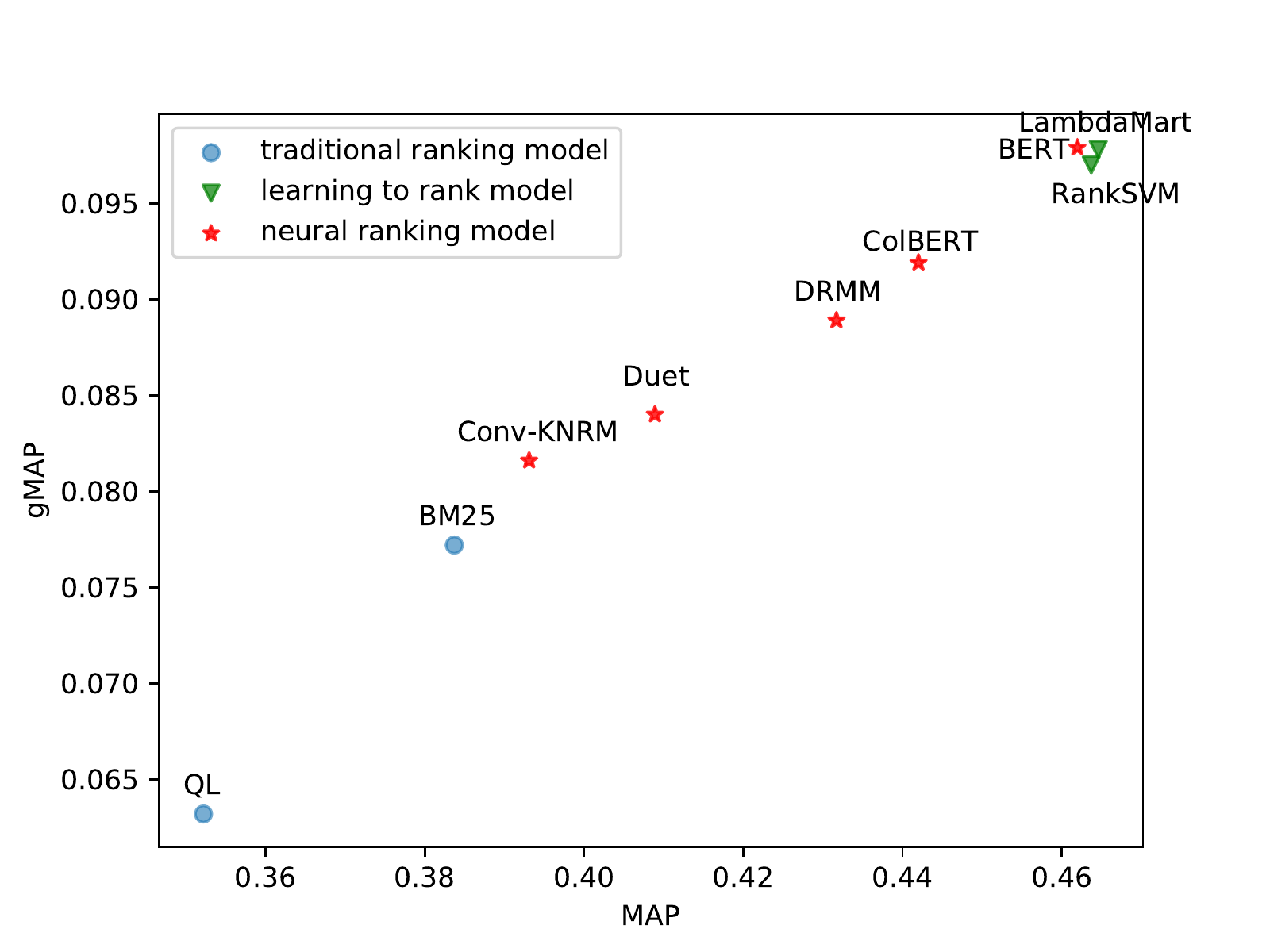} 
\end{minipage}
}
\subfigure[\%no against map]{
\begin{minipage}{6.5cm}    
\centering
\includegraphics[scale=0.44]{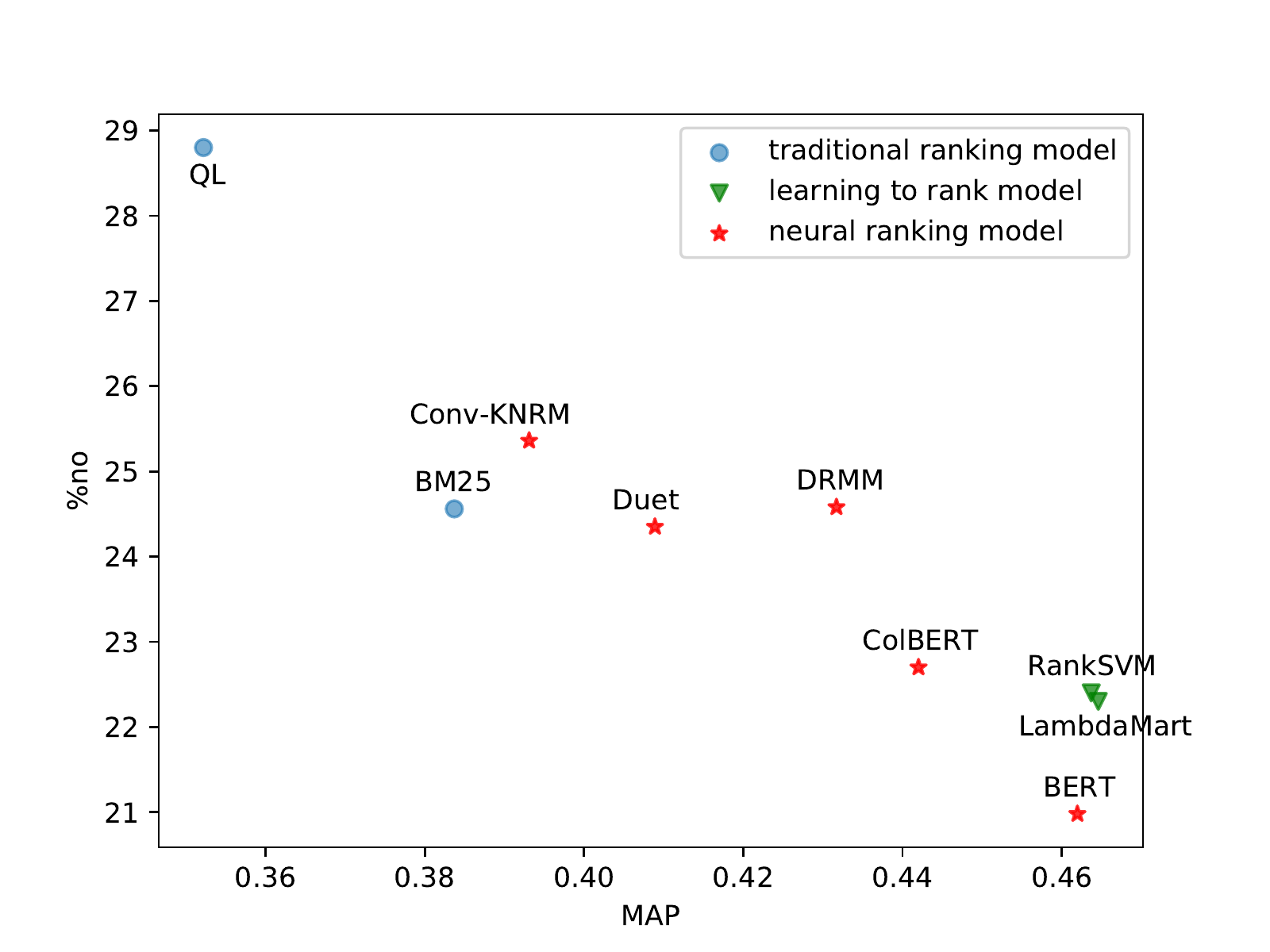}     
\end{minipage}
}
\caption{The relationship between the effectiveness (e.g., MAP) and the robustness over the poorly-performing queries (e.g., gMAP and \%no) on the MQ2007 dataset.} 
\label{figure:poorly-performing queries evaluation}  
\end{figure*}

\item{\textbf{What is the relationship between the average effectiveness and the robustness over the poorly-performing queries?}}  
Specifically, we visualize the experimental results on the MQ2007 dataset from Table \ref{table: sec3_effectiveness} and \ref{table: sec3_Poorly-performing queries evaluation} into Figure \ref{figure:poorly-performing queries evaluation}, where the horizontal axis represents the average effectiveness metric (i.e., MAP) and the vertical axis represents the robustness metric (i.e., gMAP and \%no). 
We have the following observations: 
(1) The robustness over the poorly-performing queries generally increases (i.e., the gMAP increases), with the increase of the average effectiveness (i.e., the MAP increases).   
The reason might be that the good performance over poorly-performing  queries, in turn, would improve the performance over all the queries. 
This is consistent with the finding in previous studies where the degree of ranking robustness is positively correlated with retrieval performance \cite{cikm06_ranking_robustness}. 
(2) The pre-trained models are the most robust model over the poorly-performing queries and the most effective model over all the queries under the I.I.D. setting. Future work could propose new self-supervised objectives tailored for IR that enhance model robustness.

\end{itemize}

\section{OOD Generalizability}
\label{sec:OOD}

As described above, most evaluations in IR assume that the train and test examples are I.I.D.. 
 Some advanced ranking models (e.g., pre-trained model) can achieve promising performance on some IR tasks \cite{B-PROP,spanbert,sabert_app}. 
However, in the real-world scenarios, the train and test distributions are often not identically distributed caused by the data bias \cite{torralba2011unbiased}. 
Without any special supervision, the ranking models may come at the cost of poor generalization and performance on examples not observed during training \cite{cohen2018cross}.


Therefore, in this section, we analyze the robustness of ranking models according to the transfer effectiveness on OOD examples, i.e.,  the OOD generalizability which has been formulated in Eq. (\ref{equ: OOD generalizability}). 
 Specifically, we introduce two ways to define the OOD generalizability, i.e., 1) The OOD generalizability on unforeseen query type, and 2) The OOD generalizability on unforeseen corpus.

\subsection{OOD Generalizability on Unforeseen Query Type}

In this section, we introduce the definition and metric of the OOD generalizability on unforeseen query type.

\subsubsection{Definition of OOD Generalizability on Unforeseen Query Type}

The OOD generalizability on unforeseen query type refers to the transfer effectiveness on unseen query types.    
Nowadays, users formulate their queries in the form of various types \cite{broder2002taxonomy} that can describe their information needs properly to find the right results quickly. 
When submitting a new query type to a search engine, it tends to fail due to the large gap between normal in-distribution and undesired OOD query types.  
To improve users' satisfaction, a ranking model should be robust to such new query types.

Formally, given a ranking model $f$ learned on $\{q_i,\textbf{d}_i,\textbf{y}_i\}_{i=1}^{m}$ which are drawn from the training distribution $\mathcal{G}_q$, we aim to evaluate its performance on the test examples with unforeseen query types, which are drawn from a new distribution $\mathcal{G}'_q$. 
Specifically, the OOD generalizability on unforeseen query type is defined as 
\begin{equation}
	\mathbb{E}_{(q_t',\mathbf{d}_t,\mathbf{y_t}')\sim\mathcal{G}'_q} M(\pi(q_t', \mathbf{d}_t, f), \mathbf{y}_t'),
\label{equ: OOD generalizability on unforeseen query type}
\end{equation}
where $q_t'$, $\mathbf{d}_t$ and $\mathbf{y}_t'$ denotes the query, the document list and the label with new query types. 
Note that the same document collection is used for both the training and test sets.

\subsubsection{Metric of OOD Generalizability on Unforeseen Query Type}
\label{sec:cross_type generalizability evaluation}

To measure the OOD generalizability of the ranking models on unforeseen query type, we propose one automatic metric, namely,

\begin{itemize}

\item \textbf{DR$_{OOD}$} evaluates the \textit{drop rate} between the ranking  performance $P_{OOD}$ on the OOD test set (i.e., $(q_t',\mathbf{d}_t,\mathbf{y}_t')\sim\mathcal{G}'_q$) and the ranking performance $P_{I.I.D.}$  on the I.I.D. test set (i.e., $(q_t,\mathbf{d}_t,\mathbf{y}_t)\sim\mathcal{G}_q$), which is defined as
\begin{equation} 
	DR_{OOD} = \frac{P_{OOD} - P_{I.I.D.}}{P_{I.I.D.}},
\end{equation} 
where $P_{I.I.D.}$ and $P_{OOD}$ are defined as 
\begin{equation}
\label{PIID}
	P_{I.I.D.} = \mathbb{E}_{(q_t,\mathbf{d}_t,\mathbf{y}_t)\sim\mathcal{G}_q} M(\pi(q_t, \mathbf{d}_t, f), \mathbf{y}_t),
\end{equation}
\begin{equation}
\label{POOD}
	P_{OOD} = \mathbb{E}_{(q_t',\mathbf{d}_t,\mathbf{y}_t')\sim\mathcal{G}'_q} M(\pi(q_t', \mathbf{d}_t, f), \mathbf{y}_t'),  
\end{equation}
where the effectiveness evaluation metric $M$ can be defined in different ways, such as mean reciprocal rank (MRR) \cite{MRR}, mean average precision (MAP) \cite{MAP}, normalized discounted cumulative gain (NDCG) \cite{nDCG}  and Precision (P) \cite{Precision}, with respect to the specific experimental dataset. 
The ranking model would be more robust with a higher DR$_{OOD}$.
\end{itemize}

\subsection{OOD Generalizability on Unforeseen Corpus}

In this section, we introduce the definition and metric of the OOD generalizability on unforeseen corpus.

\subsubsection{Definition of OOD Generalizability on Unforeseen Corpus}

The OOD generalizability on unforeseen corpus refers to the transfer effectiveness on unseen corpus.   
In practice, the training corpus usually has a limited volume and hardly characterizes the entire distribution.   
Chasing an evolving data distribution is costly, and even if the training corpus does not become stale, models will still encounter unexpected situations at the test time \cite{ACL2020pretrained}. 
Accordingly, training a ranking model on the given corpus that can well generalize to another new corpus is necessary.

Formally, given a ranking model $f$ learned on $\{q_i,\textbf{d}_i,\textbf{y}_i\}_{i=1}^{m}$ which are drawn from the training distribution $\mathcal{G}_d$, we aim to evaluate its performance on the test examples from unseen corpus, which are drawn from a new distribution $\mathcal{G}'_d$. 
Specifically, the OOD generalizability on unforeseen corpus is defined as 
\begin{equation}
	\mathbb{E}_{(q_t',\mathbf{d}_t',\mathbf{y_t'})\sim\mathcal{G}'_d} M(\pi(q_t', \mathbf{d}_t', f), \mathbf{y}_t'),
\label{equ: OOD generalizability on unforeseen corpus}
\end{equation}
where $q_t'$, $\mathbf{d}_t'$ and $\mathbf{y}_t'$ denotes the query, the document list and the label from unseen corpus. 

\subsubsection{Metric of OOD Generalizability on Unforeseen Corpus}

To measure the OOD generalizability on unforeseen corpus, we also employ the \textbf{DR}$_{OOD}$ metric based on the ranking  performance $P_{OOD}$ on the OOD test set (i.e., $(q_t',\mathbf{d}_t',\mathbf{y}_t')\sim\mathcal{G}'_d$) and the ranking performance $P_{I.I.D.}$  on the I.I.D. test set (i.e., $(q_t,\mathbf{d}_t,\mathbf{y}_t)\sim\mathcal{G}_d$).

\subsection{Experimental Settings}

In this section, we introduce our experimental settings, including data construction, and implementation details.

\subsubsection{Data Construction} 
For evaluation purposes, we build two benchmark dataset based on several existing IR collections. 
Specifically, we use \textbf{Robust04}, \textbf{MQ2007},  and \textbf{MS MARCO}, which have been described in Section \ref{MQ2007_data}.
In these IR collections, queries are associated with the metadata information, which helps differentiate the examples with respect to query type and corpus, respectively. We now describe the detail of the two datasets for evaluating the OOD generalizability on unforeseen query type and corpus as follows.

\begin{itemize}
 \item  \textbf{Dataset for Unforeseen Query Type}. Here, we take the MS MARCO dataset as our source data. The reason is that MS MARCO contains various types of questions, and the amount of questions are much larger than other datasets. Firstly, we leverage the official query types \footnote{https://github.com/microsoft/MSMARCO-Question-Answering}, including \textit{Location}, \textit{Numeric}, \textit{Person}, \textit{Description} and \textit{Entity}, to identify the query type for each query. 
Then, for each query type, we construct the new training/test set, by randomly sampling 15,000/300 queries from the original training/development set respectively. 
In this way, we obtain a benchmark dataset with 5 different query types. 
The detailed statistics are shown in Table \ref{table:cross-type dataset}.

\begin{table}[t]
\centering 
 \renewcommand{\arraystretch}{1.6}
 \setlength\tabcolsep{4.5pt}
 \caption{Statistics of our constructed dataset for unforeseen query type}
\begin{tabular}{lccccc} 
\toprule \hline
type & Location & Numeric & Person & Description & Entity  \\
\hline
\#Training Queries & 15,000 & 15,000 & 15,000 & 15,000 & 15,000  \\
\#Test Queries & 300 & 300 & 300  & 300 & 300 \\
Avg \#Relevant Documents &  1 & 1 & 1 & 1 & 1 \\
Avg query words & 5.68 & 6.77 & 5.96 & 5.56 & 6.27 \\

\hline \bottomrule
\end{tabular}
\label{table:cross-type dataset}
\end{table}

 \item \textbf{Dataset for Unforeseen Corpus}. We take Robust04 from news articles, MQ2007 from Gov2 documents, and MS MARCO from web documents, as the whole benchmark dataset to mimic the different data distributions in different corpora. 
As we can see, they represent different sizes and genres of heterogeneous text collections.
For Robust04 and MQ2007, we randomly divide them into a training set (80\%) and a test set (20\%). 
For MS MARCO, to obtain a similar volume of relevant query-document pairs for fair comparison, we build the new training set by sampling a quarter of queries from the original training set and directly leverage the original development set as the new test set following \cite{cohen2018cross}. 
For three corpora, we also randomly sample 20\% queries from the training set for validation, respectively. 
Thus, we can obtain a benchmark dataset with 3 different corpora. 

\end{itemize}

\begin{table}[t]
\centering
\caption{ The MRR@100 performance of different ranking models with respect to  unforeseen query type. 
The five query types in MS MARCO, i.e., Location, Numeric, Person, Description and Entity, are denoted as LOC, NUM, PER, DES and ENT, respectively. 
The column ``train'' and ``test'' denote the training and test set for the ranking models, and the MRR@100 performance is reported on the test set. 
ALL$^*$ denotes the other four query types where the test query type has been removed from the entire query set.
Significant performance degradation with respect to the corresponding  I.I.D. setting is denoted as `$-$' ($\textit{p-value} \leq 0.05$).}
 \small
  \renewcommand{\arraystretch}{1.33} 
   \setlength\tabcolsep{1.2pt} 
  \begin{tabular}{c c | c c | c c c | c c c c c c }  \toprule \hline
   Train & Test & QL & BM25 & Prank & RankSVM & LambdaMART & DSSM & DRMM & Conv-KNRM & Duet & BERT & ColBERT \\
   \hline
   
   LOC & LOC & 0.277 & 0.312 & 0.250  & 0.299 & 0.319  & 0.146 & 0.278	 & 0.168 & 0.221 & 0.379 & \textbf{0.432}\\
   NUM & LOC & 0.276 & 0.296 & 0.243  & 0.276 & 0.300 & 0.120 & 0.216$^-$ & 0.105$^-$& 0.120$^-$ & 0.295$^-$ & \textbf{0.330}$^-$\\                  
   PER & LOC & 0.267 & 0.306 & 0.272 & 0.293 & 0.310 & 0.202 & 0.232$^-$ &0.089$^-$ & 0.120$^-$ & 0.314$^-$& \textbf{0.456} \\                       
   DES & LOC & 0.274 & 0.299 & 0.242 &0.275 &0.289 & 0.176 & 0.202$ ^-$ &0.097$^-$ & 0.132$^-$ & 0.292$^-$& \textbf{0.326}$^-$ \\
   ENT & LOC & 0.276 & 0.305 & 0.265 &0.278 &0.306 & 0.091 & 0.203$^-$ &0.110$^-$ & 0.143$^-$ & 0.300$^-$ & \textbf{0.322}$^-$\\
   ALL$^*$ & LOC & 0.275 & 0.305 & 0.254 & 0.289 & 0.318 & 0.153 & 0.247 & 0.200 & 0.133 & 0.360 & \textbf{0.366} \\
   \hline
 
   NUM & NUM  & 0.249 & 0.270 & 0.223 & 0.230 & 0.278 & 0.137  & 0.190 & 0.115 & 0.130 & 0.250 & \textbf{0.325} \\
   LOC & NUM & 0.247 & \textbf{0.253} & 0.224 & 0.237 & 0.238$^-$ & 0.065 & 0.186 & 0.091 & 0.091$^-$ & 0.190$^-$ & 0.227$^-$\\                 
   PER & NUM & 0.242 & \textbf{0.265} & 0.232 & 0.241 & 0.251 & 0.124 & 0.196 & 0.084 & 0.091$^-$ & 0.181$^-$& 0.251$^-$\\                
   DES & NUM & 0.240 & \textbf{0.267} & 0.235 & 0.232 & 0.253 & 0.165  & 0.198 & 0.077$^-$  & 0.107 & 0.207 &0.253$^-$ \\              
   ENT& NUM & 0.245& \textbf{0.262} & 0.239 & 0.226 & \textbf{0.262} & 0.078 & 0.180 & 0.083 & 0.120 & 0.203  & 0.260$^-$ \\   
   ALL$^*$ & NUM & 0.245 & 0.265 & 0.235 & 0.239 & 0.267 & 0.096 & 0.209 & 0.137 & 0.097 & 0.247 & \textbf{0.285} \\         
   \hline
   
   PER & PER  & 0.283 & 0.300 & 0.261 & 0.264 & 0.313 & 0.194 & 0.242 & 0.119 & 0.134 & 0.277 & \textbf{0.398} \\
   LOC & PER & 0.271 & 0.290 & 0.258 & 0.260 & 0.284 & 0.060  & 0.206$^-$ & 0.102 & 0.110 & 0.271 & \textbf{0.339} \\
   NUM & PER & 0.274 & 0.285 & 0.230 & 0.241 & 0.306 & 0.064 & 0.198$^-$ & 0.095 & 0.101 & 0.259& \textbf{0.320}$^-$\\                   
   DES & PER & 0.273 & 0.294 & 0.254 & 0.246 & 0.289 & 0.138 & 0.200$^-$ & 0.083$^-$ & 0.089$^-$ & 0.276 &\textbf{0.349} \\               
   ENT & PER & 0.274& 0.292 & 0.242 & 0.244 & 0.291 & 0.078 & 0.203$^-$ & 0.086$^-$ & 0.119 & 0.249  & \textbf{0.336}$^-$ \\
   ALL$^*$ & PER & 0.271 & 0.296 & 0.249 & 0.248 & 0.295 & 0.074 & 0.206 & 0.132  & 0.107 & 0.308 & \textbf{0.367} \\
   \hline
 
   DES & DES & 0.236 & 0.251 & 0.211 & 0.235 & 0.283 & 0.195 & 0.233 & 0.104 & 0.151 & 0.250 & \textbf{0.332} \\
   LOC & DES & 0.232 & 0.238 & 0.215 & 0.247 & 0.268 & 0.083 & 0.190$^-$ & 0.117 & 0.107$^-$ & 0.213 & \textbf{0.291}\\                  
   NUM & DES & 0.232 & 0.238 & 0.211  & 0.234 & 0.279 & 0.098 & 0.178$^-$ & 0.082 & 0.104$^-$ & 0.220 & \textbf{0.284}\\                   
   PER & DES & 0.223 & 0.249 & 0.237  & 0.252 & 0.273 & 0.161 & 0.192$^-$ & 0.084 & 0.116 & 0.206 &\textbf{0.320} \\              
   ENT & DES & 0.232& 0.239 & 0.212 & 0.244 & 0.278 & 0.112 & 0.200$^-$ & 0.099 & 0.133 & 0.214 & \textbf{0.310}\\    
   ALL$^*$ & DES & 0.231 & 0.249 & 0.209 & 0.246 & 0.287 & 0.141 & 0.201 & 0.123 & 0.113 & 0.258 & \textbf{0.310} \\     
   \hline
   
   ENT & ENT & 0.226 & 0.270 & 0.211 & 0.235 & 0.255 & 0.104 & 0.196 & 0.114 & 0.148 & 0.232 & \textbf{0.293} \\
   LOC & ENT & 0.223 & \textbf{0.260} & 0.181 & 0.237 & 0.241 & 0.080 & 0.203 & 0.111 & 0.103$^-$ & 0.209 & \textbf{0.260}\\                      
   NUM & ENT & 0.224 & 0.251 & 0.174 & 0.231 & 0.256 & 0.073 & 0.201 & 0.089 & 0.097$^-$ & 0.229& \textbf{0.257}\\                    
   PER & ENT & 0.222 & 0.264 & 0.220 & 0.254 & 0.249 & 0.122 & 0.193 & 0.106 & 0.129 & 0.212 & \textbf{0.293} \\                
   DES & ENT & 0.222 & 0.258 & 0.180 & 0.240 & 0.231 & 0.130 & 0.207 & 0.109 & 0.121 & 0.217  & \textbf{0.275} \\   
   ALL$^*$ & ENT & 0.222 & 0.263 & 0.222 & 0.249 & 0.256 & 0.117 & 0.218 & 0.146 & 0.125 & 0.249 & \textbf{0.311} \\             

 \hline \bottomrule
    \end{tabular}
   \label{table:OOD_cross_type}
\end{table}

\subsubsection{Implementation Details}
\label{sec:OOD_dataset}

The implementation details, including the pre-processing and model implementation, are similar to that in Section \ref{Implementation}. The only expectation is that the parameters of traditional probabilistic ranking models are tuned on the corresponding development set in the two constructed benchmark dataset.

\subsection{Analysis of OOD Generalizability on Unforeseen Query Type}

\begin{figure}[t]
\centering
\includegraphics[scale=0.93]{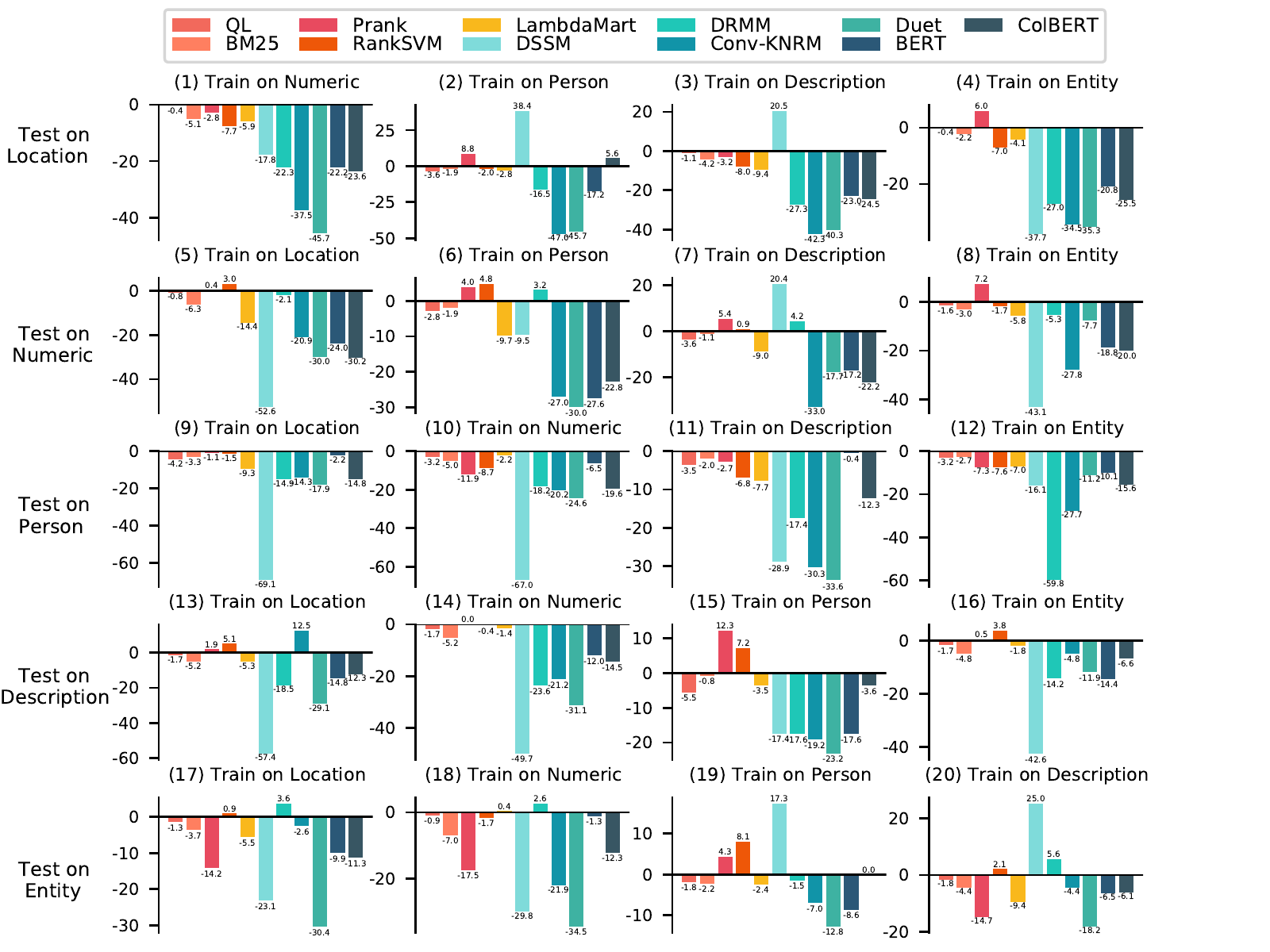}
\caption{The OOD generalizability of different ranking models on unforeseen query type in terms of DR$_{OOD}$(\%). The test query type is described on the left and the training query type is described above each subfigure.}
\label{figure: query-type OOD drop rate}
\end{figure}

In this section, we analyze the empirical results on the OOD generalizability on unforeseen query type. 
Specifically, we adopt the widely used metric for MS MARCO, i.e., \textbf{MRR@100}, as the implementation of the effectiveness evaluation metric $M$ in Eq. (\ref{PIID}) and (\ref{POOD}). 
Table \ref{table:OOD_cross_type} shows the MRR@100 performance of different ranking models under both the OOD (i.e., the training and test query types are different.) and I.I.D. (i.e., the training and test query types are the same.) settings. 
  For the OOD setting, we train the model on one query type and test it on a different query type. Since the model may be too specific to the single query type it was trained on, we also train the model on four query types and test it on the one remaining type.
Figure \ref{figure: query-type OOD drop rate} shows the DR$_{OOD}$ performance with respect to unforeseen query type.

In the following, we first give an overall analysis of different ranking models. 
Then we analyze the OOD generalizability of traditional probabilistic ranking models, LTR models and neural ranking models, respectively. 
Recall that high values of DR$_{OOD}$ attest to increased ranking robustness.

\subsubsection{\textbf{Overall analysis on all the ranking models}}

Firstly, we give an overall performance analysis on all the ranking models. We can observe that: 
(1) Under the I.I.D. setting (e.g., the training and test query type is Person), \textit{ColBERT} generally performs the best (0.398 on Person) followed by \textit{LambdaMART} (0.313 on Person), \textit{BERT} (0.277 on Person) and then \textit{RankSVM} (0.264 on Person), in terms of MRR@100.
(2) As show in Figure \ref{figure: query-type OOD drop rate}, most ranking models are not able to well generalize to OOD query types. 
Take the most effective model \textit{ColBERT} as an example, the DR$_{OOD}$ value is -30.2\% when the training query type is Location and the test query type is Numeric. 
It indicates that higher effectiveness does not reliably improve OOD generalizability. 
  (3) When being trained on four query types and tested on the one remaining type, most ranking models perform worse than the I.I.D. setting, while perform better than being trained on the single query type. It indicates that the training data for a single query type may be inadequate. When the training query type increases, the OOD generalizability improves.   

\subsubsection{\textbf{Analysis on traditional probabilistic ranking models and LTR models}}

When we look at the traditional probabilistic ranking models and LTR models, we find that:
(1) In general, traditional probabilistic ranking models achieve the highest DR$_{OOD}$ values, indicating that traditional probabilistic ranking models are the most robust models when facing unforeseen query types. 
A possible reason would be that as unsupervised methods, traditional probabilistic ranking models avoid the problem of overfitting the training data and thus have a better OOD generalizability on unforeseen query types.  
(2) The DR$_{OOD}$ values of LTR models are higher than that of neural ranking models, indicating that LTR models are more robust than neural ranking models.
The reason might be that hand-crafted features in LTR models could better generalize to unforeseen query type than automatically learned features by neural networks.

\subsubsection{\textbf{Analysis on neural ranking models}}

When we look at the neural ranking models, we find that:
(1) In general, neural ranking models achieve the lowest DR$_{OOD}$ values among all the ranking models. 
The reason might be that neural ranking models with a deeper network  architecture fit the normal in-distribution query types better, at the cost of further loss in performance on the held out OOD query types. 
It is consistent with the finding in \cite{cohen2018cross}. 
(2) Among these five neural ranking models, pre-trained models (i.e., \textit{BERT} and \textit{ColBERT}) have the best OOD generalizability on  unforeseen query type.  
For example, \textit{ColBERT} trained (fine-tuned) on Person and tested on Location, the MRR@100 value even improves 5.6\% compared with \textit{ColBERT} trained and tested on Location. 
A possible explanation would be that pre-training on the huge text corpus can improve OOD generalization. 
It is consistent with the finding in \cite{ACL2020pretrained}, which indicates that pre-trained models are more robust to OOD examples on several NLP tasks.
(3) Pre-trained models have shown great effectiveness under both the OOD and I.I.D. settings, and good robustness to OOD query types. 
Future works could design novel pre-training objectives tailored for IR that enhance ranking robustness.

\begin{table}[t]
\centering
 \caption{ The performance of different ranking models with respect to  unforeseen corpus. 
The ``train'' and ``test'' denote the training and test corpus for the ranking models, and the MAP, NDCG@10, NDCG@20, P@10, P@20, MRR@10 and MRR@100 performance are reported on the test set. 
 Significant performance degradation with respect to the corresponding I.I.D. setting is denoted as  `$-$' ($\textit{p-value} \leq 0.05$).}
  \small
  \renewcommand{\arraystretch}{1.217} 
   \setlength\tabcolsep{1pt} 
  \begin{tabular}{c c  c c c  c c c  c c }  \toprule \hline
  \multirow{4}{*}{Model} & test & \multicolumn{3}{c}{Robust04} & \multicolumn{3}{c}{MQ2007} & \multicolumn{2}{c}{MS MARCO} \\
  \cmidrule(r){3-5} \cmidrule(r){6-8} \cmidrule(r){9-10} 
 & train & MAP & NDCG@20 & P@20 & MAP & NDCG@10 & P@10 & MRR@10 & MRR@100 \\
 \hline
 \multirow{3}{*}{QL} & Robust04 & \textbf{0.2553} & 0.4575 & 0.3860 & 0.3646 & 0.3662 & 0.3349 & 0.2301 & 0.2426  \\
  & MQ2007 & 0.2492 & 0.4561 & 0.3960 & 0.3736 & 0.3675 & 0.3293 & 0.2368 & 0.2488 \\ 
  & MS MARCO & 0.2490 & 0.4525 & 0.3920 & 0.3714 & 0.3622 & 0.3240 & 0.2374 & 0.2495 \\  \hline
\multirow{3}{*}{BM25} & Robust04 & 0.2522 & 0.4514 & 0.3760 & 0.3132$^-$ & 0.3197$^-$ & 0.2843$^-$ & 0.2051$^-$ & 0.2174$^-$ \\
& MQ2007 & 0.2227  & 0.4280 & 0.3600 & 0.3863 & 0.3936 & 0.3438 & 0.2618 & 0.2732 \\
& MS MARCO & 0.1971  & 0.3995  & 0.3370 & 0.3837 & 0.3831 & 0.3391 & 0.2660 & 0.2782   \\
                       \hline
 \multirow{3}{*}{Prank} & Robust04 & 0.2186 & 0.4011 & 0.3620 & 0.3410$^-$ & 0.2847$^-$ & 0.2719 & 0.2038$^-$ & 0.2171$^-$  \\
                     
 &MQ2007 & 0.2395 & 0.4357 & 0.3830 & 0.3834 & 0.3382 & 0.3056 & 0.1531$^-$  & 0.1676$^-$  \\
                       
 &MS MARCO & 0.2162 & 0.3848 & 0.3440 & 0.3583 & 0.3076 & 0.3030 & 0.2373 & 0.2490\\
                       \hline
 \multirow{3}{*}{RankSVM} & Robust04 &  0.2501 & 0.4510 & 0.3760 & 0.3884$^-$ & 0.3502$^-$ & 0.3145$^-$ & 0.2350 & 0.2468 \\
                     
 &MQ2007 & 0.2340  & 0.4190 & 0.3680 & 0.4601 & 0.4376 & 0.3820 & 0.2021$^-$  & 0.2160$^-$ \\
                       
 &MS MARCO & 0.2523 & 0.4529 & 0.3860 & 0.3948$^-$ & 0.3652$^-$ & 0.3343$^-$ & 0.2410 & 0.2535 \\
                       \hline
 \multirow{3}{*}{LambdaMart} & Robust04 &0.2397 & 0.4357 & 0.3650 & 0.3562$^-$  & 0.3148$^-$  & 0.2970$^-$  & 0.2172$^-$  & 0.2301$^-$ \\
                       
 & MQ2007 &0.2057 & 0.3780 & 0.3300 & 0.4578 & 0.4381 & 0.3787 & 0.2068$^-$  & 0.2192$^-$  \\
                       
 &MS MARCO & 0.2174 & 0.4026 & 0.3160 & 0.3365$^-$  & 0.2829$^-$  & 0.2675$^-$  & 0.2602 & 0.2710\\
                       \hline
 \multirow{3}{*}{DSSM}& Robust04 & 0.1392 & 0.2455 & 0.2270 & 0.3158$^-$ & 0.2407$^-$ & 0.2441$^-$ & 0.1231 & 0.1350 \\
                       
&MQ2007 & 0.1229 & 0.2295 & 0.2260 & 0.3988 & 0.3530 & 0.3438 & 0.0159$^-$ & 0.0336$^-$ \\
                       
&MS MARCO & 0.1262 & 0.2360 & 0.2160 & 0.2785$^-$ & 0.1761$^-$ & 0.1902$^-$ & 0.1051 & 0.1220 \\
                       \hline
 \multirow{3}{*}{DRMM}& Robust04 &0.2541 & \textbf{0.4744} & \textbf{0.4060} & 0.4231 & 0.3859 & 0.3524 & 0.1864  & 0.2002\\
                       
&MQ2007 & 0.1829 & 0.3427$^-$  & 0.2930$^-$  & 0.4390 & 0.4038 & 0.3719 & 0.0552$^-$  & 0.0711$^-$ \\
                       
&MS MARCO & 0.1357$^-$  & 0.2445$^-$  & 0.2380$^-$  & 0.3024$^-$  & 0.2101$^-$  & 0.2257$^-$  & 0.1165 & 0.1334 \\
                       \hline
 \multirow{3}{*}{Conv-KNRM} & Robust04 & 0.1735 & 0.3084 & 0.2710 & 0.3089$^-$  & 0.2332$^-$  & 0.2479$^-$  & 0.0301$^-$  & 0.0483$^-$    \\
                       
&MQ2007 & 0.1372  & 0.2439 & 0.2210  & 0.4115 & 0.3670 & 0.3568 & 0.0123$^-$  & 0.0278$^-$   \\
                      
&MS MARCO & 0.1389 & 0.2654  & 0.2480  & 0.2665$^-$  & 0.1515$^-$  & 0.1787$^-$  & 0.1559 & 0.1716 \\
                       \hline
                   
 \multirow{3}{*}{Duet} & Robust04 & 0.1512 & 0.2592 & 0.2380 & 0.3156$^-$  & 0.2470$^-$  & 0.2509$^-$  & 0.0161$^-$  & 0.0337$^-$   \\
                    
 & MQ2007 & 0.1319  & 0.2332  & 0.2030 & 0.4126 & 0.3629 & 0.3527 & 0.0114$^-$  & 0.0270$^-$ \\
                      
&MS MARCO & 0.1338  & 0.2469  & 0.2270  & 0.2616$^-$  & 0.1442$^-$  & 0.1749$^-$  & 0.1219 & 0.1392 \\
                       \hline
 \multirow{3}{*}{BERT} & Robust04 & 0.2386 & 0.4407 & 0.3830 & 0.3772$^-$  & 0.3178$^-$  & 0.2988$^-$  & 0.1509$^-$  & 0.1667$^-$   \\
                      
& MQ2007& 0.2178  & 0.4262  & 0.3740  & 0.4656 & 0.4435 & 0.3929 & 0.0910$^-$  & 0.1091$^-$   \\
                         
& MS MARCO & 0.2091  & 0.4048  & 0.3680  & 0.3215$^-$  & 0.2488$^-$  & 0.2509$^-$  & 0.2691 & 0.2798 \\
                       \hline
 \multirow{3}{*}{ColBERT} & Robust04 & 0.2136 & 0.4126 & 0.3600 & 0.3741$^-$  & 0.3231$^-$  & 0.3133$^-$  & 0.1048$^-$  & 0.1226$^-$    \\
                      
& MQ2007 &0.1967  & 0.3785  & 0.3460  & \textbf{0.4759} & \textbf{0.4500} & \textbf{0.4059} & 0.0937$^-$  & 0.1116$^-$    \\
                         
&MS MARCO & 0.2077  & 0.4149  & 0.3580  & 0.3087$^-$  & 0.2263$^-$  & 0.2328$^-$  & \textbf{0.3279} & \textbf{0.3360} \\
                       \hline

 \hline \bottomrule
    \end{tabular}
    
   \label{table:OOD_cross_collections}
\end{table}

\begin{figure}[t]
\centering
\includegraphics[scale=0.723]{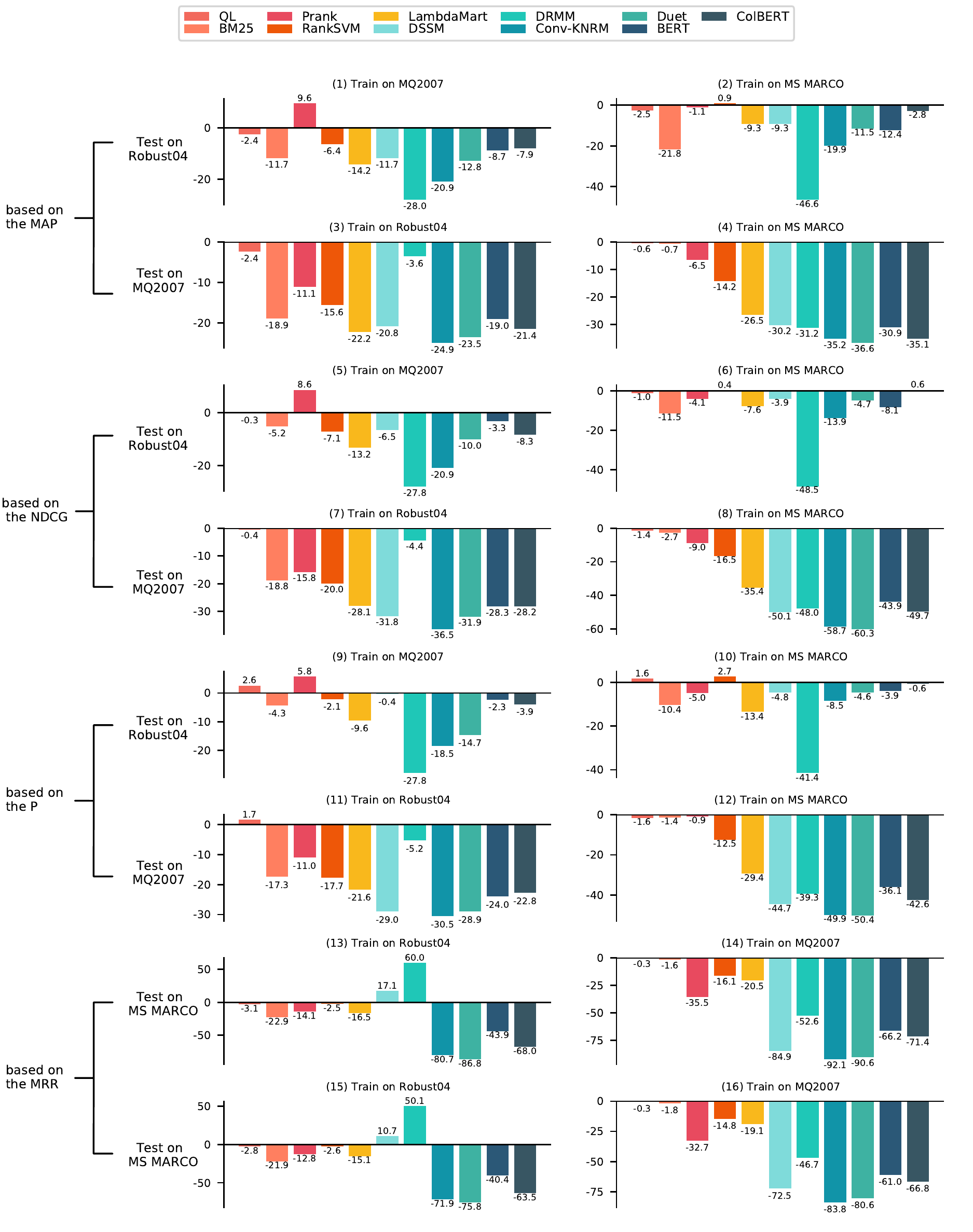}
\caption{ The OOD generalizability on unforeseen corpus in terms of DR$_{OOD}$(\%) based on the MAP, NDCG, P, and MRR respectively. For each subfigure, the test corpus is described on the left and the training corpus is described above.}
\label{figure: collection-type OOD drop rate}
\end{figure}

\subsection{Analysis of OOD Generalizability on Unforeseen Corpus}

In this section, we analyze the empirical results on the OOD generalizability on unforeseen corpus. 
Similar to the used effectiveness evaluation metric in Section \ref{average effectiveness}, we also adopt the \textbf{MAP}, \textbf{NDCG@10},   \textbf{NDCG@20}, \textbf{P@10}, \textbf{P@20}, \textbf{MRR@10} and \textbf{MRR@100} as the implementation of $M$ in Eq. (\ref{PIID}) and (\ref{POOD}). 
Table \ref{table:OOD_cross_collections} shows the MAP, NDCG@10,   NDCG@20, P@10, P@20, MRR@10 and MRR@100 performance of all the ranking models under both the OOD (i.e., the training and test corpora are different) and I.I.D. (i.e., the training and test corpora are the same) settings. 
Figure \ref{figure: collection-type OOD drop rate} shows the DR$_{OOD}$ performance based on the MAP,   NDCG, P and MRR, respectively.

In the following, we first give an overall analysis of different ranking models.  
Then we analyze the OOD generalizability of traditional probabilistic ranking models, LTR models and neural ranking models, respectively. 
Recall that high values of DR$_{OOD}$ attest to increased ranking robustness.

\subsubsection{\textbf{Overall analysis on all the ranking models}}

Firstly, we give an overall analysis on all the ranking models. 
We can observe that:
(1) Under the I.I.D. setting (e.g., the training and test corpus is MQ2007), \textit{ColBERT} generally performs the best (0.4759) followed by \textit{BERT} (0.4656) and \textit{RankSVM} (0.4601), in terms of MAP. 
(2) As show in Figure \ref{figure: collection-type OOD drop rate}, almost all DR$_{OOD}$ values are negative. 
It indicates that most ranking models are not able to well generalize to OOD examples. 
(3) Besides, the DR$_{OOD}$ value with respect to  MRR for unforeseen MS MARCO is much lower than that for unforeseen query types sampled from MS MARCO.
A possible reason is that compared with different query types sampled from the same corpus, different corpora have greater differences among samples.

\subsubsection{\textbf{Analysis on traditional probabilistic ranking models and LTR models}}
 
When we look at the traditional probabilistic ranking models and LTR models, we can find that:
(1) Under the I.I.D. setting, \textit{BM25} performs better than \textit{QL} on most corpus, and \textit{BM25} is a strong baseline which performs better than LTR models on Robust04. 
Under the OOD setting, \textit{QL} is more robust to the unforeseen corpus than \textit{BM25}.  
(2) LTR models show a good transfer effectiveness from MS MARCO to Robust04. 
For instance, the DR$_{OOD}$ value based on  MAP is close to $0$ (i.e.,  -1.1\%) when we train  \textit{Prank} on MS MARCO and then test it on Robust04.
Besides, it is surprising that when we train  \textit{RankSVM} on MS MARCO and test it on Robust04, the DR$_{OOD}$ value based on  P@20 is positive (i.e., 2.7\%).  
The reason might be that the distribution of hand-crafted features for Robust04 is similar to that for MS MARCO.  
 (3) \textit{Prank} shows a better OOD generalizability on unforeseen corpus than \textit{RankSVM} and \textit{LambdaMART} if we test them on Robust04 and MQ2007.
For example, when we train \textit{Prank} on MQ2007 and test it on Robust04, the DR$_{OOD}$ value based on NDCG@20 is even positive (i.e., 8.6\%), while \textit{LambdaMART} achieves -13.2\% DR$_{OOD}$ under the same setting.
One possible explanation is that \textit{Prank} has simpler model architecture and less parameters compared with \textit{RankSVM} and \textit{LambdaMART}, which alleviates the problem of overfitting and increases the OOD generalizability.

\subsubsection{\textbf{Analysis on neural ranking models}}

When we look at the neural ranking models, we can see that:
(1) In general, neural ranking models show the worst OOD generalizability on unforeseen corpus. 
The finding is consistent with that on unforeseen query type.  
For example, when we train \textit{Conv-KNRM} on MQ2007 and test it on MS MARCO, the DR$_{OOD}$ value based on 
 MRR@100 is even -83.8\%. 
(2) \textit{DRMM} shows a good transfer effectiveness from Robust04 to both the MS MARCO and MQ2007.  
As shown in previous studies \cite{DRMM}, \textit{DRMM} have performed quite well on Robust04. 
In this way, \textit{DRMM} trained on Robust04 may lean to generalize to other corpora.

\section{Defensive Ability against Adversarial Operations}
\label{sec:adv}

Deep neural networks have been found vulnerable to adversarial attack, where an imperceptible perturbation can trigger dramatic changes in the final result \cite{akhtar2018threat,zhou2021practical}. 
However, the vulnerability of neural ranking models remains under-explored. 
This poses a serious security risk on the practical search engines, where an imperceptible adversarial perturbation to the query/document can be sufficient to intentionally fool the neural ranking model to make wrong decisions, i.e., raise or lower the ranks of selected candidate documents. 


Therefore, in this section, we analyze the robustness of ranking models according to their ability to defend against adversarial operations, i.e., the defensive ability which has been formulated in Eq. (\ref{equ: defensive ability}). 
Specifically, we introduce two types of adversarial operations to measure the defensive ability, i.e., 1) The ability to defend against the query attack, and 2) The ability to defend against the document manipulation.

\subsection{Defensive Ability against Query Attack}

In this section, we introduce the definition and metric of the defensive ability against query attack.

\subsubsection{Definition of Defensive Ability against Query Attack
}

The defensive ability against query attack refers to that the ranking performance on the attacked queries.  
In a real-world scenario, the users' queries may be attacked unconsciously. 
For example, users may occasionally misspell or mistype a query keyword when performing a search. 
Such query typo is a longstanding real-world problem for IR, which has been extensively studied \cite{hagen2017large, zhuang2021dealing, penha2021evaluating}.
From the perspective of a user, 
if the search engine cannot tolerate the query typo, the user's satisfaction will remarkably drop.
A reliable system that always produces acceptable retrieval performance, is more preferred than another system that fails on the occasional query typo. 
Specifically, we consider the effects of misspelled or mistyped queries by exploring several types of character-level and word-level edits.

Formally, given a ranking model $f$ learned on $\{q_i,\textbf{d}_i,\textbf{y}_i\}_{i=1}^{m}$ which are drawn from the training distribution $\mathcal{G}$, we aim to evaluate its performance on the I.I.D. test queries $q_t$ attacked by a query attack function $a_q$, i.e., $a_q(q_t)$. 
Specifically, the defensive ability against query attack is defined  as
\begin{equation}
	\mathbb{E}_{(q_t,\mathbf{d}_t,\mathbf{y}_t)\sim\mathcal{G}} M(\pi(a_q(q_t), \mathbf{d}_t, f), \mathbf{y}_t),
\label{equ: defensive ability against query attack}
\end{equation} 
where $q_t, \mathbf{d}_t$ and $\mathbf{y}_t$ denotes the query, the document list and the label from the I.I.D. test set, respectively.

\subsubsection{Metric of Defensive Ability against Query Attack }

As for evaluation measure, we employ one automatic metric, namely, 
\begin{itemize}

\item \textbf{DR$_{query}$} evaluates the \textit{drop rate} between the model performance $P_o$ on the original queries (i.e., $q_t$) and the ranking performance $P_t$ on the attacked queries (i.e., $a_q(q_t)$), i.e., 
\begin{equation}
	DR_{query} = \frac{P_t - P_o}{P_o},
\end{equation}
where $P_o$ and $P_t$ are defined as
\begin{equation}
\label{P_o}
	P_o =  \mathbb{E}_{(q_t,\mathbf{d}_t,\mathbf{y}_t)\sim\mathcal{G}} M(\pi(q_t, \mathbf{d}_t, f), \mathbf{y}_t),
\end{equation}
\begin{equation}
\label{P_t}
	P_t =  \mathbb{E}_{(q_t,\mathbf{d}_t,\mathbf{y}_t)\sim\mathcal{G}} M(\pi(a_q(q_t), \mathbf{d}_t, f), \mathbf{y}_t),
\end{equation}
where the effectiveness evaluation metric $M$ can be defined in different ways, such as MRR@100, with respect to the specific experimental dataset.  
The ranking model would be more robust with a higher DR$_{query}$.

\end{itemize}

\subsection{Defensive Ability against Document Manipulation}

In this section, we introduce the definition and metric of the ability to defend against document manipulation.

\subsubsection{Definition of Defensive Ability against Document Manipulation}

The defensive ability against document manipulation refers to that the ranking performance on the adversarial documents.  
In the Web search setting, many document authors are ``ranking-incentivized'' \cite{goren2019ranking}. 
That is, they may introduce modifications to their documents, hoping to rank them higher for some queries by search engines.   
In this way, these modifications could be almost indiscernible by users and users will loose faith in the search system if they observe these rapid ranking changes.  
Thus,  a ranking model should be robust to such manipulations for maintaining high-quality search results.

Formally, given a ranking model $f$ learned on $\{q_i,\textbf{d}_i,\textbf{y}_i\}_{i=1}^{m}$ which are drawn from the training distribution $\mathcal{G}$, we aim to evaluate its performance on the I.I.D. test queries $q_t$ with the corresponding document list $\textbf{d}_t$  attacked based on a document attack function $a_d$, i.e., $a_d(\textbf{d}_t)$. 
Specifically, the defensive ability against document manipulation is defined  as
\begin{equation}
	\mathbb{E}_{(q_t,\mathbf{d}_t,\mathbf{y}_t)\sim\mathcal{G}} M(\pi(q_t, a_d(\mathbf{d}_t), f), \mathbf{y}_t),
\label{equ: defensive ability against document manipulation}
\end{equation} 
where $q_t, \mathbf{d}_t$ and $\mathbf{y}_t$ denotes the query, the document list and the label from the I.I.D. test set, respectively.

\subsubsection{Metric of Defensive Ability against Document Manipulation}

As for evaluation measure, we employ Top Change (TC) and Kendall's-$\tau$ distance (KT) following the previous work \cite{goren2018SIGIR}.

Suppose that $L^{(q_t)}$ is a ranked document list with respect to a given query $q_t$ achieved by a ranking model.  
After the document manipulation on the document list $\textbf{d}_t$, a new ranked list $L^{(q_t)\prime}$ is obtained by the ranking model.  
The key idea to quantify robustness is to measure the distance between $L^{(q_t)}$ and $L^{(q_t)\prime}$. 
Specifically, the lower the TC and KT value, the lower the ``distance''  between the two lists, i.e., the more robust the ranking model is.   

\begin{itemize}

	\item \textbf{TC} indicates the similarity between the two ranked lists based on the highest ranked document, i.e.,
	    \begin{equation}
	    	TC = \mathbb{E}_{(q_t,\mathbf{d}_t,\mathbf{y_t})\sim\mathcal{G}} \delta(L^{(q_t)}_1\neq L^{(q_t)\prime}_1),
	    \end{equation}
	    where $L^{(q_t)}_1$ and $L^{(q_t)\prime}_1$ denotes the highest ranked document in the ranked list $L^{(q_t)}$ and $L^{(q_t)\prime}$, respectively. $\delta$ is the indicator function, where the value is 1 if the highest ranked document in $L^{(q_t)}$ and $L^{(q_t)\prime}$ is different, and 0 otherwise. 
	    	    
	\item \textbf{KT} is defined as the number of discordant pairs between two paired lists normalized with respect to the number of pairs of items in a list, i.e.,
	    \begin{equation}
	    	KT = \mathbb{E}_{(q_t,\mathbf{d}_t,\mathbf{y_t})\sim\mathcal{G}} \frac{\sum_{i,j\in P}\overline{K}_{ij}(L^{(q_t)}, L^{(q_t)\prime} )}{n_{q_t}(n_{q_t}-1) / 2},
	    \end{equation}
	    where $P$ is the set of unordered pairs of distinct documents in $L^{(q_t)}$ and $L^{(q_t)\prime}$. $n_{q_t}$ is the size of the ranked list. 
	    $\overline{K}_{ij}(L^{(q_t)}, L^{(q_t)\prime})$ denotes whether the document pair $(d_{ti},d_{tj})$ is a discordant pair. A discordant pair is two documents whose relative ranking in one list is different than that in the other list. Specifically, $\overline{K}_{ij}(L^{(q_t)}, L^{(q_t)\prime})$ is defined as,
        \begin{equation}
	    	\overline{K}_{ij}(L^{(q_t)}, L^{(q_t)\prime}) = \delta((L^{(q_t)}(d_{ti}) - L^{(q_t)}(d_{tj})) (L^{(q_t)\prime}(d_{ti}) - L^{(q_t)\prime}(d_{tj})) < 0),
	    \end{equation}  
	where $L^{(q_t)}(d_{ti})$ and $L^{(q_t)\prime}(d_{ti})$ are the rankings of the document $d_{ti}$ in $L^{(q)}$ and $L^{(q)\prime}$,  respectively.

\end{itemize} 

\subsection{Experimental Settings}

\label{sec:document manipulation dataset}

We first introduce our experimental settings, including datasets and implementation details. 

\subsubsection{Datasets} 

The datasets used for evaluating the defensive ability are as follows.

To evaluate the defensive ability against query attack, we conduct experiments on the \textbf{MS MARCO} dataset, which has been described in Section \ref{sec:OOD}. 
We construct the novel training set by randomly sampling a quarter of queries from the original training set. 
We randomly sample 20\% queries from the novel training set for validation.
Besides, we directly use the original development set as the novel test set.

To evaluate the defensive ability against document manipulation, we follow the previous work \cite{goren2018SIGIR} to conduct experiments on the \textbf{ASRC} and \textbf{ClueWeb09-B} dataset. The detailed statistics of these datasets are shown in Table \ref{table: document manipulation dataset}.

\begin{itemize}
	\item \textbf{ASRC}. Adversarial Search Collection (ASRC)\footnote{https://github.com/asrcdataset/}  is an adversarial ranking dataset, which aims to analyze content-based ranking competitions so as to shed light on the strategic behavior of publishers \cite{raifer2017SIGIR}. 
The competition included 31 different repeated matches, each of which was with respect to a different TREC's ClueWeb09 query. The competition was run for eight rounds. Students are incentivized by course-grade rewards to manipulate their documents, in order to have them ranked higher in the next round.

\item \textbf{ClueWeb09-B}. ClueWeb09-B is a large Web collection with 150 queries and over 50M English documents, whose queries are accumulated from TREC Web Track 2009, 2010, and 2011. 

\end{itemize}

\begin{table}[h]
\centering 
 \renewcommand{\arraystretch}{1.4}
 \setlength\tabcolsep{10pt}
 \caption{Statistics of datasets used for evaluating the ability to defend  against document manipulation}
\begin{tabular}{lcc} 
\toprule \hline
& ASRC  & ClueWeb09-B   \\
\hline
\#Queries & 31 & 200  \\
\#Documents & 1,279 & 50M \\
Avg \#Relevant Documents &  36 & 55  \\

\hline \bottomrule
\end{tabular}
\label{table: document manipulation dataset}
\end{table}

\subsubsection{Implementation Details}

The model implementation details are similar to that in Section \ref{sec:OOD_dataset}. 
In the following, we will introduce our methods to simulate the adversarial query attack and document manipulation.

\begin{itemize}
\item \textit{Implementation of Query Attack}. The MS MARCO dataset is preprocessed in the same way as described in Section \ref{sec:OOD_dataset}. We use the official top 100 ranked documents retrieved by QL.
    Then, to implement the query attack, we simulate the query attack at the character-level and word-level.

    \quad For character-level query attack,
we follow the previous works \cite{jones2020NLProbustEncoding, pruthi2019combating}, which are inspired by psycholinguistic studies \cite{rawlinson1976psychology, davis2003psycholinguistic}. 
The psycholinguistic studies demonstrate that humans can comprehend text altered by jumbling internal characters, provided that the first and last characters of each word remain unperturbed.  

\quad Specifically, we explore to perturb queries with four types of character-level edits: (1) Add, inserting a new lower-case character internally in a word; (2) Remove, deleting an internal character of a word; (3) Substitute: substituting an internal character for any letter; (4) Swap, swapping two adjacent internal characters of a word. 
For each query in the test set, we firstly randomly choose a word and then attack it using one randomly-chosen character-level edit. 
We denote such attacked queries as \textit{1-char} attacked queries and there are 25.32\% Add, 24.96\% Remove, 23.34\% Substitute and 24.42\% Swap in the final obtained \textit{1-char} attacked queries.  

\quad Furthermore, we also randomly select one word from the \textit{1-char} attacked queries and then attack it using one randomly-chosen character-level edit \cite{pruthi2019combating}. 
We denote such attacked queries as \textit{2-char} attacked queries. 
We evaluate the performance of different ranking models on the original test queries, \textit{1-char} and \textit{2-char} attacked test queries respectively. 

\quad For word-level query attack, we explore to perturb queries with three types of word-level edits:
(1) Add, inserting a new lower-case word in a query; (2) Remove, deleting a word of a query; (3) Substitute: substituting a word with a random word.
For each query in the test set, we firstly randomly choose a word and then attack it using one randomly-chosen word-level edit.

\item \textit{Implementation of Document Manipulation}. We use the public dataset ASRC that was created as a result of an on-going ranking competition \cite{raifer2017SIGIR}, to measure the document manipulation  \cite{goren2018SIGIR}. 
Specifically, the ranking competition involved 31 repeated matches that lasted for 8 rounds, where each match was with respect to a different query.
Students in an IR course served as documents' authors.
In the first round, in addition to the query itself, students were provided with an example relevant document, and were incentivized by bonus points to the course's grade to modify their documents so as to have them ranked as high as possible in the next round.
Starting from the second round, students were presented with the ranking as well as the content of all documents submitted in the previous round in the match.
To assure the fairness of the ranking competition, students had no prior knowledge of the ranking function and all data was anonymized.
In this way, the ASRC dataset could simulate the document manipulation in the ranking competition.

\quad Since the ASRC dataset is too small to effectively train a ranking model, we firstly train the ranking models on the ClueWeb09-B, and then evaluate their defensive abilities on the ASRC \cite{goren2018SIGIR}. 
Specifically, we randomly sample 75\% queries from the ClueWeb09-B for training and leverage the remaining 25\% queries for validation. 
For the ClueWeb09-B dataset, we use the QL model to retrieve the top 100 ranked documents to build an initial document list.

\end{itemize}

\subsection{Ranking Robustness to Query Attack }

In this section, we analyze the empirical results on the defensive ability against query attack. 
Specifically, we adopt the widely used metric for MS MARCO, i.e., \textbf{MRR@100}, as the implementation of the effectiveness evaluation metric $M$ in Eq. (\ref{P_o}) and (\ref{P_t}). 
Table \ref{table:drop rate query noise}   and Table \ref{table:word-level query attack} show the MRR@100 and DR$_{query}$  performance of ranking models   against the character-level and the word-level query attack, respectively.   
  We first analyze the defensive ability of different ranking models against the character-level query attack.
Then, we analyze the defensive ability of different ranking models against the world-level query attack.
Recall that high values of DR$_{query}$ attest to increased ranking robustness.

\subsubsection{\textbf{Analysis on the defensive ability against the character-level query attack}}
 We first analyze the defensive ability against the character-level query attack. 
In the following, we first given an overall analysis of different ranking models. 
Then, we analyze traditional probabilistic ranking models, LTR models and neural ranking models, respectively. 

\label{sec: character-level query attack}

\begin{itemize}[leftmargin=*]
	\item \textbf{Overall analysis on all the ranking models.}
Firstly, we given an overall analysis on all the ranking models against the query attack. We can observe that: 
(1) In the absence of any query attack, \textit{ColBERT} performs the best (0.3360) followed by \textit{BERT} (0.2798), \textit{LambdaMART} (0.2710) and then \textit{BM25} (0.2612), in terms of MRR@100. 
However, even single-character attacks can be catastrophic, resulting in a significantly degraded performance of 26.1\%, 27.6\%, 13.2\% and 23.3\% in terms of DR$_{query}$, for \textit{ColBERT}, \textit{BERT}, \textit{LambdaMART}, and \textit{BM25}, respectively. 
(2) For all the ranking models, the MRR@100 performances on both \textit{1-char} and \textit{2-char} attacked queries are worse than that on the original queries. 
These results indicate that all the ranking models are not able to generalize to attacked queries quite well. 
(3) The absolute drop rate value between \textit{2-char} attacked queries and original queries are much higher than that between \textit{1-char} attacked queries and original queries (e.g., 46.4\% v.s. 20.8\% for \textit{RankSVM}). 
By conducting further analysis, we find that the absolute drop rate value could be larger with more characters in a query attacked. 
Therefore, it is necessary to improve the robustness of  ranking models against the query attack. 
(4) Overall, in terms of the DR$_{query}$ between \textit{1-char} attacked and original queries, the relative robustness order of defending against query attack is \textit{DRMM} < \textit{QL} < \textit{BERT} < \textit{ColBERT} < \textit{BM25} < \textit{RankSVM}  < \textit{Prank}
< \textit{LambdaMART} < \textit{Conv-KNRM} < \textit{Duet}
  < \textit{DSSM}.

\begin{table}[t]
\centering 
 \renewcommand{\arraystretch}{1.3}
 \setlength\tabcolsep{11pt}
 \caption{ The performance of different ranking models against the character-level query attack on the MS MARCO dataset. `$-$' indicates statistically significant degradation with respect to the original query performance ($\textit{p-value} \leq 0.05$).}
\begin{tabular}{lccccc} 
\toprule \hline
 &  Original & \multicolumn{2}{c}{1-char} & \multicolumn{2}{c}{2-char}  \\
 \cmidrule(r){2-2} \cmidrule(r){3-4} \cmidrule(r){5-6} 
 &  MRR@100      & MRR@100 & DR$_{query}$ & MRR@100 & DR$_{query}$\\
\hline
QL & 0.2221 & 0.1540$^-$ & -30.7\% & 0.1213$^-$ & -45.4\%   \\
BM25 & 0.2612 & 0.2004$^-$ & -23.3\%  & 0.1453$^-$ & -44.4\%  \\
\hline
Prank & 0.2311 & 0.1839$^-$ & -20.4\% & 0.1218$^-$ & -47.3\% \\
RankSVM & 0.2535 & 0.2008$^-$ & -20.8\%  & 0.1360$^-$ & -46.4\%   \\
LambdaMART & 0.2710 & 0.2353$^-$ & -13.2\% & \textbf{0.2100}$^-$ & -22.5\%   \\
\hline
DSSM & 0.1220 & 0.1210 & \textbf{-0.8\%} & 0.1209 & \textbf{-0.9\%} \\
DRMM & 0.1335 & 0.0875$^-$ & -34.5\%  & 0.0828$^-$ & -38.0\%  \\
Conv-KNRM & 0.1716 & 0.1526$^-$ & -11.1\% & 0.1359$^-$ & -20.8\% \\
Duet & 0.1392 & 0.1315 & -5.5\% & 0.1310 & -5.9\%  \\
BERT & 0.2798 & 0.2026$^-$ & -27.6\%  & 0.1376$^-$ & -50.8\%  \\
ColBERT & \textbf{0.3360} & \textbf{0.2605}$^-$ & -26.1\% & 0.1640$^-$ & -51.2\% \\

\hline \bottomrule
\end{tabular}
\label{table:drop rate query noise}
\end{table}

    \item \textbf{Analysis on traditional probabilistic ranking models and LTR models.}
When we look at the traditional probabilistic ranking models and LTR models, we find that: 
(1) Traditional probabilistic ranking models are less robust than LTR models under \textit{1-char} attack (e.g., -30.7\% of \textit{QL} v.s. -13.2\% of \textit{LambdaMART}). 
Specifically, the adversarial edits might flip words either to a different word in the vocabulary or, more often, to the out-of-vocabulary token UNK. 
Consequently, adversarial edits can degrade ranking models by transforming the informative words in a query to UNK.  
For traditional probabilistic ranking models, they emphasize too much on exact matching signals between the query and the document, and treat each unique character combination differently, resulting in the vulnerability to query attack. 
(2) For the LTR models, \textit{LambdaMART} is more robust than \textit{RankSVM}   and \textit{Prank}, especially against the \textit{2-char} attack. 
One possible explanation is that as a listwise LTR method, \textit{LambdaMART} can make use of the whole ranked document list, which can better reduce the effect of character-level adversarial attacks with more context information than the pairwise LTR method \textit{RankSVM}   and the pointwise LTR method \textit{Prank}. 

    \item \textbf{Analysis on neural ranking models.}
When we look at the neural ranking models, we find that: 
 (1) \textit{DSSM} is the most robust model against both the \textit{1-char} and \textit{2-char} query attack, but performs poorly on the original query set. 
The reason might be that DSSM utilizes a character-level n-gram based word hashing, which is more robust to misspelling problems than just treating each word as the basic semantic units. 
(2) \textit{Duet} is the  second
 most robust model against both the \textit{1-char} and \textit{2-char} query attack, 
   while also performs poorly on the original query set.  
Specifically, \textit{Duet} leverages both the local and distributed representations for text matching.  
In the distributed representation, an activation pattern that has some errors or other differences from past data can still be mapped to 
the query, using a similarity function. 
In this way, \textit{Duet} is robust to noise and has the ability to generalize \cite{Duet}. 
(3) Since   \textit{DSSM} and \textit{Duet} have shown strong robustness, a promising direction to design a robust neural ranking model   against the misspelling lies in applying the  character-level operations.
For instance, we could combine pre-training objectives with the   n-gram word hashing layer to simultaneously achieve good ranking effectiveness and robustness.
(4) \textit{DRMM} is the least robust model against the \textit{1-char} attacked queries. 
The result indicates that the adversarial edits significantly degrade the \textit{DRMM} model which directly uses GloVe \cite{pennington2014glove} word embeddings, by transforming the informative words to UNK.  
Besides, for \textit{DRMM}, the improvement of the absolute drop rate value with respect to \textit{2-char} queries over \textit{1-char} queries is only 3.5\%. 
It is an interesting finding that while \textit{DRMM} is susceptible to adversarial typos, it shows some robustness against the further attacks. 
(5) When we take a look at the pre-trained models (i.e., \textit{BERT} and \textit{ColBERT}), the phenomenon is absolutely different with \textit{DRMM}. 
For example, for \textit{ColBERT}, the improvement of absolute drop rate value with respect to \textit{2-char} queries over \textit{1-char} queries is 25.1\%, which is the highest among all the ranking models. 
These results indicate that the pre-trained models are less robust than other ranking models against the query attack. 
One possible explanation could be that such pre-trained models apply WordPiece \cite{WordPiece} tokenization, where a word is broken down into more than one sub-words. 
In this way, an attacked word in a query may be decomposed into more than one  attacked sub-words, which have a more significant impact on the performance.   

\end{itemize}

\begin{table}[t]
\centering 
 \renewcommand{\arraystretch}{1.3}
 \setlength\tabcolsep{15pt}
 \caption{ The performance of different ranking models against the word-level query attack on the MS MARCO dataset. `$-$' indicates statistically significant degradation with respect to the original query performance ($\textit{p-value} \leq 0.05$).}
\begin{tabular}{lccc} 
\toprule \hline
 &  Original & \multicolumn{2}{c}{word-level query attack}   \\
 \cmidrule(r){2-2} \cmidrule(r){3-4}
 &  MRR@100      & MRR@100 & DR$_{query}$ \\
\hline
QL & 0.2221 &  0.1433$^-$ & -35.5\% \\
BM25 & 0.2612 & 0.1915$^-$  & -24.3\% \\
\hline
Prank & 0.2311 & 0.1976$^-$ & -14.5\% \\
RankSVM & 0.2535 & 0.2184$^-$ & -13.8\% \\
LambdaMART & 0.2710 &  \textbf{0.2483}$^-$ &  -8.4\%  \\
\hline
DSSM & 0.1220 & 0.1208 & \textbf{-1.0\%} \\
DRMM & 0.1335 & 0.0898$^-$  & -32.7\%   \\
Conv-KNRM & 0.1716 & 0.1528$^-$  & -11.0\% \\
Duet  & 0.1392 & 0.1325 & -4.8\%  \\
BERT  & 0.2798 & 0.1931$^-$ & -31.0\%   \\
ColBERT & \textbf{0.3360}& 0.2221$^-$ & -33.9\% \\
 
\hline \bottomrule
\end{tabular}
\label{table:word-level query attack}
\end{table}

\subsubsection{\textbf{Analysis on the defensive ability against the word-level query attack}}
  We analyze the defensive ability against the word-level query attack. From the results, we can observe that:
(1) Overall, for all the ranking models, the MRR@100 performance on the word-level query attack is worse than that on the original queries. The results indicate that all the ranking models are not able to generalize to word-level attacked queries quite well.
(2) In terms of the DR$_{query}$ between \textit{word-level} attacked and original queries, the relative robustness order of defending against word-level query attack is \textit{QL} < \textit{ColBERT} < \textit{DRMM} < \textit{BERT} < \textit{BM25} < \textit{Prank} < \textit{RankSVM}
< \textit{Conv-KNRM} < \textit{LambdaMart} < \textit{Duet} < \textit{DSSM}.
(3) Traditional probabilistic ranking models are less robust than LTR models under \textit{word-level} attack (e.g., -35.5\% of \textit{QL} v.s. -8.4\% of \textit{LambdaMART}). Meanwhile, for the LTR models, \textit{LambdaMART} is more robust than \textit{RankSVM} and \textit{Prank}. The reason is similar to what we have analyzed in Section \ref{sec: character-level query attack}.
(4) It is surprising to find that \textit{DSSM} and \textit{Duet} are the most robust ranking models against the \textit{word-level} query attack. In addition, the DR$_{query}$ of \textit{Conv-KNRM} is also relatively high (e.g., -11.0\%). 
    The three ranking models all apply the character-level n-gram operations. The results indicates that ranking models which contains a character-level operation will be more robust than those ranking models which treat each word as the basic semantic units.
(5) When we take a look at pre-trained models (i.e., BERT and ColBERT), we could find they are not robust compared with other neural ranking models. For example, the DR$_{query}$ of \textit{ColBERT} is -33.9\%, which is the highest among all the neural ranking models. 
    One possible explanation could be that WordPiece \cite{WordPiece} tokenization decompose an attacked word into more attacked sub-words, which have a more significant impact on the performance.

\subsection{Ranking Robustness under Document Manipulation}

In this section, we analyze the empirical results on the defensive ability against document manipulation. 
We first measure the average effectiveness of different ranking models and then   show the robustness evaluation results of different models in terms of TC and KT.

\begin{figure}[t]
\centering
\includegraphics[scale=0.57]{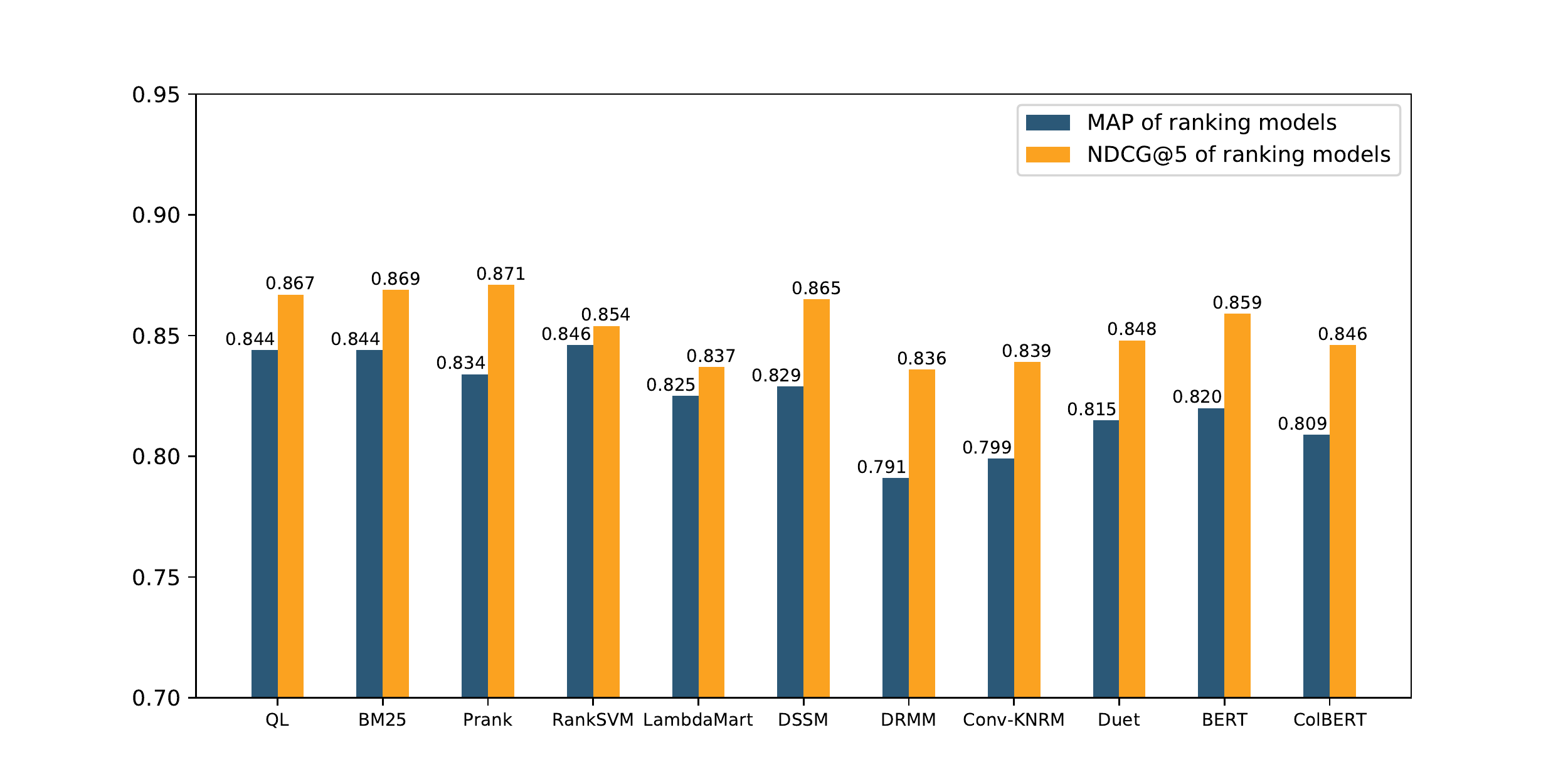}
\caption{The ranking effectiveness of ranking models on the ASRC dataset in terms of MAP and NDCG@5.}
\label{figure: document manipulation effectiveness}
\end{figure}

\subsubsection{\textbf{Analysis on the effectiveness of ranking models}}

We first analyze the average effectiveness of ranking models on the ASRC dataset. 
Specifically, we use the mean average precision (MAP) \cite{MAP} and normalized discounted cumulative gain at rank  5 (NDCG@5)  \cite{nDCG} to evaluate ranking effectiveness \cite{goren2018SIGIR}.  
The results are shown in Figure \ref{figure: document manipulation effectiveness}.

From the results we observe that: 
(1) All the ranking models can generally achieve good effectiveness on the ASRC dataset. 
The high MAP and NDCG values can be attributed to the fact that most documents generated by the students were judged to be relevant (e.g., 1113 out of 1279 documents are relevant) \cite{goren2018SIGIR}.
(2)   Overall, traditional probabilistic ranking models perform better than neural ranking models and LTR models. 
(3)   Among all the neural ranking models,  \textit{DSSM} performs the best. 
(4)   \textit{BERT} performs second only to DSSM in neural ranking models. 
Specifically, we train the ranking models on the ClueWeb09-B dataset and then evaluate them on the ASRC dataset, i.e., under the OOD setting. 
As noted in Section \ref{sec:OOD}, we have found that \textit{BERT} shows greater robustness than other neural ranking models in terms of the OOD generalizability, which is quite consistent with the result on the ASRC.  
(5) Overall, the relative average effectiveness order in terms of MAP is \textit{DRMM} < \textit{Conv-KNRM} < \textit{ColBERT} < \textit{Duet} < \textit{BERT} < \textit{LambdaMART} <   \textit{DSSM} < \textit{Prank} <   \textit{QL} = \textit{BM25} < \textit{RankSVM}.

\subsubsection{\textbf{Analysis on the robustness of ranking models}}

\begin{figure}[h]
\centering
\includegraphics[scale=0.57]{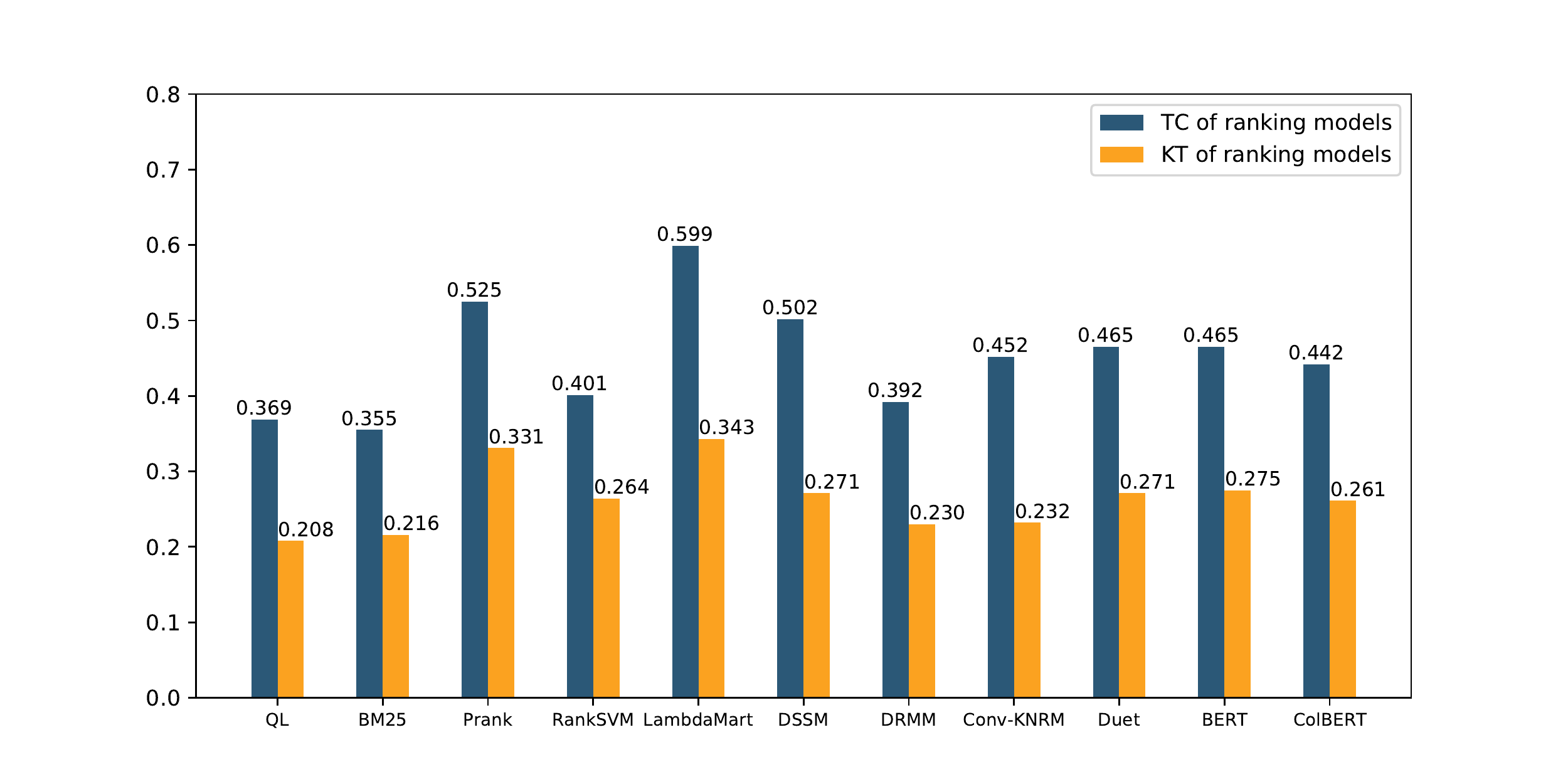}
\caption{The ranking robustness of ranking models on the ASRC dataset against  document manipulation in terms of TC and KT.}
\label{figure: document manipulation robustness}
\end{figure}

Here, we analyze the robustness of ranking models against document manipulation on the ASRC dataset.  
The results are shown in Figure \ref{figure: document manipulation robustness}. Recall that high values of KT and TC attest to decreased ranking robustness, as these are measures of inter-list distance.  

From the results we observe that: 
(1) For the traditional probabilistic ranking models, \textit{QL} and \textit{BM25} achieve the lowest TC and KT performance among all the ranking models, indicating that traditional probabilistic ranking models are the most robust to defend against document manipulations. 
In the generation process of the ASRC dataset, students were asked to modify relevant documents so as to have them ranked as high as possible in the next round. 
This suggests that traditional probabilistic ranking models have the lowest sensitivity to the difference between relevant documents.    
(2) For the LTR models, the ranked lists induced by \textit{RankSVM}    and \textit{Prank} are more robust than those induced by \textit{LambdaMART}. 
Since \textit{RankSVM}    and \textit{Prank} is linear and \textit{LambdaMART} is not, the variance of \textit{RankSVM}    and \textit{Prank} is in general lower, and we saw that its ranking robustness is higher.
  Meanwhile, the ranked lists induced by \textit{Prank} is less robust than those induced by \textit{RankSVM}, indicating that pointwise learning objective may suffer more from the document manipulation compared with pairwise learning objective. 
(3) For the neural ranking models, they achieve comparable results with LTR models against document manipulation. 
The reason might be that, flipping original words in a document to new words for higher rank, both the statistical features and word embeddings change significantly, resulting in the performance drop. 
(4) Overall, the relative robustness order of defending against document manipulation (e.g., in terms of TC) is \textit{LambdaMART} <   \textit{Prank} < \textit{DSSM} < \textit{BERT} = \textit{Duet} < \textit{Conv-KNRM} < \textit{ColBERT} < \textit{RankSVM} < \textit{DRMM} < \textit{QL} < \textit{BM25}.

\section{Related Work}
\label{sec:related work}
In this section, we briefly review the works related to our study,
including the performance variance over different queries, the OOD generalizability and the adversarial operation. 

\subsection{Performance Variance}
The performance variance over different queries has been studied as an early exploration of robustness in IR \cite{trec2003overview, trec2004overview, trec2005overview, collins2009reducing, zighelnic2008query, zhang2013bias, zhang2014bias_qikan}. 
The key idea is to pay more attention to the difference on ranking performance, rather than the average effectivenss performance which is considered by most ranking models.

Early works \cite{trec2003overview, trec2004overview, trec2005overview} mainly focused on the poorly performing queries of ranking systems. 
They evaluated the ranking robustness by proposing new metrics which emphasized the  poorly performing queries.
Beyond the widely-used average performance metrics such as the mean average precision (MAP) and the average of precision at rank 10 (P@10), \citet{trec2003overview} proposed the percentage of queries with no relevant in the top 10 retrieved (\%no) and the area underneath the MAP(X) vs. X curve to test the robustness of a ranking model.
Besides, \citet{trec2004overview} proposed the gMAP, which is a variant of the traditional MAP measure that used a geometric mean rather than an arithmetic mean to average individual query results, to evaluate the ranking robustness.
The proposed gMAP showed promise of giving appropriate emphasis to poorly performing queries while being more stable at equal query set sizes.
In addition, \citet{trec2005overview} used the document from AQUAINT Corpus of English News Text to test poorly performing queries and summarized the results of the three-year run of the Robust track.

Besides, some works \cite{collins2009reducing, zighelnic2008query} proposed the evaluation metrics to analyze the the performance variance over queries. 
For example, \citet{collins2009reducing} proposed the \textit{R-Loss} and \textit{Robustness Index} to measure the robustness on the query expansion. 
\textit{R-Loss} computed the averaged net loss of relevant documents in the retrieved document, due to the failure on query expansion. 
And \textit{Robustness Index} was defined as the difference between the number of improved queries and the number of hurt queries. The improved/hurt queries refer to the queries whose performance is improved/hurt over the original model.
Besides, \citet{zighelnic2008query} proposed $<Init$ to calculate the percentage of queries for which the retrieval performance of a query model is worse than that of the original query model.

However, such robustness metrics were often defined separately from the effectiveness metric (e.g., MAP).
There was a lack of a unified formulation to allow them to be evaluated in an integrated manner.
Later, a series of methods were proposed based from the perspective of bias-variance decomposition in IR to solve the problem.
\citet{zhang2013bias, zhang2014bias_qikan} analyzed the IR robustness from the perspective of bias-variance decomposition of IR evaluation. They proposed the variance of AP to analyze the robustness of ranking models in a unified framework with ranking effectiveness.
\citet{cormack2019quantifying} quantified bias and variance of system rankings by treating rankings as embeddings in a Euclidean space. They then showed that shallow pooling has substantially higher bias and insubstantially lower variance than probability-proportional-to-size sampling.

Recently, \citet{shivaswamy2021bias} defined notions of bias and variance directly on pairwise ordering of items. 
They showed that ranking disagreements between true orderings and a ranking function could be decomposed into bias and variance components.
This decomposition is similar to the squared loss and other losses that have been previously studied. 
\subsection{OOD Generalizability}

The OOD generalizability refers to the model's ability to generalize to various new test sets. 
We first briefly review the works which study the OOD generalizability in Computer Vision (CV) and Natural language processing (NLP). 
Then we review the recent works in the IR field. 

Extensive prior works \cite{hendrycks2020many, ACL2020pretrained, hendrycks2019scaling, koh2020wilds} have studied the OOD generalizability on CV and NLP. 
These works aimed to fairly measure the neural models' OOD generalizability in benchmark dataset of image/text classification. 
\citet{hendrycks2020many} systematically studied seven robustness hypotheses on image classification and proposed three new robustness benchmarks to analyze them. The authors showed that using large models and synthetic data augmentation could improve robustness on real-world distribution shifts.
\citet{ACL2020pretrained} systematically measured OOD generalization for seven NLP datasets by constructing a new robustness benchmark with realistic distribution shifts. By measuring OOD generalization and OOD detection of different neural models, they showed that pre-trained transformers are more effective at detecting 	OOD examples.
\citet{koh2020wilds} focused on the real world distribution shifts and summarized them into two types: domain generalization and subpopulation shift. The authors then proposed eight benchmark datasets in the wild to evaluate these distribution shifts.

Recently, since deep learning methods became more prevalent to solve the ranking problems, many works began to study the problem of OOD generalizalibility in IR.
Such works mainly focused on the cross domain adaption for neural ranking models.
\citet{cohen2018cross} proposed cross domain regularization on the ranking models to improve their performance on unforeseen domains. 
They adopted the adversarial learning by using an adversarial discriminator and the gradient reversal layers \cite{ganin2016domain} to train their neural ranking model on a small set of domains.
The effectiveness of their proposed architecture was evaluated on cross domain question answering data.
\citet{long2018deep} also leveraged the adversarial learning framework by devising a domain discriminator to solve the problem of domain adaptation. 
They evaluated the effectiveness of their proposed method on three domain transfer tasks, including cross-domain digits retrieval, image to image and image to videos transfers, on several benchmarks.
\citet{mao2018deep} utilized the multi-layer joint kernelized mean distance to measure the distance between the target data and the source data. They measured the distance based on deep features extracted from the deep networks. 
The target data which was found by their method was then iteratively added to the training data.
In ad-hoc retrieval, \citet{yilmaz2019cross} aggregated sentence-level evidence to fine-tune BERT models.
The methods were proposed to solve the challenges of exceeding document length and document-level relevance judgments for BERT.
By this way, the fine-tuned BERT models could capture cross-domain notions of relevance and can be directly used for ranking documents from other domains.

\subsection{Adversarial Operation}

Deep neural networks have been found vulnerable to adversarial operations \cite{szegedy2014ICLR, goodfellow2015ICLR}.
Specifically, the key idea of adversarial operation is to find a minimal perturbation in the data that can maximize the risk of making wrong predictions. 

Early works have extensively studied the adversarial operation in CV \cite{goodfellow2015ICLR, miyato2016adversarial, madry2017towards, JSMA_0}.
Adversarial operation can be classified as white-box adversarial attacks and black-box adversarial attacks, which means that adversaries have full access to target models (e.g., model's architecture and parameters) or no knowledge about target models, respectively.
Fast gradient sign method (FGSM) \cite{goodfellow2015ICLR} and its variants (e.g., FGM \cite{miyato2016adversarial} and PGD \cite{madry2017towards}) were classical white-box gradient-based attack methods. These methods added the noise to the whole image based on the gradient of loss.
On the contrary, another popular approach Jacobian-based saliency map (JSMA) \cite{JSMA_0, JSMA_1, JSMA_2} only perturbed one pixel at a time. It chose the pixel with the highest saliency score, which was defined as the gradient of the target class multiplied by the sum of the gradients of other classes. As a result, JSMA produced adversarial examples which increased the probability of the target class while decreased others.
While the above works focused on the gradient information of models, \citet{transferability} utilized the transferability of adversarial examples to train substitute models, in order to conduct a black-box attack.
Another representative black-box attack was score-based method \cite{chen2017zoo, li2018nattack}, which utilized the output score of model to conduct attack.

Adversarial operations have also been conducted to the NLP field, including text classification \cite{22gao2018black, ebrahimi2018NLProbust, 25sato2018interpretable, 26liang2017deep}, sentiment analysis \cite{25sato2018interpretable,26liang2017deep,20li2018textbugger, 23alzantot2018generating} and natural language inference \cite{23alzantot2018generating, 28minervini2018adversarially}.
Different from CV where the image is represented as continuous data, the main problem of generating adversarial text in NLP is the discrete input space. 
To address the problem, \citet{gong2018adversarial} proposed to attack in the embedding space. They used FGSM to produce perturbations in word embedding and used nearest neighbor search to find the closest words. 
However, this approach treated all tokens as equally vulnerable and replace all tokens with their nearest neighbors, which led to non-sensical, word-salad outputs \cite{xu2021grey}.
A number of works \cite{jia2017NLProbust, samanta2017towards, 26liang2017deep, jin2020bert} utilized the ideas of JSMA (e.g., find important pixel) to solve this problem.
For example, \citet{jin2020bert} first found the most important words by sorting all the original words in terms of their importance score. The importance score was calculated as the prediction change before and after deleting the word. Then the word was replaced by synonym which has the most similar semantic meaning with the original word. The candidate synonyms were selected to force the target model to make wrong predictions.
This method was more efficient since it attacked the most importance words. 
Meanwhile, it was classified as a black-box method since it only required the prediction score of the model.

In IR, early works on adversarial information retrieval mainly focused on identifying and addressing different types of spam \cite{castillo2010adversialwebsearch}. 
Recently, \citet{raifer2017SIGIR} studied the ranking competition on the web retrieval, which means that the document authors manipulate their documents intentionally to have them highly ranked. 
They constructed the ASRC dataset to simulate the ranking competition and used game theory to analyze the strategies employed by document authors.
Following the above work, \citet{goren2018SIGIR} defined the ranking robustness and analyzed the robustness of ranking functions based on the ASRC dataset. 
  For the work related to the query attack, query spelling correction has been a lively research topic since the mid 2000's, especially in the NLP community \cite{ahmad2005learning, cucerzan2004spelling}. The Microsoft Speller Challenge held in the year 2011 \cite{wang2011review} also attracted much attention for the problem of query spelling correction. 
Hagen \cite{hagen2017large} presented a new large-scale collection of 54,772 queries with manually annotated spelling corrections.
A recent work \cite{zhuang2021dealing} proposed to train a more robust Dense Retriever and BERT re-ranker which are robust to typos in queries.
Overall, the research area on the adversarial operation in IR is largely unexplored up till now, and more efforts are expected in this direction in the short future.

%
\section{Conclusion and Future Work}
\label{sec:con}

In this paper, we systematically analyzed the robustness of several representative neural ranking models against traditional probabilistic ranking models and LTR models. 
Specifically, we proposed three ways to define the robustness, i.e., the performance variance under I.I.D. settings, the OOD generalizability and the defensive ability against adversarial operations. 
The latter two definitions were further specialized into two different perspectives respectively.  
Overall, the results showed that neural ranking models are in general not robust as compared with other IR models in terms of  3 out of 5 robustness tasks. 
The exception is that the pre-trained ranking models achieve the best robustness from the perspective of the performance variance, while DSSM, Duet and Conv-KNRM model show robustness in terms of the defensive ability against query attacks. 
More research efforts are needed to develop robust neural ranking models for IR. 
Future research efforts could explore novel pre-training objectives and   model architectures with character-level operations that enhance the robustness of ranking models.
The analysis of the ranking robustness is a foundation for designing ideal ranking models in real world applications. 
We believe the way we study the robustness (definition and metric) as well as our findings would be beneficial to the IR community.

In the future work, we would like to apply the findings to improve the robustness of existing ranking models. 
We would also try to design new robust neural ranking models based on the findings in this work.
Moreover, it is valuable to define a unified formulation based on the different perspectives of the robustness to analyze the ranking models comprehensively.

\normalem
\bibliographystyle{ACM-Reference-Format}
\bibliography{acmart}


\begin{thebibliography}{104}


\ifx \showCODEN    \undefined \def \showCODEN     #1{\unskip}     \fi
\ifx \showDOI      \undefined \def \showDOI       #1{#1}\fi
\ifx \showISBNx    \undefined \def \showISBNx     #1{\unskip}     \fi
\ifx \showISBNxiii \undefined \def \showISBNxiii  #1{\unskip}     \fi
\ifx \showISSN     \undefined \def \showISSN      #1{\unskip}     \fi
\ifx \showLCCN     \undefined \def \showLCCN      #1{\unskip}     \fi
\ifx \shownote     \undefined \def \shownote      #1{#1}          \fi
\ifx \showarticletitle \undefined \def \showarticletitle #1{#1}   \fi
\ifx \showURL      \undefined \def \showURL       {\relax}        \fi
\providecommand\bibfield[2]{#2}
\providecommand\bibinfo[2]{#2}
\providecommand\natexlab[1]{#1}
\providecommand\showeprint[2][]{arXiv:#2}

\bibitem[\protect\citeauthoryear{Adhikari, Ram, Tang, and Lin}{Adhikari
  et~al\mbox{.}}{2019}]%
        {bert_cls_app}
\bibfield{author}{\bibinfo{person}{Ashutosh Adhikari}, \bibinfo{person}{Achyudh
  Ram}, \bibinfo{person}{Raphael Tang}, {and} \bibinfo{person}{Jimmy Lin}.}
  \bibinfo{year}{2019}\natexlab{}.
\newblock \showarticletitle{Docbert: Bert for document classification}.
\newblock \bibinfo{journal}{\emph{arXiv preprint arXiv:1904.08398}}
  (\bibinfo{year}{2019}).
\newblock


\bibitem[\protect\citeauthoryear{Ahmad and Kondrak}{Ahmad and Kondrak}{2005}]%
        {ahmad2005learning}
\bibfield{author}{\bibinfo{person}{Farooq Ahmad} {and}
  \bibinfo{person}{Grzegorz Kondrak}.} \bibinfo{year}{2005}\natexlab{}.
\newblock \showarticletitle{Learning a spelling error model from search query
  logs}. In \bibinfo{booktitle}{\emph{Proceedings of Human Language Technology
  Conference and Conference on Empirical Methods in Natural Language
  Processing}}. \bibinfo{pages}{955--962}.
\newblock


\bibitem[\protect\citeauthoryear{Akhtar and Mian}{Akhtar and Mian}{2018}]%
        {akhtar2018threat}
\bibfield{author}{\bibinfo{person}{Naveed Akhtar} {and} \bibinfo{person}{Ajmal
  Mian}.} \bibinfo{year}{2018}\natexlab{}.
\newblock \showarticletitle{Threat of adversarial attacks on deep learning in
  computer vision: A survey}.
\newblock \bibinfo{journal}{\emph{Ieee Access}}  \bibinfo{volume}{6}
  (\bibinfo{year}{2018}), \bibinfo{pages}{14410--14430}.
\newblock


\bibitem[\protect\citeauthoryear{Alzantot, Sharma, Elgohary, Ho, Srivastava,
  and Chang}{Alzantot et~al\mbox{.}}{2018}]%
        {23alzantot2018generating}
\bibfield{author}{\bibinfo{person}{Moustafa Alzantot},
  \bibinfo{person}{Yash~Sharma Sharma}, \bibinfo{person}{Ahmed Elgohary},
  \bibinfo{person}{Bo-Jhang Ho}, \bibinfo{person}{Mani Srivastava}, {and}
  \bibinfo{person}{Kai-Wei Chang}.} \bibinfo{year}{2018}\natexlab{}.
\newblock \showarticletitle{Generating Natural Language Adversarial Examples}.
  In \bibinfo{booktitle}{\emph{Proceedings of the 2018 Conference on Empirical
  Methods in Natural Language Processing}}.
\newblock


\bibitem[\protect\citeauthoryear{Broder}{Broder}{2002}]%
        {broder2002taxonomy}
\bibfield{author}{\bibinfo{person}{Andrei Broder}.}
  \bibinfo{year}{2002}\natexlab{}.
\newblock \showarticletitle{A taxonomy of web search}. In
  \bibinfo{booktitle}{\emph{ACM Sigir forum}}, Vol.~\bibinfo{volume}{36}. ACM
  New York, NY, USA, \bibinfo{pages}{3--10}.
\newblock


\bibitem[\protect\citeauthoryear{Burges, Shaked, Renshaw, Lazier, Deeds,
  Hamilton, and Hullender}{Burges et~al\mbox{.}}{2005}]%
        {nDCG}
\bibfield{author}{\bibinfo{person}{Chris Burges}, \bibinfo{person}{Tal Shaked},
  \bibinfo{person}{Erin Renshaw}, \bibinfo{person}{Ari Lazier},
  \bibinfo{person}{Matt Deeds}, \bibinfo{person}{Nicole Hamilton}, {and}
  \bibinfo{person}{Greg Hullender}.} \bibinfo{year}{2005}\natexlab{}.
\newblock \showarticletitle{Learning to rank using gradient descent}. In
  \bibinfo{booktitle}{\emph{Proceedings of the 22nd international conference on
  Machine learning}}. \bibinfo{pages}{89--96}.
\newblock


\bibitem[\protect\citeauthoryear{Burges}{Burges}{2010}]%
        {LambdaMART}
\bibfield{author}{\bibinfo{person}{Christopher~JC Burges}.}
  \bibinfo{year}{2010}\natexlab{}.
\newblock \showarticletitle{From ranknet to lambdarank to lambdamart: An
  overview}.
\newblock \bibinfo{journal}{\emph{Learning}} \bibinfo{volume}{11},
  \bibinfo{number}{23-581} (\bibinfo{year}{2010}), \bibinfo{pages}{81}.
\newblock


\bibitem[\protect\citeauthoryear{Castillo and Davison}{Castillo and
  Davison}{2011}]%
        {castillo2010adversialwebsearch}
\bibfield{author}{\bibinfo{person}{Carlos Castillo} {and}
  \bibinfo{person}{Brian~D Davison}.} \bibinfo{year}{2011}\natexlab{}.
\newblock \showarticletitle{Adversarial Web Search}.
\newblock \bibinfo{journal}{\emph{Foundations and Trends in Information
  Retrieval}} \bibinfo{volume}{4}, \bibinfo{number}{5} (\bibinfo{year}{2011}),
  \bibinfo{pages}{377--486}.
\newblock


\bibitem[\protect\citeauthoryear{Chen, Zhang, Sharma, Yi, and Hsieh}{Chen
  et~al\mbox{.}}{2017}]%
        {chen2017zoo}
\bibfield{author}{\bibinfo{person}{Pin-Yu Chen}, \bibinfo{person}{Huan Zhang},
  \bibinfo{person}{Yash Sharma}, \bibinfo{person}{Jinfeng Yi}, {and}
  \bibinfo{person}{Cho-Jui Hsieh}.} \bibinfo{year}{2017}\natexlab{}.
\newblock \showarticletitle{Zoo: Zeroth order optimization based black-box
  attacks to deep neural networks without training substitute models}. In
  \bibinfo{booktitle}{\emph{Proceedings of the 10th ACM workshop on artificial
  intelligence and security}}. \bibinfo{pages}{15--26}.
\newblock


\bibitem[\protect\citeauthoryear{Cohen, Mitra, Hofmann, and Croft}{Cohen
  et~al\mbox{.}}{2018}]%
        {cohen2018cross}
\bibfield{author}{\bibinfo{person}{Daniel Cohen}, \bibinfo{person}{Bhaskar
  Mitra}, \bibinfo{person}{Katja Hofmann}, {and} \bibinfo{person}{W~Bruce
  Croft}.} \bibinfo{year}{2018}\natexlab{}.
\newblock \showarticletitle{Cross domain regularization for neural ranking
  models using adversarial learning}. In \bibinfo{booktitle}{\emph{The 41st
  International ACM SIGIR Conference on Research \& Development in Information
  Retrieval}}. \bibinfo{pages}{1025--1028}.
\newblock


\bibitem[\protect\citeauthoryear{Collins-Thompson}{Collins-Thompson}{2009}]%
        {collins2009reducing}
\bibfield{author}{\bibinfo{person}{Kevyn Collins-Thompson}.}
  \bibinfo{year}{2009}\natexlab{}.
\newblock \showarticletitle{Reducing the risk of query expansion via robust
  constrained optimization}. In \bibinfo{booktitle}{\emph{Proceedings of the
  18th ACM conference on Information and knowledge management}}.
  \bibinfo{pages}{837--846}.
\newblock


\bibitem[\protect\citeauthoryear{Combey, Loison, Faucher, and Hajri}{Combey
  et~al\mbox{.}}{2020}]%
        {JSMA_2}
\bibfield{author}{\bibinfo{person}{Th{\'e}o Combey},
  \bibinfo{person}{Ant{\'o}nio Loison}, \bibinfo{person}{Maxime Faucher}, {and}
  \bibinfo{person}{Hatem Hajri}.} \bibinfo{year}{2020}\natexlab{}.
\newblock \showarticletitle{Probabilistic Jacobian-based Saliency Maps
  Attacks}.
\newblock \bibinfo{journal}{\emph{Machine Learning and Knowledge Extraction}}
  \bibinfo{volume}{2}, \bibinfo{number}{4} (\bibinfo{year}{2020}),
  \bibinfo{pages}{558--578}.
\newblock


\bibitem[\protect\citeauthoryear{Cormack and Grossman}{Cormack and
  Grossman}{2019}]%
        {cormack2019quantifying}
\bibfield{author}{\bibinfo{person}{Gordon~V Cormack} {and}
  \bibinfo{person}{Maura~R Grossman}.} \bibinfo{year}{2019}\natexlab{}.
\newblock \showarticletitle{Quantifying Bias and Variance of System Rankings}.
  In \bibinfo{booktitle}{\emph{Proceedings of the 42nd International ACM SIGIR
  Conference on Research and Development in Information Retrieval}}.
  \bibinfo{pages}{1089--1092}.
\newblock


\bibitem[\protect\citeauthoryear{Crammer, Singer, et~al\mbox{.}}{Crammer
  et~al\mbox{.}}{2001}]%
        {Prank}
\bibfield{author}{\bibinfo{person}{Koby Crammer}, \bibinfo{person}{Yoram
  Singer}, {et~al\mbox{.}}} \bibinfo{year}{2001}\natexlab{}.
\newblock \showarticletitle{Pranking with ranking.}. In
  \bibinfo{booktitle}{\emph{Nips}}, Vol.~\bibinfo{volume}{1}.
  \bibinfo{pages}{641--647}.
\newblock


\bibitem[\protect\citeauthoryear{Croft, Metzler, and Strohman}{Croft
  et~al\mbox{.}}{2010}]%
        {Precision}
\bibfield{author}{\bibinfo{person}{W~Bruce Croft}, \bibinfo{person}{Donald
  Metzler}, {and} \bibinfo{person}{Trevor Strohman}.}
  \bibinfo{year}{2010}\natexlab{}.
\newblock \bibinfo{booktitle}{\emph{Search engines: Information retrieval in
  practice}}. Vol.~\bibinfo{volume}{520}.
\newblock \bibinfo{publisher}{Addison-Wesley Reading}.
\newblock


\bibitem[\protect\citeauthoryear{Cucerzan and Brill}{Cucerzan and
  Brill}{2004}]%
        {cucerzan2004spelling}
\bibfield{author}{\bibinfo{person}{Silviu Cucerzan} {and} \bibinfo{person}{Eric
  Brill}.} \bibinfo{year}{2004}\natexlab{}.
\newblock \showarticletitle{Spelling correction as an iterative process that
  exploits the collective knowledge of web users}. In
  \bibinfo{booktitle}{\emph{Proceedings of the 2004 Conference on Empirical
  Methods in Natural Language Processing}}. \bibinfo{pages}{293--300}.
\newblock


\bibitem[\protect\citeauthoryear{Dai and Callan}{Dai and Callan}{2019}]%
        {BERT_FIRSTP}
\bibfield{author}{\bibinfo{person}{Zhuyun Dai} {and} \bibinfo{person}{Jamie
  Callan}.} \bibinfo{year}{2019}\natexlab{}.
\newblock \showarticletitle{Deeper Text Understanding for IR with Contextual
  Neural Language Modeling}. In \bibinfo{booktitle}{\emph{Proceedings of the
  42nd International ACM SIGIR Conference on Research and Development in
  Information Retrieval}}. ACM, \bibinfo{pages}{985--988}.
\newblock


\bibitem[\protect\citeauthoryear{Dai, Xiong, Callan, and Liu}{Dai
  et~al\mbox{.}}{2018}]%
        {Conv-KNRM}
\bibfield{author}{\bibinfo{person}{Zhuyun Dai}, \bibinfo{person}{Chenyan
  Xiong}, \bibinfo{person}{Jamie Callan}, {and} \bibinfo{person}{Zhiyuan Liu}.}
  \bibinfo{year}{2018}\natexlab{}.
\newblock \showarticletitle{Convolutional neural networks for soft-matching
  n-grams in ad-hoc search}. In \bibinfo{booktitle}{\emph{Proceedings of the
  eleventh ACM international conference on web search and data mining}}.
  \bibinfo{pages}{126--134}.
\newblock


\bibitem[\protect\citeauthoryear{Davis}{Davis}{2003}]%
        {davis2003psycholinguistic}
\bibfield{author}{\bibinfo{person}{Matt Davis}.}
  \bibinfo{year}{2003}\natexlab{}.
\newblock \bibinfo{title}{Psycholinguistic evidence on scrambled letters in
  reading}.
\newblock
\newblock


\bibitem[\protect\citeauthoryear{Ebrahimi, Rao, Lowd, and Dou}{Ebrahimi
  et~al\mbox{.}}{2018}]%
        {ebrahimi2018NLProbust}
\bibfield{author}{\bibinfo{person}{Javid Ebrahimi}, \bibinfo{person}{Anyi Rao},
  \bibinfo{person}{Daniel Lowd}, {and} \bibinfo{person}{Dejing Dou}.}
  \bibinfo{year}{2018}\natexlab{}.
\newblock \showarticletitle{HotFlip: White-Box Adversarial Examples for Text
  Classification}. In \bibinfo{booktitle}{\emph{Proceedings of the 56th Annual
  Meeting of the Association for Computational Linguistics (Volume 2: Short
  Papers)}}. \bibinfo{pages}{31--36}.
\newblock


\bibitem[\protect\citeauthoryear{Fan, Guo, Lan, Xu, Zhai, and Cheng}{Fan
  et~al\mbox{.}}{2018}]%
        {HiNT}
\bibfield{author}{\bibinfo{person}{Yixing Fan}, \bibinfo{person}{Jiafeng Guo},
  \bibinfo{person}{Yanyan Lan}, \bibinfo{person}{Jun Xu},
  \bibinfo{person}{Chengxiang Zhai}, {and} \bibinfo{person}{Xueqi Cheng}.}
  \bibinfo{year}{2018}\natexlab{}.
\newblock \showarticletitle{Modeling diverse relevance patterns in ad-hoc
  retrieval}. In \bibinfo{booktitle}{\emph{The 41st international ACM SIGIR
  conference on research \& development in information retrieval}}.
  \bibinfo{pages}{375--384}.
\newblock


\bibitem[\protect\citeauthoryear{Ganin, Ustinova, Ajakan, Germain, Larochelle,
  Laviolette, Marchand, and Lempitsky}{Ganin et~al\mbox{.}}{2016}]%
        {ganin2016domain}
\bibfield{author}{\bibinfo{person}{Yaroslav Ganin}, \bibinfo{person}{Evgeniya
  Ustinova}, \bibinfo{person}{Hana Ajakan}, \bibinfo{person}{Pascal Germain},
  \bibinfo{person}{Hugo Larochelle}, \bibinfo{person}{Fran{\c{c}}ois
  Laviolette}, \bibinfo{person}{Mario Marchand}, {and} \bibinfo{person}{Victor
  Lempitsky}.} \bibinfo{year}{2016}\natexlab{}.
\newblock \showarticletitle{Domain-adversarial training of neural networks}.
\newblock \bibinfo{journal}{\emph{The journal of machine learning research}}
  \bibinfo{volume}{17}, \bibinfo{number}{1} (\bibinfo{year}{2016}),
  \bibinfo{pages}{2096--2030}.
\newblock


\bibitem[\protect\citeauthoryear{Gao, Lanchantin, Soffa, and Qi}{Gao
  et~al\mbox{.}}{2018}]%
        {22gao2018black}
\bibfield{author}{\bibinfo{person}{Ji Gao}, \bibinfo{person}{Jack Lanchantin},
  \bibinfo{person}{Mary~Lou Soffa}, {and} \bibinfo{person}{Yanjun Qi}.}
  \bibinfo{year}{2018}\natexlab{}.
\newblock \showarticletitle{Black-box generation of adversarial text sequences
  to evade deep learning classifiers}. In \bibinfo{booktitle}{\emph{2018 IEEE
  Security and Privacy Workshops (SPW)}}. IEEE, \bibinfo{pages}{50--56}.
\newblock


\bibitem[\protect\citeauthoryear{Gong, Wang, Li, Song, and Ku}{Gong
  et~al\mbox{.}}{2018}]%
        {gong2018adversarial}
\bibfield{author}{\bibinfo{person}{Zhitao Gong}, \bibinfo{person}{Wenlu Wang},
  \bibinfo{person}{Bo Li}, \bibinfo{person}{Dawn Song}, {and}
  \bibinfo{person}{Wei-Shinn Ku}.} \bibinfo{year}{2018}\natexlab{}.
\newblock \showarticletitle{Adversarial texts with gradient methods}.
\newblock \bibinfo{journal}{\emph{arXiv preprint arXiv:1801.07175}}
  (\bibinfo{year}{2018}).
\newblock


\bibitem[\protect\citeauthoryear{Goodfellow, Shlens, and Szegedy}{Goodfellow
  et~al\mbox{.}}{2014}]%
        {goodfellow2015ICLR}
\bibfield{author}{\bibinfo{person}{Ian~J Goodfellow}, \bibinfo{person}{Jonathon
  Shlens}, {and} \bibinfo{person}{Christian Szegedy}.}
  \bibinfo{year}{2014}\natexlab{}.
\newblock \showarticletitle{Explaining and harnessing adversarial examples}.
\newblock \bibinfo{journal}{\emph{arXiv preprint arXiv:1412.6572}}
  (\bibinfo{year}{2014}).
\newblock


\bibitem[\protect\citeauthoryear{Goren}{Goren}{2019}]%
        {goren2019ranking}
\bibfield{author}{\bibinfo{person}{Gregory Goren}.}
  \bibinfo{year}{2019}\natexlab{}.
\newblock \showarticletitle{Ranking Robustness In Adversarial Retrieval
  Settings}. In \bibinfo{booktitle}{\emph{Proceedings of the 42nd International
  ACM SIGIR Conference on Research and Development in Information Retrieval}}.
  \bibinfo{pages}{1446--1446}.
\newblock


\bibitem[\protect\citeauthoryear{Goren, Kurland, Tennenholtz, and Raiber}{Goren
  et~al\mbox{.}}{2018}]%
        {goren2018SIGIR}
\bibfield{author}{\bibinfo{person}{Gregory Goren}, \bibinfo{person}{Oren
  Kurland}, \bibinfo{person}{Moshe Tennenholtz}, {and} \bibinfo{person}{Fiana
  Raiber}.} \bibinfo{year}{2018}\natexlab{}.
\newblock \showarticletitle{Ranking robustness under adversarial document
  manipulations}. In \bibinfo{booktitle}{\emph{The 41st International ACM SIGIR
  Conference on Research \& Development in Information Retrieval}}.
  \bibinfo{pages}{395--404}.
\newblock


\bibitem[\protect\citeauthoryear{Gu, Li, Liu, Ling, Su, Wei, and Zhu}{Gu
  et~al\mbox{.}}{2020}]%
        {sabert_app}
\bibfield{author}{\bibinfo{person}{Jia-Chen Gu}, \bibinfo{person}{Tianda Li},
  \bibinfo{person}{Quan Liu}, \bibinfo{person}{Zhen-Hua Ling},
  \bibinfo{person}{Zhiming Su}, \bibinfo{person}{Si Wei}, {and}
  \bibinfo{person}{Xiaodan Zhu}.} \bibinfo{year}{2020}\natexlab{}.
\newblock \showarticletitle{Speaker-aware bert for multi-turn response
  selection in retrieval-based chatbots}. In
  \bibinfo{booktitle}{\emph{Proceedings of the 29th ACM International
  Conference on Information \& Knowledge Management}}.
  \bibinfo{pages}{2041--2044}.
\newblock


\bibitem[\protect\citeauthoryear{Guo, Fan, Ai, and Croft}{Guo
  et~al\mbox{.}}{2016}]%
        {DRMM}
\bibfield{author}{\bibinfo{person}{Jiafeng Guo}, \bibinfo{person}{Yixing Fan},
  \bibinfo{person}{Qingyao Ai}, {and} \bibinfo{person}{W~Bruce Croft}.}
  \bibinfo{year}{2016}\natexlab{}.
\newblock \showarticletitle{A deep relevance matching model for ad-hoc
  retrieval}. In \bibinfo{booktitle}{\emph{Proceedings of the 25th ACM
  International on Conference on Information and Knowledge Management}}.
  \bibinfo{pages}{55--64}.
\newblock


\bibitem[\protect\citeauthoryear{Guo, Fan, Ji, and Cheng}{Guo
  et~al\mbox{.}}{2019}]%
        {matchzoo}
\bibfield{author}{\bibinfo{person}{Jiafeng Guo}, \bibinfo{person}{Yixing Fan},
  \bibinfo{person}{Xiang Ji}, {and} \bibinfo{person}{Xueqi Cheng}.}
  \bibinfo{year}{2019}\natexlab{}.
\newblock \showarticletitle{MatchZoo: A Learning, Practicing, and Developing
  System for Neural Text Matching}. In \bibinfo{booktitle}{\emph{Proceedings of
  the 42Nd International ACM SIGIR Conference on Research and Development in
  Information Retrieval}} \emph{(\bibinfo{series}{SIGIR'19})}.
  \bibinfo{publisher}{ACM}, \bibinfo{address}{New York, NY, USA},
  \bibinfo{pages}{1297--1300}.
\newblock
\showISBNx{978-1-4503-6172-9}
\urldef\tempurl%
\url{https://doi.org/10.1145/3331184.3331403}
\showDOI{\tempurl}


\bibitem[\protect\citeauthoryear{Hagen, Potthast, Gohsen, Rathgeber, and
  Stein}{Hagen et~al\mbox{.}}{2017}]%
        {hagen2017large}
\bibfield{author}{\bibinfo{person}{Matthias Hagen}, \bibinfo{person}{Martin
  Potthast}, \bibinfo{person}{Marcel Gohsen}, \bibinfo{person}{Anja Rathgeber},
  {and} \bibinfo{person}{Benno Stein}.} \bibinfo{year}{2017}\natexlab{}.
\newblock \showarticletitle{A large-scale query spelling correction corpus}. In
  \bibinfo{booktitle}{\emph{Proceedings of the 40th International ACM SIGIR
  Conference on Research and Development in Information Retrieval}}.
  \bibinfo{pages}{1261--1264}.
\newblock


\bibitem[\protect\citeauthoryear{Hendrycks, Basart, Mazeika, Mostajabi,
  Steinhardt, and Song}{Hendrycks et~al\mbox{.}}{2019}]%
        {hendrycks2019scaling}
\bibfield{author}{\bibinfo{person}{Dan Hendrycks}, \bibinfo{person}{Steven
  Basart}, \bibinfo{person}{Mantas Mazeika}, \bibinfo{person}{Mohammadreza
  Mostajabi}, \bibinfo{person}{Jacob Steinhardt}, {and} \bibinfo{person}{Dawn
  Song}.} \bibinfo{year}{2019}\natexlab{}.
\newblock \showarticletitle{Scaling Out-of-Distribution Detection for
  Real-World Settings}.
\newblock \bibinfo{journal}{\emph{arXiv preprint arXiv:1911.11132}}
  (\bibinfo{year}{2019}).
\newblock


\bibitem[\protect\citeauthoryear{Hendrycks, Basart, Mu, Kadavath, Wang,
  Dorundo, Desai, Zhu, Parajuli, Guo, et~al\mbox{.}}{Hendrycks
  et~al\mbox{.}}{2021}]%
        {hendrycks2020many}
\bibfield{author}{\bibinfo{person}{Dan Hendrycks}, \bibinfo{person}{Steven
  Basart}, \bibinfo{person}{Norman Mu}, \bibinfo{person}{Saurav Kadavath},
  \bibinfo{person}{Frank Wang}, \bibinfo{person}{Evan Dorundo},
  \bibinfo{person}{Rahul Desai}, \bibinfo{person}{Tyler Zhu},
  \bibinfo{person}{Samyak Parajuli}, \bibinfo{person}{Mike Guo},
  {et~al\mbox{.}}} \bibinfo{year}{2021}\natexlab{}.
\newblock \showarticletitle{The many faces of robustness: A critical analysis
  of out-of-distribution generalization}. In
  \bibinfo{booktitle}{\emph{Proceedings of the IEEE/CVF International
  Conference on Computer Vision}}. \bibinfo{pages}{8340--8349}.
\newblock


\bibitem[\protect\citeauthoryear{Hendrycks, Liu, Wallace, Dziedzic, Krishnan,
  and Song}{Hendrycks et~al\mbox{.}}{2020}]%
        {ACL2020pretrained}
\bibfield{author}{\bibinfo{person}{Dan Hendrycks}, \bibinfo{person}{Xiaoyuan
  Liu}, \bibinfo{person}{Eric Wallace}, \bibinfo{person}{Adam Dziedzic},
  \bibinfo{person}{Rishabh Krishnan}, {and} \bibinfo{person}{Dawn Song}.}
  \bibinfo{year}{2020}\natexlab{}.
\newblock \showarticletitle{Pretrained Transformers Improve Out-of-Distribution
  Robustness}. In \bibinfo{booktitle}{\emph{Proceedings of the 58th Annual
  Meeting of the Association for Computational Linguistics}}.
  \bibinfo{pages}{2744--2751}.
\newblock


\bibitem[\protect\citeauthoryear{Hofst{\"a}tter, Zlabinger, and
  Hanbury}{Hofst{\"a}tter et~al\mbox{.}}{2020}]%
        {hofstatter2020interpretable}
\bibfield{author}{\bibinfo{person}{Sebastian Hofst{\"a}tter},
  \bibinfo{person}{Markus Zlabinger}, {and} \bibinfo{person}{Allan Hanbury}.}
  \bibinfo{year}{2020}\natexlab{}.
\newblock \showarticletitle{Interpretable \& Time-Budget-Constrained
  Contextualization for Re-Ranking}.
\newblock In \bibinfo{booktitle}{\emph{ECAI 2020}}. \bibinfo{publisher}{IOS
  Press}, \bibinfo{pages}{513--520}.
\newblock


\bibitem[\protect\citeauthoryear{Huang, He, Gao, Deng, Acero, and Heck}{Huang
  et~al\mbox{.}}{2013}]%
        {DSSM}
\bibfield{author}{\bibinfo{person}{Po-Sen Huang}, \bibinfo{person}{Xiaodong
  He}, \bibinfo{person}{Jianfeng Gao}, \bibinfo{person}{Li Deng},
  \bibinfo{person}{Alex Acero}, {and} \bibinfo{person}{Larry Heck}.}
  \bibinfo{year}{2013}\natexlab{}.
\newblock \showarticletitle{Learning deep structured semantic models for web
  search using clickthrough data}. In \bibinfo{booktitle}{\emph{Proceedings of
  the 22nd ACM international conference on Information \& Knowledge
  Management}}. \bibinfo{pages}{2333--2338}.
\newblock


\bibitem[\protect\citeauthoryear{Huston and Croft}{Huston and Croft}{2014}]%
        {huston2014parameters}
\bibfield{author}{\bibinfo{person}{Samuel Huston} {and}
  \bibinfo{person}{W~Bruce Croft}.} \bibinfo{year}{2014}\natexlab{}.
\newblock \showarticletitle{Parameters learned in the comparison of retrieval
  models using term dependencies}.
\newblock \bibinfo{journal}{\emph{Ir, University of Massachusetts}}
  (\bibinfo{year}{2014}).
\newblock


\bibitem[\protect\citeauthoryear{Jia and Liang}{Jia and Liang}{2017}]%
        {jia2017NLProbust}
\bibfield{author}{\bibinfo{person}{Robin Jia} {and} \bibinfo{person}{Percy
  Liang}.} \bibinfo{year}{2017}\natexlab{}.
\newblock \showarticletitle{Adversarial Examples for Evaluating Reading
  Comprehension Systems}. In \bibinfo{booktitle}{\emph{Proceedings of the 2017
  Conference on Empirical Methods in Natural Language Processing}}.
  \bibinfo{pages}{2021--2031}.
\newblock


\bibitem[\protect\citeauthoryear{Jin, Jin, Zhou, and Szolovits}{Jin
  et~al\mbox{.}}{2020}]%
        {jin2020bert}
\bibfield{author}{\bibinfo{person}{Di Jin}, \bibinfo{person}{Zhijing Jin},
  \bibinfo{person}{Joey~Tianyi Zhou}, {and} \bibinfo{person}{Peter Szolovits}.}
  \bibinfo{year}{2020}\natexlab{}.
\newblock \showarticletitle{Is bert really robust? a strong baseline for
  natural language attack on text classification and entailment}. In
  \bibinfo{booktitle}{\emph{Proceedings of the AAAI conference on artificial
  intelligence}}, Vol.~\bibinfo{volume}{34}. \bibinfo{pages}{8018--8025}.
\newblock


\bibitem[\protect\citeauthoryear{Joachims}{Joachims}{2002}]%
        {RankSVM}
\bibfield{author}{\bibinfo{person}{Thorsten Joachims}.}
  \bibinfo{year}{2002}\natexlab{}.
\newblock \showarticletitle{Optimizing search engines using clickthrough data}.
  In \bibinfo{booktitle}{\emph{Proceedings of the eighth ACM SIGKDD
  international conference on Knowledge discovery and data mining}}.
  \bibinfo{pages}{133--142}.
\newblock


\bibitem[\protect\citeauthoryear{Joachims}{Joachims}{2006}]%
        {svm_rank}
\bibfield{author}{\bibinfo{person}{Thorsten Joachims}.}
  \bibinfo{year}{2006}\natexlab{}.
\newblock \showarticletitle{Training linear SVMs in linear time}. In
  \bibinfo{booktitle}{\emph{Proceedings of the 12th ACM SIGKDD international
  conference on Knowledge discovery and data mining}}.
  \bibinfo{pages}{217--226}.
\newblock


\bibitem[\protect\citeauthoryear{Jones, Jia, Raghunathan, and Liang}{Jones
  et~al\mbox{.}}{2020}]%
        {jones2020NLProbustEncoding}
\bibfield{author}{\bibinfo{person}{Erik Jones}, \bibinfo{person}{Robin Jia},
  \bibinfo{person}{Aditi Raghunathan}, {and} \bibinfo{person}{Percy Liang}.}
  \bibinfo{year}{2020}\natexlab{}.
\newblock \showarticletitle{Robust Encodings: A Framework for Combating
  Adversarial Typos}. In \bibinfo{booktitle}{\emph{Proceedings of the 58th
  Annual Meeting of the Association for Computational Linguistics}}.
  \bibinfo{pages}{2752--2765}.
\newblock


\bibitem[\protect\citeauthoryear{Joshi, Chen, Liu, Weld, Zettlemoyer, and
  Levy}{Joshi et~al\mbox{.}}{2020}]%
        {spanbert}
\bibfield{author}{\bibinfo{person}{Mandar Joshi}, \bibinfo{person}{Danqi Chen},
  \bibinfo{person}{Yinhan Liu}, \bibinfo{person}{Daniel~S Weld},
  \bibinfo{person}{Luke Zettlemoyer}, {and} \bibinfo{person}{Omer Levy}.}
  \bibinfo{year}{2020}\natexlab{}.
\newblock \showarticletitle{Spanbert: Improving pre-training by representing
  and predicting spans}.
\newblock \bibinfo{journal}{\emph{Transactions of the Association for
  Computational Linguistics}}  \bibinfo{volume}{8} (\bibinfo{year}{2020}),
  \bibinfo{pages}{64--77}.
\newblock


\bibitem[\protect\citeauthoryear{Kaneko and Komachi}{Kaneko and
  Komachi}{2019}]%
        {bert_st_app}
\bibfield{author}{\bibinfo{person}{Masahiro Kaneko} {and}
  \bibinfo{person}{Mamoru Komachi}.} \bibinfo{year}{2019}\natexlab{}.
\newblock \showarticletitle{Multi-head multi-layer attention to deep language
  representations for grammatical error detection}.
\newblock \bibinfo{journal}{\emph{Computaci{\'o}n y Sistemas}}
  \bibinfo{volume}{23}, \bibinfo{number}{3} (\bibinfo{year}{2019}),
  \bibinfo{pages}{883--891}.
\newblock


\bibitem[\protect\citeauthoryear{Kenton and Toutanova}{Kenton and
  Toutanova}{2019}]%
        {BERT}
\bibfield{author}{\bibinfo{person}{Jacob Devlin Ming-Wei~Chang Kenton} {and}
  \bibinfo{person}{Lee~Kristina Toutanova}.} \bibinfo{year}{2019}\natexlab{}.
\newblock \showarticletitle{BERT: Pre-training of Deep Bidirectional
  Transformers for Language Understanding}. In
  \bibinfo{booktitle}{\emph{Proceedings of NAACL-HLT}}.
  \bibinfo{pages}{4171--4186}.
\newblock


\bibitem[\protect\citeauthoryear{Khattab and Zaharia}{Khattab and
  Zaharia}{2020}]%
        {ColBERT}
\bibfield{author}{\bibinfo{person}{Omar Khattab} {and} \bibinfo{person}{Matei
  Zaharia}.} \bibinfo{year}{2020}\natexlab{}.
\newblock \showarticletitle{Colbert: Efficient and effective passage search via
  contextualized late interaction over bert}. In
  \bibinfo{booktitle}{\emph{Proceedings of the 43rd International ACM SIGIR
  Conference on Research and Development in Information Retrieval}}.
  \bibinfo{pages}{39--48}.
\newblock


\bibitem[\protect\citeauthoryear{Kingma and Ba}{Kingma and Ba}{2014}]%
        {kingma2014adam}
\bibfield{author}{\bibinfo{person}{Diederik~P Kingma} {and}
  \bibinfo{person}{Jimmy Ba}.} \bibinfo{year}{2014}\natexlab{}.
\newblock \showarticletitle{Adam: A method for stochastic optimization}.
\newblock \bibinfo{journal}{\emph{arXiv preprint arXiv:1412.6980}}
  (\bibinfo{year}{2014}).
\newblock


\bibitem[\protect\citeauthoryear{Koh, Sagawa, Marklund, Xie, Zhang,
  Balsubramani, Hu, Yasunaga, Phillips, Gao, et~al\mbox{.}}{Koh
  et~al\mbox{.}}{2021}]%
        {koh2020wilds}
\bibfield{author}{\bibinfo{person}{Pang~Wei Koh}, \bibinfo{person}{Shiori
  Sagawa}, \bibinfo{person}{Henrik Marklund}, \bibinfo{person}{Sang~Michael
  Xie}, \bibinfo{person}{Marvin Zhang}, \bibinfo{person}{Akshay Balsubramani},
  \bibinfo{person}{Weihua Hu}, \bibinfo{person}{Michihiro Yasunaga},
  \bibinfo{person}{Richard~Lanas Phillips}, \bibinfo{person}{Irena Gao},
  {et~al\mbox{.}}} \bibinfo{year}{2021}\natexlab{}.
\newblock \showarticletitle{Wilds: A benchmark of in-the-wild distribution
  shifts}. In \bibinfo{booktitle}{\emph{International Conference on Machine
  Learning}}. PMLR, \bibinfo{pages}{5637--5664}.
\newblock


\bibitem[\protect\citeauthoryear{Krovetz}{Krovetz}{2000}]%
        {krovetz_stemmer}
\bibfield{author}{\bibinfo{person}{Robert Krovetz}.}
  \bibinfo{year}{2000}\natexlab{}.
\newblock \showarticletitle{Viewing morphology as an inference process}.
\newblock \bibinfo{journal}{\emph{Artificial intelligence}}
  \bibinfo{volume}{118}, \bibinfo{number}{1-2} (\bibinfo{year}{2000}),
  \bibinfo{pages}{277--294}.
\newblock


\bibitem[\protect\citeauthoryear{Li, Ji, Du, Li, and Wang}{Li
  et~al\mbox{.}}{2019a}]%
        {20li2018textbugger}
\bibfield{author}{\bibinfo{person}{J Li}, \bibinfo{person}{S Ji},
  \bibinfo{person}{T Du}, \bibinfo{person}{B Li}, {and} \bibinfo{person}{T
  Wang}.} \bibinfo{year}{2019}\natexlab{a}.
\newblock \showarticletitle{TextBugger: Generating Adversarial Text Against
  Real-world Applications}. In \bibinfo{booktitle}{\emph{26th Annual Network
  and Distributed System Security Symposium}}.
\newblock


\bibitem[\protect\citeauthoryear{Li, Li, Wang, Zhang, and Gong}{Li
  et~al\mbox{.}}{2019b}]%
        {li2018nattack}
\bibfield{author}{\bibinfo{person}{Yandong Li}, \bibinfo{person}{Lijun Li},
  \bibinfo{person}{Liqiang Wang}, \bibinfo{person}{Tong Zhang}, {and}
  \bibinfo{person}{Boqing Gong}.} \bibinfo{year}{2019}\natexlab{b}.
\newblock \showarticletitle{Nattack: Learning the distributions of adversarial
  examples for an improved black-box attack on deep neural networks}. In
  \bibinfo{booktitle}{\emph{International Conference on Machine Learning}}.
  PMLR, \bibinfo{pages}{3866--3876}.
\newblock


\bibitem[\protect\citeauthoryear{Liang, Li, Su, Bian, Li, and Shi}{Liang
  et~al\mbox{.}}{2018}]%
        {26liang2017deep}
\bibfield{author}{\bibinfo{person}{Bin Liang}, \bibinfo{person}{Hongcheng Li},
  \bibinfo{person}{Miaoqiang Su}, \bibinfo{person}{Pan Bian},
  \bibinfo{person}{Xirong Li}, {and} \bibinfo{person}{Wenchang Shi}.}
  \bibinfo{year}{2018}\natexlab{}.
\newblock \showarticletitle{Deep text classification can be fooled}. In
  \bibinfo{booktitle}{\emph{Proceedings of the 27th International Joint
  Conference on Artificial Intelligence}}. \bibinfo{pages}{4208--4215}.
\newblock


\bibitem[\protect\citeauthoryear{Lin}{Lin}{2019}]%
        {neural_hype}
\bibfield{author}{\bibinfo{person}{Jimmy Lin}.}
  \bibinfo{year}{2019}\natexlab{}.
\newblock \showarticletitle{The neural hype and comparisons against weak
  baselines}. In \bibinfo{booktitle}{\emph{ACM SIGIR Forum}},
  Vol.~\bibinfo{volume}{52}. ACM New York, NY, USA, \bibinfo{pages}{40--51}.
\newblock


\bibitem[\protect\citeauthoryear{Liu}{Liu}{2011}]%
        {liu2011learning}
\bibfield{author}{\bibinfo{person}{Tie-Yan Liu}.}
  \bibinfo{year}{2011}\natexlab{}.
\newblock \showarticletitle{Learning to rank for information retrieval}.
\newblock  (\bibinfo{year}{2011}).
\newblock


\bibitem[\protect\citeauthoryear{Long, Yao, Dai, Tian, Luo, and Mei}{Long
  et~al\mbox{.}}{2018}]%
        {long2018deep}
\bibfield{author}{\bibinfo{person}{Fuchen Long}, \bibinfo{person}{Ting Yao},
  \bibinfo{person}{Qi Dai}, \bibinfo{person}{Xinmei Tian},
  \bibinfo{person}{Jiebo Luo}, {and} \bibinfo{person}{Tao Mei}.}
  \bibinfo{year}{2018}\natexlab{}.
\newblock \showarticletitle{Deep domain adaptation hashing with adversarial
  learning}. In \bibinfo{booktitle}{\emph{The 41st International ACM SIGIR
  Conference on Research \& Development in Information Retrieval}}.
  \bibinfo{pages}{725--734}.
\newblock


\bibitem[\protect\citeauthoryear{Ma, Guo, Zhang, Fan, Li, and Cheng}{Ma
  et~al\mbox{.}}{2021}]%
        {B-PROP}
\bibfield{author}{\bibinfo{person}{Xinyu Ma}, \bibinfo{person}{Jiafeng Guo},
  \bibinfo{person}{Ruqing Zhang}, \bibinfo{person}{Yixing Fan},
  \bibinfo{person}{Yingyan Li}, {and} \bibinfo{person}{Xueqi Cheng}.}
  \bibinfo{year}{2021}\natexlab{}.
\newblock \bibinfo{booktitle}{\emph{B-PROP: Bootstrapped Pre-Training with
  Representative Words Prediction for Ad-Hoc Retrieval}}.
\newblock \bibinfo{publisher}{Association for Computing Machinery},
  \bibinfo{address}{New York, NY, USA}, \bibinfo{pages}{1513–1522}.
\newblock
\urldef\tempurl%
\url{https://doi.org/10.1145/3404835.3462869}
\showURL{%
\tempurl}


\bibitem[\protect\citeauthoryear{Madry, Makelov, Schmidt, Tsipras, and
  Vladu}{Madry et~al\mbox{.}}{2018}]%
        {madry2017towards}
\bibfield{author}{\bibinfo{person}{Aleksander Madry},
  \bibinfo{person}{Aleksandar Makelov}, \bibinfo{person}{Ludwig Schmidt},
  \bibinfo{person}{Dimitris Tsipras}, {and} \bibinfo{person}{Adrian Vladu}.}
  \bibinfo{year}{2018}\natexlab{}.
\newblock \showarticletitle{Towards Deep Learning Models Resistant to
  Adversarial Attacks}. In \bibinfo{booktitle}{\emph{International Conference
  on Learning Representations}}.
\newblock


\bibitem[\protect\citeauthoryear{Mao, Shen, and Chung}{Mao
  et~al\mbox{.}}{2018}]%
        {mao2018deep}
\bibfield{author}{\bibinfo{person}{Sitong Mao}, \bibinfo{person}{Xiao Shen},
  {and} \bibinfo{person}{Fu-lai Chung}.} \bibinfo{year}{2018}\natexlab{}.
\newblock \showarticletitle{Deep domain adaptation based on multi-layer joint
  kernelized distance}. In \bibinfo{booktitle}{\emph{The 41st International ACM
  SIGIR Conference on Research \& Development in Information Retrieval}}.
  \bibinfo{pages}{1049--1052}.
\newblock


\bibitem[\protect\citeauthoryear{Minervini and Riedel}{Minervini and
  Riedel}{2018}]%
        {28minervini2018adversarially}
\bibfield{author}{\bibinfo{person}{Pasquale Minervini} {and}
  \bibinfo{person}{Sebastian Riedel}.} \bibinfo{year}{2018}\natexlab{}.
\newblock \showarticletitle{Adversarially Regularising Neural NLI Models to
  Integrate Logical Background Knowledge}. In
  \bibinfo{booktitle}{\emph{Proceedings of the 22nd Conference on Computational
  Natural Language Learning}}. \bibinfo{pages}{65--74}.
\newblock


\bibitem[\protect\citeauthoryear{Mitra, Diaz, and Craswell}{Mitra
  et~al\mbox{.}}{2017}]%
        {Duet}
\bibfield{author}{\bibinfo{person}{Bhaskar Mitra}, \bibinfo{person}{Fernando
  Diaz}, {and} \bibinfo{person}{Nick Craswell}.}
  \bibinfo{year}{2017}\natexlab{}.
\newblock \showarticletitle{Learning to match using local and distributed
  representations of text for web search}. In
  \bibinfo{booktitle}{\emph{Proceedings of the 26th International Conference on
  World Wide Web}}. \bibinfo{pages}{1291--1299}.
\newblock


\bibitem[\protect\citeauthoryear{Miyato, Dai, and Goodfellow}{Miyato
  et~al\mbox{.}}{2016}]%
        {miyato2016adversarial}
\bibfield{author}{\bibinfo{person}{Takeru Miyato}, \bibinfo{person}{Andrew~M
  Dai}, {and} \bibinfo{person}{Ian Goodfellow}.}
  \bibinfo{year}{2016}\natexlab{}.
\newblock \showarticletitle{Adversarial training methods for semi-supervised
  text classification}.
\newblock \bibinfo{journal}{\emph{arXiv preprint arXiv:1605.07725}}
  (\bibinfo{year}{2016}).
\newblock


\bibitem[\protect\citeauthoryear{Naseer, Khan, Rahman, and Porikli}{Naseer
  et~al\mbox{.}}{2018}]%
        {transferability}
\bibfield{author}{\bibinfo{person}{Muzammal Naseer}, \bibinfo{person}{Salman~H
  Khan}, \bibinfo{person}{Shafin Rahman}, {and} \bibinfo{person}{Fatih
  Porikli}.} \bibinfo{year}{2018}\natexlab{}.
\newblock \showarticletitle{Task-generalizable Adversarial Attack based on
  Perceptual Metric}.
\newblock \bibinfo{journal}{\emph{arXiv preprint arXiv:1811.09020}}
  (\bibinfo{year}{2018}).
\newblock


\bibitem[\protect\citeauthoryear{Nogueira and Cho}{Nogueira and Cho}{2019}]%
        {monoBERT}
\bibfield{author}{\bibinfo{person}{Rodrigo Nogueira} {and}
  \bibinfo{person}{Kyunghyun Cho}.} \bibinfo{year}{2019}\natexlab{}.
\newblock \showarticletitle{Passage Re-ranking with BERT}.
\newblock \bibinfo{journal}{\emph{arXiv preprint arXiv:1901.04085}}
  (\bibinfo{year}{2019}).
\newblock


\bibitem[\protect\citeauthoryear{Olson and Delen}{Olson and Delen}{2008}]%
        {Recall}
\bibfield{author}{\bibinfo{person}{David~L Olson} {and} \bibinfo{person}{Dursun
  Delen}.} \bibinfo{year}{2008}\natexlab{}.
\newblock \bibinfo{booktitle}{\emph{Advanced data mining techniques}}.
\newblock \bibinfo{publisher}{Springer Science \& Business Media}.
\newblock


\bibitem[\protect\citeauthoryear{Pang, Lan, Guo, Xu, Xu, and Cheng}{Pang
  et~al\mbox{.}}{2017}]%
        {deeprank}
\bibfield{author}{\bibinfo{person}{Liang Pang}, \bibinfo{person}{Yanyan Lan},
  \bibinfo{person}{Jiafeng Guo}, \bibinfo{person}{Jun Xu},
  \bibinfo{person}{Jingfang Xu}, {and} \bibinfo{person}{Xueqi Cheng}.}
  \bibinfo{year}{2017}\natexlab{}.
\newblock \showarticletitle{Deeprank: A new deep architecture for relevance
  ranking in information retrieval}. In \bibinfo{booktitle}{\emph{Proceedings
  of the 2017 ACM on Conference on Information and Knowledge Management}}.
  \bibinfo{pages}{257--266}.
\newblock


\bibitem[\protect\citeauthoryear{Papernot, McDaniel, Jha, Fredrikson, Celik,
  and Swami}{Papernot et~al\mbox{.}}{2016}]%
        {JSMA_0}
\bibfield{author}{\bibinfo{person}{Nicolas Papernot}, \bibinfo{person}{Patrick
  McDaniel}, \bibinfo{person}{Somesh Jha}, \bibinfo{person}{Matt Fredrikson},
  \bibinfo{person}{Z~Berkay Celik}, {and} \bibinfo{person}{Ananthram Swami}.}
  \bibinfo{year}{2016}\natexlab{}.
\newblock \showarticletitle{The limitations of deep learning in adversarial
  settings}. In \bibinfo{booktitle}{\emph{2016 IEEE European symposium on
  security and privacy (EuroS\&P)}}. IEEE, \bibinfo{pages}{372--387}.
\newblock


\bibitem[\protect\citeauthoryear{Penha, C{\^a}mara, and Hauff}{Penha
  et~al\mbox{.}}{2021}]%
        {penha2021evaluating}
\bibfield{author}{\bibinfo{person}{Gustavo Penha}, \bibinfo{person}{Arthur
  C{\^a}mara}, {and} \bibinfo{person}{Claudia Hauff}.}
  \bibinfo{year}{2021}\natexlab{}.
\newblock \showarticletitle{Evaluating the Robustness of Retrieval Pipelines
  with Query Variation Generators}.
\newblock \bibinfo{journal}{\emph{arXiv preprint arXiv:2111.13057}}
  (\bibinfo{year}{2021}).
\newblock


\bibitem[\protect\citeauthoryear{Pennington, Socher, and Manning}{Pennington
  et~al\mbox{.}}{2014}]%
        {pennington2014glove}
\bibfield{author}{\bibinfo{person}{Jeffrey Pennington},
  \bibinfo{person}{Richard Socher}, {and} \bibinfo{person}{Christopher~D
  Manning}.} \bibinfo{year}{2014}\natexlab{}.
\newblock \showarticletitle{Glove: Global vectors for word representation}. In
  \bibinfo{booktitle}{\emph{Proceedings of the 2014 conference on empirical
  methods in natural language processing (EMNLP)}}.
  \bibinfo{pages}{1532--1543}.
\newblock


\bibitem[\protect\citeauthoryear{Peters, Neumann, Iyyer, Gardner, Clark, Lee,
  and Zettlemoyer}{Peters et~al\mbox{.}}{2018}]%
        {elmo}
\bibfield{author}{\bibinfo{person}{Matthew~E. Peters}, \bibinfo{person}{Mark
  Neumann}, \bibinfo{person}{Mohit Iyyer}, \bibinfo{person}{Matt Gardner},
  \bibinfo{person}{Christopher Clark}, \bibinfo{person}{Kenton Lee}, {and}
  \bibinfo{person}{Luke Zettlemoyer}.} \bibinfo{year}{2018}\natexlab{}.
\newblock \showarticletitle{Deep Contextualized Word Representations}. In
  \bibinfo{booktitle}{\emph{Proceedings of the 2018 Conference of the North
  {A}merican Chapter of the Association for Computational Linguistics: Human
  Language Technologies, Volume 1 (Long Papers)}}.
  \bibinfo{publisher}{Association for Computational Linguistics},
  \bibinfo{address}{New Orleans, Louisiana}, \bibinfo{pages}{2227--2237}.
\newblock
\urldef\tempurl%
\url{https://doi.org/10.18653/v1/N18-1202}
\showDOI{\tempurl}


\bibitem[\protect\citeauthoryear{Pruthi, Dhingra, and Lipton}{Pruthi
  et~al\mbox{.}}{2019}]%
        {pruthi2019combating}
\bibfield{author}{\bibinfo{person}{Danish Pruthi}, \bibinfo{person}{Bhuwan
  Dhingra}, {and} \bibinfo{person}{Zachary~C Lipton}.}
  \bibinfo{year}{2019}\natexlab{}.
\newblock \showarticletitle{Combating Adversarial Misspellings with Robust Word
  Recognition}. In \bibinfo{booktitle}{\emph{Proceedings of the 57th Annual
  Meeting of the Association for Computational Linguistics}}.
  \bibinfo{pages}{5582--5591}.
\newblock


\bibitem[\protect\citeauthoryear{Qin and Liu}{Qin and Liu}{2013}]%
        {letor4.0}
\bibfield{author}{\bibinfo{person}{Tao Qin} {and} \bibinfo{person}{Tie-Yan
  Liu}.} \bibinfo{year}{2013}\natexlab{}.
\newblock \showarticletitle{Introducing LETOR 4.0 datasets}.
\newblock \bibinfo{journal}{\emph{arXiv preprint arXiv:1306.2597}}
  (\bibinfo{year}{2013}).
\newblock


\bibitem[\protect\citeauthoryear{Qin, Yan, Zhuang, Tay, Pasumarthi, Wang,
  Bendersky, and Najork}{Qin et~al\mbox{.}}{2020}]%
        {qin2021neural}
\bibfield{author}{\bibinfo{person}{Zhen Qin}, \bibinfo{person}{Le Yan},
  \bibinfo{person}{Honglei Zhuang}, \bibinfo{person}{Yi Tay},
  \bibinfo{person}{Rama~Kumar Pasumarthi}, \bibinfo{person}{Xuanhui Wang},
  \bibinfo{person}{Michael Bendersky}, {and} \bibinfo{person}{Marc Najork}.}
  \bibinfo{year}{2020}\natexlab{}.
\newblock \showarticletitle{Are Neural Rankers still Outperformed by Gradient
  Boosted Decision Trees?}. In \bibinfo{booktitle}{\emph{International
  Conference on Learning Representations}}.
\newblock


\bibitem[\protect\citeauthoryear{Radev, Qi, Wu, and Fan}{Radev
  et~al\mbox{.}}{2002}]%
        {MRR}
\bibfield{author}{\bibinfo{person}{Dragomir~R Radev}, \bibinfo{person}{Hong
  Qi}, \bibinfo{person}{Harris Wu}, {and} \bibinfo{person}{Weiguo Fan}.}
  \bibinfo{year}{2002}\natexlab{}.
\newblock \showarticletitle{Evaluating Web-based Question Answering Systems.}.
  In \bibinfo{booktitle}{\emph{LREC}}. Citeseer.
\newblock


\bibitem[\protect\citeauthoryear{Raifer, Raiber, Tennenholtz, and
  Kurland}{Raifer et~al\mbox{.}}{2017}]%
        {raifer2017SIGIR}
\bibfield{author}{\bibinfo{person}{Nimrod Raifer}, \bibinfo{person}{Fiana
  Raiber}, \bibinfo{person}{Moshe Tennenholtz}, {and} \bibinfo{person}{Oren
  Kurland}.} \bibinfo{year}{2017}\natexlab{}.
\newblock \showarticletitle{Information retrieval meets game theory: The
  ranking competition between documents' authors}. In
  \bibinfo{booktitle}{\emph{Proceedings of the 40th International ACM SIGIR
  Conference on Research and Development in Information Retrieval}}.
  \bibinfo{pages}{465--474}.
\newblock


\bibitem[\protect\citeauthoryear{Rawlinson}{Rawlinson}{2007}]%
        {rawlinson1976psychology}
\bibfield{author}{\bibinfo{person}{Graham Rawlinson}.}
  \bibinfo{year}{2007}\natexlab{}.
\newblock \showarticletitle{The significance of letter position in word
  recognition}.
\newblock \bibinfo{journal}{\emph{IEEE Aerospace and Electronic Systems
  Magazine}} \bibinfo{volume}{22}, \bibinfo{number}{1} (\bibinfo{year}{2007}),
  \bibinfo{pages}{26--27}.
\newblock


\bibitem[\protect\citeauthoryear{Robertson and Walker}{Robertson and
  Walker}{1994}]%
        {bm25}
\bibfield{author}{\bibinfo{person}{Stephen~E Robertson} {and}
  \bibinfo{person}{Steve Walker}.} \bibinfo{year}{1994}\natexlab{}.
\newblock \showarticletitle{Some simple effective approximations to the
  2-poisson model for probabilistic weighted retrieval}. In
  \bibinfo{booktitle}{\emph{SIGIR'94}}. Springer, \bibinfo{pages}{232--241}.
\newblock


\bibitem[\protect\citeauthoryear{Samanta and Mehta}{Samanta and Mehta}{2017}]%
        {samanta2017towards}
\bibfield{author}{\bibinfo{person}{Suranjana Samanta} {and}
  \bibinfo{person}{Sameep Mehta}.} \bibinfo{year}{2017}\natexlab{}.
\newblock \showarticletitle{Towards crafting text adversarial samples}.
\newblock \bibinfo{journal}{\emph{arXiv preprint arXiv:1707.02812}}
  (\bibinfo{year}{2017}).
\newblock


\bibitem[\protect\citeauthoryear{Sato, Suzuki, Shindo, and Matsumoto}{Sato
  et~al\mbox{.}}{2018}]%
        {25sato2018interpretable}
\bibfield{author}{\bibinfo{person}{Motoki Sato}, \bibinfo{person}{Jun Suzuki},
  \bibinfo{person}{Hiroyuki Shindo}, {and} \bibinfo{person}{Yuji Matsumoto}.}
  \bibinfo{year}{2018}\natexlab{}.
\newblock \showarticletitle{Interpretable Adversarial Perturbation in Input
  Embedding Space for Text}. In \bibinfo{booktitle}{\emph{IJCAI}}.
\newblock


\bibitem[\protect\citeauthoryear{Sch{\"u}tze, Manning, and
  Raghavan}{Sch{\"u}tze et~al\mbox{.}}{2008}]%
        {MAP}
\bibfield{author}{\bibinfo{person}{Hinrich Sch{\"u}tze},
  \bibinfo{person}{Christopher~D Manning}, {and} \bibinfo{person}{Prabhakar
  Raghavan}.} \bibinfo{year}{2008}\natexlab{}.
\newblock \bibinfo{booktitle}{\emph{Introduction to information retrieval}}.
  Vol.~\bibinfo{volume}{39}.
\newblock \bibinfo{publisher}{Cambridge University Press Cambridge}.
\newblock


\bibitem[\protect\citeauthoryear{Shafique, Naseer, Theocharides, Kyrkou, Mutlu,
  Orosa, and Choi}{Shafique et~al\mbox{.}}{2020}]%
        {shafique2020robust}
\bibfield{author}{\bibinfo{person}{Muhammad Shafique}, \bibinfo{person}{Mahum
  Naseer}, \bibinfo{person}{Theocharis Theocharides}, \bibinfo{person}{Christos
  Kyrkou}, \bibinfo{person}{Onur Mutlu}, \bibinfo{person}{Lois Orosa}, {and}
  \bibinfo{person}{Jungwook Choi}.} \bibinfo{year}{2020}\natexlab{}.
\newblock \showarticletitle{Robust machine learning systems: Challenges,
  current trends, perspectives, and the road ahead}.
\newblock \bibinfo{journal}{\emph{IEEE Design \& Test}} \bibinfo{volume}{37},
  \bibinfo{number}{2} (\bibinfo{year}{2020}), \bibinfo{pages}{30--57}.
\newblock


\bibitem[\protect\citeauthoryear{Shivaswamy and Chandrashekar}{Shivaswamy and
  Chandrashekar}{2021}]%
        {shivaswamy2021bias}
\bibfield{author}{\bibinfo{person}{Pannaga Shivaswamy} {and}
  \bibinfo{person}{Ashok Chandrashekar}.} \bibinfo{year}{2021}\natexlab{}.
\newblock \showarticletitle{Bias-Variance Decomposition for Ranking}. In
  \bibinfo{booktitle}{\emph{Proceedings of the 14th ACM International
  Conference on Web Search and Data Mining}}. \bibinfo{pages}{472--480}.
\newblock


\bibitem[\protect\citeauthoryear{Szegedy, Zaremba, Sutskever, Estrach, Erhan,
  Goodfellow, and Fergus}{Szegedy et~al\mbox{.}}{2014}]%
        {szegedy2014ICLR}
\bibfield{author}{\bibinfo{person}{Christian Szegedy},
  \bibinfo{person}{Wojciech Zaremba}, \bibinfo{person}{Ilya Sutskever},
  \bibinfo{person}{Joan~Bruna Estrach}, \bibinfo{person}{Dumitru Erhan},
  \bibinfo{person}{Ian Goodfellow}, {and} \bibinfo{person}{Robert Fergus}.}
  \bibinfo{year}{2014}\natexlab{}.
\newblock \showarticletitle{Intriguing properties of neural networks}. In
  \bibinfo{booktitle}{\emph{2nd International Conference on Learning
  Representations, ICLR 2014}}.
\newblock


\bibitem[\protect\citeauthoryear{Torralba and Efros}{Torralba and
  Efros}{2011}]%
        {torralba2011unbiased}
\bibfield{author}{\bibinfo{person}{Antonio Torralba} {and}
  \bibinfo{person}{Alexei~A Efros}.} \bibinfo{year}{2011}\natexlab{}.
\newblock \showarticletitle{Unbiased look at dataset bias}. In
  \bibinfo{booktitle}{\emph{CVPR 2011}}. IEEE, \bibinfo{pages}{1521--1528}.
\newblock


\bibitem[\protect\citeauthoryear{Tripathi, Muhr, Brunner, Jodlbauer, Dehmer,
  and Emmert-Streib}{Tripathi et~al\mbox{.}}{2021}]%
        {ml_complex_robust}
\bibfield{author}{\bibinfo{person}{Shailesh Tripathi}, \bibinfo{person}{David
  Muhr}, \bibinfo{person}{Manuel Brunner}, \bibinfo{person}{Herbert Jodlbauer},
  \bibinfo{person}{Matthias Dehmer}, {and} \bibinfo{person}{Frank
  Emmert-Streib}.} \bibinfo{year}{2021}\natexlab{}.
\newblock \showarticletitle{Ensuring the Robustness and Reliability of
  Data-Driven Knowledge Discovery Models in Production and Manufacturing}.
\newblock \bibinfo{journal}{\emph{Frontiers in Artificial Intelligence}}
  \bibinfo{volume}{4} (\bibinfo{year}{2021}), \bibinfo{pages}{22}.
\newblock


\bibitem[\protect\citeauthoryear{Voorhees}{Voorhees}{2003}]%
        {trec2003overview}
\bibfield{author}{\bibinfo{person}{Ellen~M. Voorhees}.}
  \bibinfo{year}{2003}\natexlab{}.
\newblock \showarticletitle{Overview of the {TREC} 2003 Robust Retrieval
  Track}. In \bibinfo{booktitle}{\emph{Proceedings of The Twelfth Text
  REtrieval Conference, {TREC} 2003, Gaithersburg, Maryland, USA, November
  18-21, 2003}}.
\newblock


\bibitem[\protect\citeauthoryear{Voorhees}{Voorhees}{2004}]%
        {trec2004overview}
\bibfield{author}{\bibinfo{person}{Ellen~M. Voorhees}.}
  \bibinfo{year}{2004}\natexlab{}.
\newblock \showarticletitle{Overview of the {TREC} 2004 Robust Track}. In
  \bibinfo{booktitle}{\emph{Proceedings of the Thirteenth Text REtrieval
  Conference, {TREC} 2004, Gaithersburg, Maryland, USA, November 16-19, 2004}}.
\newblock


\bibitem[\protect\citeauthoryear{Voorhees}{Voorhees}{2006}]%
        {trec2005overview}
\bibfield{author}{\bibinfo{person}{Ellen~M. Voorhees}.}
  \bibinfo{year}{2006}\natexlab{}.
\newblock \showarticletitle{Overview of the TREC 2005 Robust Retrieval Track}.
\newblock \bibinfo{journal}{\emph{TREC 2005 Proceedings}}
  (\bibinfo{year}{2006}).
\newblock


\bibitem[\protect\citeauthoryear{Wang and Pedersen}{Wang and Pedersen}{2011}]%
        {wang2011review}
\bibfield{author}{\bibinfo{person}{Kuansan Wang} {and} \bibinfo{person}{Jan
  Pedersen}.} \bibinfo{year}{2011}\natexlab{}.
\newblock \showarticletitle{Review of MSR-Bing web scale speller challenge}. In
  \bibinfo{booktitle}{\emph{Proceedings of the 34th international ACM SIGIR
  conference on Research and development in Information Retrieval}}.
  \bibinfo{pages}{1339--1340}.
\newblock


\bibitem[\protect\citeauthoryear{Wiyatno and Xu}{Wiyatno and Xu}{2018}]%
        {JSMA_1}
\bibfield{author}{\bibinfo{person}{Rey Wiyatno} {and} \bibinfo{person}{Anqi
  Xu}.} \bibinfo{year}{2018}\natexlab{}.
\newblock \showarticletitle{Maximal jacobian-based saliency map attack}.
\newblock \bibinfo{journal}{\emph{arXiv preprint arXiv:1808.07945}}
  (\bibinfo{year}{2018}).
\newblock


\bibitem[\protect\citeauthoryear{Wu, Schuster, Chen, Le, Norouzi, Macherey,
  Krikun, Cao, Gao, Macherey, et~al\mbox{.}}{Wu et~al\mbox{.}}{2016}]%
        {WordPiece}
\bibfield{author}{\bibinfo{person}{Yonghui Wu}, \bibinfo{person}{Mike
  Schuster}, \bibinfo{person}{Zhifeng Chen}, \bibinfo{person}{Quoc~V Le},
  \bibinfo{person}{Mohammad Norouzi}, \bibinfo{person}{Wolfgang Macherey},
  \bibinfo{person}{Maxim Krikun}, \bibinfo{person}{Yuan Cao},
  \bibinfo{person}{Qin Gao}, \bibinfo{person}{Klaus Macherey}, {et~al\mbox{.}}}
  \bibinfo{year}{2016}\natexlab{}.
\newblock \showarticletitle{Google's neural machine translation system:
  Bridging the gap between human and machine translation}.
\newblock \bibinfo{journal}{\emph{arXiv preprint arXiv:1609.08144}}
  (\bibinfo{year}{2016}).
\newblock


\bibitem[\protect\citeauthoryear{Xiong, Dai, Callan, Liu, and Power}{Xiong
  et~al\mbox{.}}{2017}]%
        {KNRM}
\bibfield{author}{\bibinfo{person}{Chenyan Xiong}, \bibinfo{person}{Zhuyun
  Dai}, \bibinfo{person}{Jamie Callan}, \bibinfo{person}{Zhiyuan Liu}, {and}
  \bibinfo{person}{Russell Power}.} \bibinfo{year}{2017}\natexlab{}.
\newblock \showarticletitle{End-to-end neural ad-hoc ranking with kernel
  pooling}. In \bibinfo{booktitle}{\emph{Proceedings of the 40th International
  ACM SIGIR conference on research and development in information retrieval}}.
  \bibinfo{pages}{55--64}.
\newblock


\bibitem[\protect\citeauthoryear{Xu, Zhong, Yepes, and Lau}{Xu
  et~al\mbox{.}}{2021}]%
        {xu2021grey}
\bibfield{author}{\bibinfo{person}{Ying Xu}, \bibinfo{person}{Xu Zhong},
  \bibinfo{person}{Antonio~Jimeno Yepes}, {and} \bibinfo{person}{Jey~Han Lau}.}
  \bibinfo{year}{2021}\natexlab{}.
\newblock \showarticletitle{Grey-box Adversarial Attack And Defence For
  Sentiment Classification}. In \bibinfo{booktitle}{\emph{Proceedings of the
  2021 Conference of the North American Chapter of the Association for
  Computational Linguistics: Human Language Technologies}}.
  \bibinfo{pages}{4078--4087}.
\newblock


\bibitem[\protect\citeauthoryear{Yang, Fang, and Lin}{Yang
  et~al\mbox{.}}{2018}]%
        {yang2018anserini}
\bibfield{author}{\bibinfo{person}{Peilin Yang}, \bibinfo{person}{Hui Fang},
  {and} \bibinfo{person}{Jimmy Lin}.} \bibinfo{year}{2018}\natexlab{}.
\newblock \showarticletitle{Anserini: Reproducible ranking baselines using
  Lucene}.
\newblock \bibinfo{journal}{\emph{Journal of Data and Information Quality
  (JDIQ)}} \bibinfo{volume}{10}, \bibinfo{number}{4} (\bibinfo{year}{2018}),
  \bibinfo{pages}{1--20}.
\newblock


\bibitem[\protect\citeauthoryear{Yang, Xie, Tan, Xiong, Li, and Lin}{Yang
  et~al\mbox{.}}{2019}]%
        {bert_qa_app}
\bibfield{author}{\bibinfo{person}{Wei Yang}, \bibinfo{person}{Yuqing Xie},
  \bibinfo{person}{Luchen Tan}, \bibinfo{person}{Kun Xiong},
  \bibinfo{person}{Ming Li}, {and} \bibinfo{person}{Jimmy Lin}.}
  \bibinfo{year}{2019}\natexlab{}.
\newblock \showarticletitle{Data augmentation for bert fine-tuning in
  open-domain question answering}.
\newblock \bibinfo{journal}{\emph{arXiv preprint arXiv:1904.06652}}
  (\bibinfo{year}{2019}).
\newblock


\bibitem[\protect\citeauthoryear{Yates, Arora, Zhang, Yang, Jose, and
  Lin}{Yates et~al\mbox{.}}{2020}]%
        {yates2020capreolus}
\bibfield{author}{\bibinfo{person}{Andrew Yates}, \bibinfo{person}{Siddhant
  Arora}, \bibinfo{person}{Xinyu Zhang}, \bibinfo{person}{Wei Yang},
  \bibinfo{person}{Kevin~Martin Jose}, {and} \bibinfo{person}{Jimmy Lin}.}
  \bibinfo{year}{2020}\natexlab{}.
\newblock \showarticletitle{Capreolus: A toolkit for end-to-end neural ad hoc
  retrieval}. In \bibinfo{booktitle}{\emph{Proceedings of the 13th
  International Conference on Web Search and Data Mining}}.
  \bibinfo{pages}{861--864}.
\newblock


\bibitem[\protect\citeauthoryear{Yilmaz, Yang, Zhang, and Lin}{Yilmaz
  et~al\mbox{.}}{2019}]%
        {yilmaz2019cross}
\bibfield{author}{\bibinfo{person}{Zeynep~Akkalyoncu Yilmaz},
  \bibinfo{person}{Wei Yang}, \bibinfo{person}{Haotian Zhang}, {and}
  \bibinfo{person}{Jimmy Lin}.} \bibinfo{year}{2019}\natexlab{}.
\newblock \showarticletitle{Cross-domain modeling of sentence-level evidence
  for document retrieval}. In \bibinfo{booktitle}{\emph{Proceedings of the 2019
  Conference on Empirical Methods in Natural Language Processing and the 9th
  International Joint Conference on Natural Language Processing
  (EMNLP-IJCNLP)}}. \bibinfo{pages}{3481--3487}.
\newblock


\bibitem[\protect\citeauthoryear{Zhai and Lafferty}{Zhai and Lafferty}{2017}]%
        {ql_dir}
\bibfield{author}{\bibinfo{person}{Chengxiang Zhai} {and} \bibinfo{person}{John
  Lafferty}.} \bibinfo{year}{2017}\natexlab{}.
\newblock \showarticletitle{A study of smoothing methods for language models
  applied to ad hoc information retrieval}. In \bibinfo{booktitle}{\emph{ACM
  SIGIR Forum}}, Vol.~\bibinfo{volume}{51}. ACM New York, NY, USA,
  \bibinfo{pages}{268--276}.
\newblock


\bibitem[\protect\citeauthoryear{Zhang, Harman, Ma, and Liu}{Zhang
  et~al\mbox{.}}{2020}]%
        {zhang2020robustnessml}
\bibfield{author}{\bibinfo{person}{Jie~M Zhang}, \bibinfo{person}{Mark Harman},
  \bibinfo{person}{Lei Ma}, {and} \bibinfo{person}{Yang Liu}.}
  \bibinfo{year}{2020}\natexlab{}.
\newblock \showarticletitle{Machine learning testing: Survey, landscapes and
  horizons}.
\newblock \bibinfo{journal}{\emph{IEEE Transactions on Software Engineering}}
  (\bibinfo{year}{2020}).
\newblock


\bibitem[\protect\citeauthoryear{Zhang, Song, Wang, and Hou}{Zhang
  et~al\mbox{.}}{2013}]%
        {zhang2013bias}
\bibfield{author}{\bibinfo{person}{Peng Zhang}, \bibinfo{person}{Dawei Song},
  \bibinfo{person}{Jun Wang}, {and} \bibinfo{person}{Yuexian Hou}.}
  \bibinfo{year}{2013}\natexlab{}.
\newblock \showarticletitle{Bias-variance decomposition of ir evaluation}. In
  \bibinfo{booktitle}{\emph{Proceedings of the 36th international ACM SIGIR
  conference on Research and development in information retrieval}}.
  \bibinfo{pages}{1021--1024}.
\newblock


\bibitem[\protect\citeauthoryear{Zhang, Song, Wang, and Hou}{Zhang
  et~al\mbox{.}}{2014}]%
        {zhang2014bias_qikan}
\bibfield{author}{\bibinfo{person}{Peng Zhang}, \bibinfo{person}{Dawei Song},
  \bibinfo{person}{Jun Wang}, {and} \bibinfo{person}{Yuexian Hou}.}
  \bibinfo{year}{2014}\natexlab{}.
\newblock \showarticletitle{Bias--variance analysis in estimating true query
  model for information retrieval}.
\newblock \bibinfo{journal}{\emph{Information processing \& management}}
  \bibinfo{volume}{50}, \bibinfo{number}{1} (\bibinfo{year}{2014}),
  \bibinfo{pages}{199--217}.
\newblock


\bibitem[\protect\citeauthoryear{Zhou, Wang, Niu, Zhang, Xu, Zheng, and
  Hua}{Zhou et~al\mbox{.}}{2021}]%
        {zhou2021practical}
\bibfield{author}{\bibinfo{person}{Mo Zhou}, \bibinfo{person}{Le Wang},
  \bibinfo{person}{Zhenxing Niu}, \bibinfo{person}{Qilin Zhang},
  \bibinfo{person}{Yinghui Xu}, \bibinfo{person}{Nanning Zheng}, {and}
  \bibinfo{person}{Gang Hua}.} \bibinfo{year}{2021}\natexlab{}.
\newblock \showarticletitle{Practical Relative Order Attack in Deep Ranking}.
  In \bibinfo{booktitle}{\emph{Proceedings of the IEEE/CVF International
  Conference on Computer Vision}}. \bibinfo{pages}{16413--16422}.
\newblock


\bibitem[\protect\citeauthoryear{Zhou and Croft}{Zhou and Croft}{2006}]%
        {cikm06_ranking_robustness}
\bibfield{author}{\bibinfo{person}{Yun Zhou} {and} \bibinfo{person}{W~Bruce
  Croft}.} \bibinfo{year}{2006}\natexlab{}.
\newblock \showarticletitle{Ranking robustness: a novel framework to predict
  query performance}. In \bibinfo{booktitle}{\emph{Proceedings of the 15th ACM
  international conference on Information and knowledge management}}.
  \bibinfo{pages}{567--574}.
\newblock


\bibitem[\protect\citeauthoryear{Zhuang and Zuccon}{Zhuang and Zuccon}{2021}]%
        {zhuang2021dealing}
\bibfield{author}{\bibinfo{person}{Shengyao Zhuang} {and}
  \bibinfo{person}{Guido Zuccon}.} \bibinfo{year}{2021}\natexlab{}.
\newblock \showarticletitle{Dealing with Typos for BERT-based Passage Retrieval
  and Ranking}. In \bibinfo{booktitle}{\emph{Proceedings of the 2021 Conference
  on Empirical Methods in Natural Language Processing}}.
  \bibinfo{pages}{2836--2842}.
\newblock


\bibitem[\protect\citeauthoryear{Zighelnic and Kurland}{Zighelnic and
  Kurland}{2008}]%
        {zighelnic2008query}
\bibfield{author}{\bibinfo{person}{Liron Zighelnic} {and} \bibinfo{person}{Oren
  Kurland}.} \bibinfo{year}{2008}\natexlab{}.
\newblock \showarticletitle{Query-drift prevention for robust query expansion}.
  In \bibinfo{booktitle}{\emph{Proceedings of the 31st annual international ACM
  SIGIR conference on Research and development in information retrieval}}.
  \bibinfo{pages}{825--826}.
\newblock


\end{thebibliography}

%
%
%
%
%
%
%
%

\end{document}